%% file: draft.tex
\documentclass[11pt,a4paper]{article}
\usepackage[utf8]{inputenc}
\usepackage[left=2.3cm,right=2.3cm,top=3cm,bottom=3.5cm]{geometry}

\usepackage{amsmath}
\usepackage{amsfonts}
\usepackage{amssymb}
\usepackage{mathrsfs}
\usepackage{mathtools}
\usepackage{braket}
\usepackage{slashed}

\usepackage{array}
\usepackage{pifont}
\newcommand{\xmark}{\ding{55}}

\usepackage{graphicx}

\usepackage{xcolor}
\usepackage{color}
\definecolor{nicered}{rgb}{0.7,0.1,0.1}
\definecolor{nicegreen}{rgb}{0.1,0.5,.1}

\usepackage{hyperref}
\hypersetup{colorlinks,citecolor=nicegreen,linkcolor=nicered}
\hypersetup{colorlinks=true}

\usepackage{subcaption}

\usepackage{stackrel}

\usepackage{multirow}

\usepackage{float}

\usepackage{authblk}

\usepackage{comment}
\usepackage{bm}
\usepackage[numbers, sort&compress]{natbib}
\usepackage{url}

\usepackage[compat=1.1.0]{tikz-feynman}

\usepackage{relsize}

\newcommand{\nl}{{\mathnormal l}}
\newcommand{\R}{{\mathcal R}}

\newcommand{\cD}{\mathcal{D}}
\newcommand{\cO}{\mathcal{O}}

\title{
The Electric Dipole Moment of the electron in the decoupling limit of the aligned Two-Higgs Doublet Model
}
\author[a]{Juan~Manuel~D\'{a}vila\thanks{juandai@ific.uv.es}}
\author[a]{Anirban~Karan\thanks{kanirban@ific.uv.es}}
\author[a, b]{Emilie~Passemar\thanks{passemar@ific.uv.es}}
\author[a]{Antonio~Pich\thanks{pich@ific.uv.es}}
\author[c]{Luiz~Vale~Silva\thanks{luiz.valesilva@uchceu.es}}

\affil[a]{\it Departament de F\'{i}sica Te\`{o}rica, Instituto de F\'{i}sica Corpuscular,

Universitat de Val\`encia -- Consejo Superior de Investigaciones Cient\'{i}ficas,

Parc Cient\'{i}fic, Catedr\'{a}tico Jos\'{e} Beltr\'{a}n 2, E-46980 Paterna, Valencia, Spain}

\affil[b]{\it Physics Department, Indiana University, Bloomington, Indiana 47405, USA}

\affil[c]{\it Departamento de Matem\'{a}ticas, F\'{i}sica y Ciencias Tecnol\'{o}gicas,

Universidad Cardenal Herrera-CEU, CEU Universities,

46115 Alfara del Patriarca, Val\`{e}ncia, Spain}

\date{}

\tikzfeynmanset{/tikzfeynman/warn luatex=false}

\newcommand\aprx{\stackrel{\mathclap{\normalfont\mbox{\footnotesize{dec.}}}}{\mathlarger{\approx}}}

\allowdisplaybreaks

\begin{document}

\maketitle

\noindent
\textbf{Abstract.}
We present a discussion of model-independent contributions to the EDM of the electron. 
We focus on those contributions that emerge from a heavy scalar sector that is linearly realized. In particular, we explore the decoupling limit of the aligned 2HDM. 
In this model, Barr-Zee diagrams with a fermion loop produce logarithmically-enhanced contributions that are proportional to potentially large new sources of CP violation. In the decoupling limit these contributions are generated by effective dimension-6 operators via the mixing of four-fermion operators into electroweak dipole operators.
These logarithmic contributions are not present in more constrained versions of the 2HDM where a $\mathcal Z_2$ symmetry is imposed, which then controls the basis of effective operators needed to describe the new physics contributions to the electron EDM. Thus, the $\mathcal Z_2$ symmetry provides a suppression mechanism.
In the course of the comparison of the results from the aligned 2HDM with the leading logarithms from SMEFT, we needed to specify or correct signs of expressions found in the literature.
We then study how the experimental bounds on the electron EDM constrain the set of parameters of the aligned 2HDM.

\section{Introduction}

Phenomena sensitive to Charge-Parity (CP) violation provide a powerful test of the Standard Model (SM) structure, both its gauge and matter field contents and properties.
Electric Dipole Moments (EDMs) play a crucial role in searching for
New Physics (NP) sources of CP violation, 
since experimental sensitivities have achieved exquisite levels, and SM contributions are substantially suppressed.
Indeed, SM contributions to the electron EDM (eEDM) or quark EDMs do not appear at two loops in the perturbative expansion of electroweak (EW) and strong couplings, being further suppressed in the loop counting.
For an introduction to the physics of EDMs, see for instance the reviews in Refs.~\cite{Bernreuther:1990jx,Pospelov:2005pr,Engel:2013lsa}.

NP couplings that display violation of CP symmetry can be investigated in a model-independent way under broad assumptions, when the underlying NP sector is heavy, by exploiting the SM Effective Field Theory (SMEFT),
see e.g. Refs.~\cite{Dwivedi:2015nta,Ferreira:2016jea,Cirigliano:2019vfc,Haisch:2019xyi,Rossia:2024rfo,Asteriadis:2024xts,Thomas:2024dwd} where operators of the (schematic) classes $X^3$ and $X^2 H^2$ are studied, $X$ being a field strength tensor and $H$ the SM-like Higgs doublet field.
Sticking to operators of dimension 6 and beyond the bosonic operators discussed in the latter references, other operators are also relevant when addressing CP-violating phenomena, namely, $ \psi^2 H^3$, $\psi^2 X H$, $ \psi^2 H^2 D $ and $\psi^4 $, where $D$ is the covariant derivative and $\psi$ a fermion. (We implicitly refer to operators of the Warsaw basis \cite{Grzadkowski:2010es}.)
The impact of EDMs in constraining the Wilson coefficients of these operators has been the subject of numerous studies \cite{Hisano:2012cc,Alonso:2013hga,Engel:2013lsa,Chien:2015xha,Cirigliano:2016njn,Cirigliano:2016nyn,Eeg:2016fsy,Panico:2018hal,Jager:2019wkc,Silva:2020vyx,Aebischer:2021uvt,Haisch:2021hcg,Kley:2021yhn,Brod:2022bww,Dermisek:2023nhe,Fajfer:2023gie,Miralles:2024huv}; in much the same spirit of studying the category of $ \psi^2 H^3$ operators, one can investigate beyond-the-SM couplings of dimension 4 of the SM-like Higgs to fermions \cite{Brod:2013cka,Brod:2018pli,Brod:2018lbf,Brod:2023wsh}.

Beyond a purely effective description, a new scalar sector introduces a rich phenomenology in the context of CP violation. In particular, the Kobayashi-Maskawa picture can change substantially due to new complex phases entering from the scalar sector, see e.g. Ref.~\cite{Nierste:2019fbx}. Multiple contributions of the scalar sector to EDMs are possible. It is well known that two-loop contributions can be more important than one-loop contributions due to the extra Yukawa suppression in the latter case \cite{Weinberg:1989dx,Barr:1990vd}.
Various extensions of the SM introduce new scalars that can contribute to EDMs, such as Two-Higgs Doublet Models (2HDMs), Supersymmetric Models \cite{Pilaftsis:2002fe,Demir:2003js,Ning:2025zfh}, Left-Right Models \cite{Zhang:2007da,Bernard:2015boz,Bertolini:2019out,Bernard:2020cyi}, minimal extensions with leptoquarks \cite{Dekens:2018bci}, and the Manohar-Wise Model \cite{Gisbert:2021htg}.
When there is a large separation of energy scales between the EW scale and the characteristic energy scale of NP, the
Effective Field Theory (EFT) induced by the heavy scalar sector of the theory can be exploited; see Refs.~\cite{Davidson:2016utf,Panico:2018hal,Altmannshofer:2020shb} for discussions analogous to the one performed hereafter.

The 2HDM \cite{Branco:2011iw,Gunion:1989we,Ivanov:2017dad}, which extends the SM field content with an additional scalar doublet, is one of the simplest extensions of the SM. Due to the presence of three neutral and one pair of charged scalars, the 2HDM exhibits very rich phenomenology such as dark matter aspects \cite{LopezHonorez:2006gr,Belyaev:2016lok,Tsai:2019eqi}, new sources of CP violation \cite{Gunion:2005ja,Wu:1994ja,Keus:2015hva,Chen:2017com,Iguro:2019zlc,Chun:2019oix,Cheung:2020ugr,Darvishi:2023fjh}, axion-like phenomenology \cite{Kim:1986ax, Espriu:2015mfa, Celis:2014zaa}, neutrino mass generation \cite{Ma:2006km, Hirsch:2013ola}, electroweak baryogenesis \cite{Turok:1990zg, Cline:2011mm,Fuyuto:2015jha,Enomoto:2021dkl}, stability of the scalar potential \cite{Ferreira:2015rha,Das:2015mwa,Schuh:2018hig}, etc. Moreover, it can provide an effective low-energy framework for various models with larger symmetry groups (e.g. supersymmetry).

In general, the 2HDM suffers from tree-level Flavour-Changing Neutral Currents (FCNCs), which are tightly constrained experimentally. Usually, a discrete $\mathcal Z_2$ symmetry is imposed on the Lagrangian so that each type of right-handed (quark or lepton) fermion only couples to one scalar doublet, and so tree-level FCNCs vanish \cite{Glashow:1976nt,Paschos:1976ay}. Nevertheless, the `Aligned Two-Higgs Doublet Model' (A2HDM) solves this issue with a much weaker requirement: flavour alignment of Yukawa couplings, i.e. the Yukawa interactions of both scalar doublets have the same structure in flavour space 
\cite{Pich:2009sp, Pich:2010ic}. In this model highly suppressed FCNCs appear at higher perturbative order through minimal flavour violation only, which makes the model \textit{radiatively secure} \cite{Pich:2009sp,Pich:2010ic,Ferreira:2010xe,Braeuninger:2010td,Bijnens:2011gd,Botella:2015yfa,Han:2015yys,Penuelas:2017ikk,Gori:2017qwg,Jung:2010ik,Li:2014fea,Abbas:2015cua}. Additionally, the A2HDM provides a generic framework for different 2HDM cases; various $\mathcal Z_2$-symmetric versions of the 2HDM can be considered as special cases of the A2HDM \cite{Pich:2009sp}. This model provides quite compelling phenomenology both at low-energy flavour experiments and high-energy colliders \cite{Abbas:2015cua, Celis:2013ixa,Celis:2013rcs,Jung:2010ik,Jung:2010ab,Jung:2012vu,Li:2014fea,Kanemura:2022cth, Kanemura:2020ibp,Iguro:2023tbk,Ilisie:2014hea,Abbas:2018pfp,Cai:2024xjq,Connell:2023jqq,Eberhardt:2020dat,Karan:2023kyj,Karan:2023xze,Karan:2024kgr,Coutinho:2024vzm,Banik:2024ugs,Coutinho:2024zyp,Heo:2025vkz}.
An even more general 2HDM (i.e. G2HDM) flavour setup is still possible, and can be probed by FCNC observables, see e.g. Refs.~\cite{Eeg:2019eei,Altmannshofer:2025nsl}.

It is a well-known fact that the CP violation arising from the CKM matrix of the SM is insufficient to explain the observed \textit{baryon asymmetry of the universe} \cite{Huet:1994jb}.
Recently, it has been argued that real 2HDM scenarios suffer from a theoretical inconsistency due to quark-induced divergent CP-violating amplitudes
\cite{deLima:2024lfc,Fontes:2021znm,deLima:2024hnk}. We thus focus on 2HDM scenarios that can introduce new sources of CP violation. Usually, in these 2HDM scenarios new CP-violating terms arise from the scalar potential only (through the quartic interaction parameter $\widetilde{\lambda}_5$ and the soft $\mathcal{Z}_2$-breaking parameter $m_{12}^2$ in the case of the Complex 2HDM, C2HDM; the scalar potential of the $\mathcal{Z}_2$-symmetric case instead respects the CP symmetry).
Nevertheless, the A2HDM can introduce new sources of CP violation in both the scalar and Yukawa sectors \cite{Pich:2009sp}. This CP violation in the Yukawa sector generates new two-loop contributions to the eEDM through for instance the charged-current fermion-loop Barr-Zee diagram involving $H^+\bar ff'$ couplings, which is completely absent in the C2HDM.\footnote{New contributions to the Weinberg operator are also possible in presence of a CP violating $H^+\bar ff'$ coupling \cite{Dicus:1989va}.}

In the 2HDM, the minimization of the scalar potential determines the mass scale of the first scalar doublet in terms of the EW vacuum expectation value $v$. However, the mass scale of the second doublet, which governs the masses of the new scalars, remains an independent parameter in this model. Therefore, by keeping this mass parameter much larger than $v$, a scenario commonly referred to as the \textit{decoupling limit}, the new physics scale is effectively separated from the EW scale of the SM.
%
The decoupling limit of conventional 2HDMs is studied outside the context of EDMs in Refs.~\cite{Egana-Ugrinovic:2015vgy,Crivellin:2016ihg,Dawson:2022cmu,Dawson:2023ebe,DasBakshi:2024krs,Dermisek:2024ohe,Dermisek:2024btn}.
Since taking the decoupling limit is consistent with the EW symmetry, we will be able to analyze this limit in terms of the SMEFT framework.\footnote{A discussion of HEFT is found in Ref.~\cite{Davila:2023fkk}, where the couplings of the Higgs to the charged and neutral heavy gauge bosons are analyzed.}

In this work, we analyze large logarithmic contributions to the eEDM in the decoupling limit of the A2HDM, namely, $ \log^2 (m_{EW} / M) $ and $ \log (m_{EW} / M) $ contributions, where $m_{EW}$ is the EW scale or the top-quark mass, and $M$ is the characteristic energy scale of the NP phenomena. We explicitly show that these logarithmic contributions are reproduced in a model-independent way by Renormalization-Group-Equation (RGE) effects calculated within the SMEFT, as expected.
The successful comparison between the A2HDM and SMEFT required revisiting calculations in these two frameworks to specify or correct relative and overall signs found in the literature \cite{Bowser-Chao:1997kjp,Jung:2013hka,Panico:2018hal,EliasMiro:2020tdv}.
The reproduction of the logarithmic pattern by RGEs is true for double logarithms enhanced by the squared top-quark mass, $ m_t^2 / M^2 \times \log^2 (m_t / M) $, and also for single logarithms proportional to the squared bottom-quark mass, $ m_b^2 / M^2 \times \log (m_t / M) $, both originating from Barr-Zee diagrams with a fermion loop.
(As explained later in the text, fully discussing $ m_t^2 / M^2 \times \log (m_t / M) $ terms would require
analyzing finite contributions in SMEFT, as well as two-loop anomalous-dimension matrix elements that have not been computed so far in the literature.)
These large logarithms in CP-violating observables are a specific feature of 
the A2HDM, being absent in the more studied 2HDMs where a $\mathcal Z_2$ symmetry is enforced, and thus where possible logarithmic effects from fermion loops are pushed to higher orders in the EFT, $1/M^2$ counting (nowhere in the following text we discuss in details SMEFT operators of dimension other than six).
This is so because some parameters become real under the effect of the $\mathcal Z_2$ symmetry, in particular the charged scalar couplings to fermions are real under this symmetry.
The underlying diagrams in the decoupling limit of the A2HDM result from the exchange of both neutral and charged heavy scalar fields which, when combined, are encoded by the SMEFT operators $ Q_{\ell e q u}^{(1)} $ and $Q_{\ell e d q}$ that are at the origin of the model-independent logarithmic effects $ m_t^2 / M^2 \times \log^2 (m_t / M) $ and $ m_b^2 / M^2 \times \log (m_t / M) $, respectively.
Our analysis does not include the operators $Q_{\ell e}$ because tau loops do not introduce
logarithmically enhanced contributions to EDMs in the decoupling limit of the A2HDM,
since in this model the alignment requirement relates CP-violating couplings across lepton generations;
in more general versions of 2HDMs, this purely leptonic operator has to be considered, since it is at the origin of $ m_\tau^2 / M^2 \times \log (m_{EW} / M) $ contributions \cite{Altmannshofer:2025nsl}.\footnote{Contributions of this type are present, for instance, in the generalized version of the A2HDM with family-dependent alignment parameters $\varsigma_f$  \cite{Penuelas:2017ikk}.}
Besides large logarithms from fermion-loop Barr-Zee diagrams,
large logarithms are also induced by Barr-Zee diagrams
having a light scalar or gauge boson loop, that can be accounted for in a model-independent way in terms of the SMEFT operator $ Q_{e H} $ as shown in Ref.~\cite{Altmannshofer:2020shb}, see also Ref.~\cite{Egana-Ugrinovic:2018fpy}.
Operators of the category $ H^2 X^2 $ will not be discussed hereafter, since the model does not introduce heavy fermions, which could lead to the required dual field strength tensor structures.

When studying CP violation in the leptonic sector,
operators other than the electron dipole must also be discussed.
This is so because four-fermion semileptonic operators generated by the exchange of heavy scalars induce at low energies CP-violating couplings of electrons to nucleons \cite{Pilaftsis:2002fe,Demir:2003js,Pospelov:2005pr,Ellis:2008zy,Yamanaka:2017mef,Yanase:2018qqq,Ardu:2025rqy}, for instance when the coupling of the heavy scalar to leptons violates CP symmetry.
For the heavy quarks (top, bottom, and charm), this leads to lepton-gluonic operators that go beyond the Barr-Zee mechanism and could be of help in constraining NP couplings.
This is an interesting possibility, leveraged by the fact that these contributions are generated at one-loop order in the perturbative counting.
The case of this latter lepton-gluonic operator illustrates well the need for operators of dimension higher than 6 in the low-energy EFT (see also Ref.~\cite{Davidson:2016utf,Dawson:2022cmu}).
%
The systematic analysis of effects proportional to powers of $ \log (m_{low} / m_{EW}) $, where $ m_{low} \ll m_{EW} $, in the A2HDM goes beyond the scope of the present discussion, that focuses on the features of the decoupling limit.
We note, however, that a similar suppression mechanism in the decoupling limit operating for $\mathcal Z_2$-symmetric 2HDMs to eliminate large logarithms in Barr-Zee diagrams is also at play for these new lepton-gluonic contributions.
%
We also note that their discussion is more involved, since both lepton-gluonic operators and
operators involving light quarks require dealing with the non-perturbative regime of strong interactions; see Refs.~\cite{Shifman:1978zn,Donoghue:1990xh,Prades:1990vn,Celis:2013xja,Cheung:2019bkw}.

The full model-independent analysis of CP violation in the quark sector requires the consideration of a larger set of SMEFT operators. In addition to the four-quark operator $Q_{quqd}^{(1)}$ and the dimension-6 Yukawa operators $Q_{uH}$ and $Q_{dH}$, the chromoelectric \cite{Bhattacharya:2015rsa} and the Weinberg gluon operators are also important (for a discussion about chromomagnetic operators, not relevant for CP violation, see Ref.~\cite{Bisal:2024nbb}). We delegate their discussion to future work.
A global fit analysis of low-energy parameters is discussed in Ref.~\cite{Degenkolb:2024eve}.

This text is organized as follows.
In Section~\ref{sec:A2HDM} we introduce the A2HDM, discussing its sources of CP violation and its decoupling limit.
In Section~\ref{sec:calculation_A2HDM} we discuss the eEDM and the different contributions to it from the A2HDM.
Section~\ref{sec:SMEFT} discusses the effective operators that emerge in the decoupling limit of the A2HDM, and the computation of logarithmic contributions within SMEFT based on available results of anomalous-dimension matrix elements at one and two loops.
Finally, we also show in Section~\ref{sec:SMEFT} the consistency between the decoupling limit of the A2HDM and SMEFT.
A phenomenological discussion is given in Section~\ref{sec:phenomenology}.
Conclusions are found in Section~\ref{sec:conclusions}.
A series of appendices provides more technical details; in particular, we rederive Barr-Zee contributions with a fermion loop, pointing out an inconsistent relative sign in some previous calculations; we find agreement with the very recent Ref.~\cite{Altmannshofer:2025nsl} in this respect.

\section{The aligned two-Higgs doublet model}\label{sec:A2HDM}

In the 2HDM the SM is extended with a second complex scalar doublet having hypercharge $Y=1/2$. After EW symmetry breaking, both doublets acquire complex vacuum expectation values (VEVs). Nevertheless, using a suitable $SU(2)_L \otimes U(1)_Y$ transformation, it is always possible to rotate the scalar doublets to the so-called ``Higgs-basis'' where only the first doublet acquires a nonzero real VEV. In this basis, the scalar doublets take the following form:
\begin{equation}
\label{eq:Phi}
\Phi_1=\frac{1}{\sqrt 2}\begin{pmatrix}
\sqrt 2\;G^+\\
S_1+v+i\, G^0
\end{pmatrix}\, ,\qquad\qquad \Phi_2=\frac{1}{\sqrt 2}\begin{pmatrix}
\sqrt 2\;H^+\\
S_2+i\, S_3
\end{pmatrix}\, ,
\end{equation}
where $v=246$ GeV is the VEV of the $\Phi_1$ scalar. In the process of generating masses to the $W^\pm$ and $Z$ bosons, the components $G^\pm$ and $G^0$ act as Goldstone bosons. Thus, the scalar sector of this model contains one pair of charged scalars $H^\pm$, two CP-even scalars $S_1$, $S_2$ and one CP-odd pseudoscalar $S_3$. Respecting the SM gauge symmetries, the most general scalar potential takes the form:
\begin{align}
\label{eq:pot}
V=&\mu_1\,\Phi_1^\dagger \Phi_1+\mu_2\,\Phi_2^\dagger \Phi_2+ \Big[\mu_3\,\Phi_1^\dagger \Phi_2+\mathrm{h.c.}\Big]+\frac{\lambda_1}{2}\,(\Phi_1^\dagger \Phi_1)^2+\frac{\lambda_2}{2}\,(\Phi_2^\dagger \Phi_2)^2+\lambda_3\,(\Phi_1^\dagger \Phi_1)(\Phi_2^\dagger \Phi_2)\nonumber\\
&+\lambda_4\,(\Phi_1^\dagger \Phi_2)(\Phi_2^\dagger \Phi_1)+\Big[\Big(\frac{\lambda_5}{2}\,\Phi_1^\dagger \Phi_2+\lambda_6 \,\Phi_1^\dagger \Phi_1 +\lambda_7 \,\Phi_2^\dagger \Phi_2\Big)(\Phi_1^\dagger \Phi_2)+ \mathrm{h.c.}\Big]\, ,
\end{align}
where the parameters $\lambda_{\{5,6,7\}}$ and $\mu_3$ can have complex values. Depending on the parameters of the scalar potential, the neutral scalars $S_1$, $S_2$ and $S_3$ mix with each other through an orthogonal matrix $\mathcal R$ and produce the mass eigenstates $H_j\in\{H_1, H_2, H_3\}$, where the lightest state $H_1$ is identified with the SM-like Higgs $h$. When some of the scalar potential parameters have complex values, CP symmetry gets violated and hence the mass eigenstates do not possess any definite CP quantum number.

The situation where the mass parameter of the second doublet $\Phi_2$ becomes very large compared to the VEV of $\Phi_1$, i.e. $\sqrt{\mu_2}\gg v$ (or more specifically $\sqrt{\mu_2}\gg v\,\sqrt{\lambda_i}$), is called the \textit{decoupling limit}. This scenario can be thought of as the doublet $\Phi_2$ sitting at a very high energy scale ($\sqrt{\mu_2}$) and hence decoupling from the SM which is at a much lower energy scale ($v\,\sqrt{\lambda_i}$). Using the masses of the particles as independent parameters, the condition for achieving the decoupling limit can be stated as $M_{\{H^\pm,\,H_2,\,H_3\}}\approx M \gg m_h$. It is important to mention that two of the mixing angles (that mix $S_1$ with $S_2$ and $S_3$) of the matrix $\mathcal R$ automatically tend to zero in the decoupling limit.

In addition to the usual mass terms for the fermions $f\in\{u,d,l\}$, the interaction part of the Yukawa Lagrangian in the A2HDM becomes:
\begin{equation}
\label{eq:Yukawa}
-\mathcal L_Y=\sum_{j,f}\Bigg(\frac{y_f^{H
_j}}{v}\Bigg)\,H_j\,\bar f M_f \mathcal{P}_R f+\frac{\sqrt 2 H^+}{v} \Big[\bar u\,\big\{\varsigma_d V M_d \mathcal{P}_R-\varsigma_u M_u^\dagger V\mathcal{P}_L\big\}\, d+\varsigma_\nl\, \bar \nu M_\nl \mathcal P_R\,\nl\Big] + \mathrm{h.c.}\, ,
\end{equation}
where $\mathcal P_{L,R}$ are chirality projection operators, $M_f$ are the diagonal fermionic mass matrices, and $V$ is the usual CKM matrix. The complex parameters $\varsigma_f$ are called the flavour alignment parameters from which the name A2HDM originates. The Yukawa couplings of fermions with the neutral scalars $(y_f^{H_j})$ are given by:
\begin{equation}
y_{d,l}^{H_j}=\R_{j1}+ (\R_{j2}+i \R_{j3}) \varsigma_{d,l}^{}\qquad \text{and} \qquad y_{u}^{H_j}=\R_{j1}+ (\R_{j2}-i \R_{j3}) \varsigma_{u}^{*}~.
\end{equation}
The usual 2HDM scenarios can easily be retrieved from the A2HDM Lagrangian by imposing the following conditions on the alignment parameters:
\begin{align}
\label{eq:THDM_types}
&\text{Type I:\;\;} \varsigma_{u}=\varsigma_d=\varsigma_\nl=\cot\beta,\quad \text{Type II:\;\;} \varsigma_{u}=-\frac{1}{\varsigma_d}=-\frac{1}{\varsigma_\nl}=\cot\beta\, ,\quad 
\text{Inert:\;\;}\varsigma_{u}=\varsigma_d=\varsigma_\nl=0\, ,
\nonumber\\
&\text{Type X:\;\;} \varsigma_{u}=\varsigma_d=-\frac{1}{\varsigma_\nl}=\cot\beta 
\qquad \text{and}\qquad
\text{Type Y:\;\;} \varsigma_{u}=-\frac{1}{\varsigma_d}=\varsigma_\nl=\cot\beta \,,
\end{align}
along with vanishing $\widetilde{\lambda}_6$ and $\widetilde{\lambda}_7$ terms in the $\mathcal Z_2$-symmetric basis,\footnote{We note that the explicit implementation of the $\mathcal Z_2$ symmetry is dependent on the scalar basis.} while soft breaking of $\mathcal{Z}_2$ can be implemented with a $m_{12}^2$ term.
Appendices~\ref{sec:pot} and \ref{sec:Yuk} discuss the scalar potential and the Yukawa interactions of this model in more detail, while Appendix~\ref{app:decoupling_limit} discusses the decoupling limit.

If the demand of alignment is protected by a $\mathcal Z_2$ symmetry (i.e. the usual 2HDM types), then the alignment remains stable under renormalization \cite{Ferreira:2010xe}. More generally, in the leptonic sector, the alignment is also respected to all orders in perturbation theory. But higher-order quantum corrections in the quark sector create in general a small misalignment, and hence loop-suppressed FCNC effects.
Nonetheless, the flavour symmetries embedded in the A2HDM constrain the possible FCNC structures, suppressing strongly their effects
\cite{Pich:2009sp,Pich:2010ic,Jung:2010ik};
the amount of misalignment generated through the running of the couplings remains well below the current experimental limits \cite{Braeuninger:2010td,Bijnens:2011gd,Penuelas:2017ikk,Gori:2017qwg}.

\section{Full calculation of the eEDM in the A2HDM}\label{sec:calculation_A2HDM}

The EDM of the electron $d_e$ can be defined as the coefficient of the following dimension five operator (see Appendix~\ref{sec:LeEDM}) in the effective Lagrangian at a very low energy scale ($\mu\sim m_e$):

\begin{equation}
	\mathcal L \supset -\frac{i}{2} d_e (\mu)\, \bar \psi_e\, \sigma^{\mu \nu}\gamma^5 \, \psi_e\, F_{\mu\nu}
\end{equation}
where $\psi_e$ is the electron Dirac spinor and $F^{\mu\nu}$ is the field strength tensor of the photon. In the SM, the CP violation in the quark sector from the CKM matrix can induce EDMs for leptons. However, a nonzero value for the eEDM appears first at 
four loops in perturbation theory~\cite{Pospelov:1991zt}. A much larger estimate has been obtained when taking into account long-distance hadronic effects, giving
$d_e^{SM} = 5.8 \times 10^{-40}\, e\,\text{cm}$ (and also $d_\mu^{SM} = 1.4 \times 10^{-38}\, e\,\text{cm}$ and $d_\tau^{SM} = -7.3 \times 10^{-38}\, e\,\text{cm}$) \cite{Yamaguchi:2020eub,Yamaguchi:2020dsy}, with an estimated error bar of 70\%.
If neutrinos are assumed to be Majorana particles, new CP-violating phases emerge in the lepton sector which give rise to an eEDM at the two-loop level \cite{Ng:1995cs}. Nevertheless, using the type-I seesaw, it is found that after fine-tuning of parameters $d_e$ can reach a maximum value of $10^{-33}\,e\,\text{cm}$ \cite{Archambault:2004td}, which is well below the experimentally probed value (at 90\% C.L.) \cite{Roussy:2022cmp}:
\begin{equation}\label{eq:current_exp_value}
   |d_e^{exp}| \, < \, 4.1\times 10^{-30} \, e\,\text{cm}.
\end{equation}
Thus, the simple extension of the neutrino sector is not very efficient to saturate the current experimental bound.

It is worth noting that the quoted experimental bound on $d_e$ has been extracted from a diatomic molecule (HfF${}^+$), assuming that the eEDM is the only source of CP violation. Diatomic molecules are also sensitive to a CP-odd pseudoscalar-scalar electron-nucleon coupling $C_S$. It has been estimated that this effective interaction produces a CP-violating effect equivalent in size to $d_e^{SM} \sim 10^{-38}\, e\,\text{cm}$ \cite{Pospelov:2013sca}.
Recently identified contributions pushed this value to a much higher level, namely, $d_e^{SM} \sim 10^{-35}\, e\,\text{cm}$ \cite{Ema:2022yra}.
The two effects can be disentangled, combining EDM data from HfF${}^+$ \cite{Roussy:2022cmp} and ThO \cite{ACME:2018yjb}, which gives the weaker bound 
$ |d_e^{exp}| \, < \, 2.1\times 10^{-29} \, e\,\text{cm}$ (90\% C.L.).

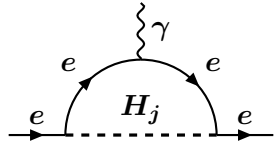
\begin{figure}[h!]
	\centering
	\begin{tikzpicture}[thick]
		\begin{feynman}
			\vertex (a1);
			\vertex [left=0.75cm of a1] (a0);
			\vertex [right=2cm of a1] (a2);
			\vertex [right=0.75cm of a2] (a3);
			\vertex [above right=1.41cm of a1] (b1);
			\vertex [above=0.75cm of b1] (b2);
			\diagram*{{(a0)--[small,fermion,edge label=$\bm e$](a1)--[scalar,very thick,edge label=$\bm{H_j}$](a2)--[small,fermion,edge label=$\bm e$](a3)},
			(a1)--[quarter left, small,fermion,edge label=$\bm e$](b1),
			(b1)--[quarter left, small,fermion,edge label=$\bm e$](a2),
			(b1)--[boson,edge label'=$\bm \gamma$](b2)};
		\end{feynman}
	\end{tikzpicture}
	\caption{One-loop diagram for the eEDM in the A2HDM.}
	\label{fig:one-loop}
\end{figure}

In the A2HDM, the eEDM starts getting nonzero contributions at one loop. It is evident from Eq.~\eqref{eq:Yukawa} that each fermion-scalar interaction vertex brings a factor $(m_f/v)$ where $m_f$ is the mass of the fermion. Therefore,
the one-loop contribution to $d_e$ will be proportional to $m_e^3/(v^2\,M_{H_j}^2)\propto G_F\,m_e\,(m_e^2/M_{H_j}^2)$; two factors of $(m_e/v)$ come from two $\bar e e H_j$ vertices and the remaining $(m_e/M_{H_j}^2)$ factor arises from the scalar in the loop and the chirality flip.
This one-loop contribution, mediated through neutral scalars (see Fig.~\ref{fig:one-loop}), is expressed as:\footnote{In the decoupling limit there is no large logarithm $ \log ( M^2/m_{EW}^2 ) $ associated to the one-loop expression, up to $1/M^2$.}

\begin{equation}
   \frac{d_e^{\mathrm{one-loop}}}{e} \, = \, \frac{m_e}{16 \pi^2 v^2} \sum_i \text{Re}(y_e^{H_i}) \text{Im}(y_e^{H_i}) \frac{m_e^2}{M_{H_i}^2} \left( 3 + 2 \log \left( \frac{m_e^2}{M_{H_i}^2} \right) \right).
\end{equation}

\renewcommand{\arraystretch}{1.2}
\begin{table}[h!]
	\centering
	\begin{tabular}{|c|c c c c|}
		\hline\hline
		\multicolumn{2}{|c}{\multirow{2}{*}{Classes of diagrams}}&Fermion&Charged Higgs&Gauge boson\\
		\multicolumn{2}{|c}{ }&loop&loop&loop\\
		\hline\hline
		\multirow{12}{*}{\rotatebox{90}{\Large{Barr-Zee}}}&Charged current&\multirow{4}{*}{$d_{e,f}^{CC}$}&\multirow{4}{*}{$d_{e,H^\pm}^{CC}$}&\multirow{4}{*}{$d_{e,W}^{CC}$}\\
		&
\begin{tikzpicture}[thick]
	\begin{feynman}
		\vertex (a0);
		\vertex [right=1.8cm of a0](a1);
		\vertex at ($(a0)!0.15!(a1)$) (ax);
		\vertex at ($(a0)!0.85!(a1)$) (ay);
		\vertex [small,blob, fill=gray!50!white ] at ($(a0)!0.5!(a1)+(0,0.6cm)$) (bx) {};
		\vertex [above=0.7cm of bx] (by);
		\diagram*{(a0)--[small,fermion](a1),
			(ax)--[boson,edge label={\scriptsize$\bm W$}](bx)--[scalar,edge label={\scriptsize$\bm{H^\pm}$}](ay),
			(bx) -- [boson](by)};
	\end{feynman}
\end{tikzpicture}&&&\\[1.0mm]
&Neutral current&\multirow{4}{*}{$d_{e,f}^{NC}$}&\multirow{4}{*}{$d_{e,H^\pm}^{NC}$}&\multirow{4}{*}{$d_{e,W}^{NC}$}\\
&
\begin{tikzpicture}[thick]
	\begin{feynman}
		\vertex (a0);
		\vertex [right=1.8cm of a0](a1);
		\vertex at ($(a0)!0.15!(a1)$) (ax);
		\vertex at ($(a0)!0.85!(a1)$) (ay);
		\vertex [small,blob, fill=gray!50!white ] at ($(a0)!0.5!(a1)+(0,0.6cm)$) (bx) {};
		\vertex [above=0.7cm of bx] (by);
		\diagram*{(a0)--[small,fermion](a1),
			(ax)--[boson,edge label={\scriptsize$\bm{Z}$}](bx)--[scalar,edge label={\scriptsize$\bm{H_j^{}}$}](ay),
			(bx) -- [boson](by)};
	\end{feynman}
\end{tikzpicture}&&&\\[1.0mm]
&Electromagnetic &\multirow{4}{*}{$d_{e,f}^{EM}$}&\multirow{4}{*}{$d_{e,H^\pm}^{EM}$}&\multirow{4}{*}{$d_{e,W}^{EM}$}\\
&
\begin{tikzpicture}[thick]
\begin{feynman}
	\vertex (a0);
	\vertex [right=1.8cm of a0](a1);
	\vertex at ($(a0)!0.15!(a1)$) (ax);
	\vertex at ($(a0)!0.85!(a1)$) (ay);
	\vertex [small,blob, fill=gray!50!white ] at ($(a0)!0.5!(a1)+(0,0.6cm)$) (bx) {};
	\vertex [above=0.7cm of bx] (by);
	\diagram*{(a0)--[small,fermion](a1),
		(ax)--[boson,edge label={\scriptsize$\bm{\gamma}$}](bx)--[scalar,edge label={\scriptsize$\bm{H_j^{}}$}](ay),
		(bx) -- [boson](by)};
\end{feynman}
\end{tikzpicture}&&&\\
		\hline\hline
		\multirow{8}{*}{\rotatebox{90}{\Large{Kite}}}&Charged current&\multirow{4.5}{*}{---}&\multirow{4.5}{*}{---}&\multirow{4.5}{*}{$d_{e,\mathrm{Kite}}^{CC}$}\\
		& \hspace*{7mm}
		\begin{tikzpicture}[thick,baseline=(cy.base)]
			\begin{feynman}
				\vertex (a0);
				\vertex [right=1.8cm of a0](a1);
				\vertex at ($(a0)!0.5!(a1)+(0,0.9cm)$) (bx);
				\vertex at ($(a0)!0.2!(bx)$) (ax);
				\vertex at ($(bx)!0.8!(a1)$) (ay);
				\vertex at ($(ax)!0.4!(ay)$) (cx);
				\vertex at ($(bx)!0.4!(ay)$) (cy);
				\vertex [above=0.5cm of bx] (by);
				\diagram*{(a0)--[small,fermion](bx)--[small,fermion](a1),
					(ax)--[scalar,edge label'={\scriptsize$\bm{H_j^{}}$}](cx)--[boson,edge label'={\scriptsize$\bm{W}$}](ay),
					(cx)--[boson](cy),
					(bx) -- [boson](by)};
			\end{feynman}
		\end{tikzpicture} + ... &&&\\[3mm]
		&Neutral current&\multirow{4.5}{*}{---}&\multirow{4.5}{*}{---}&\multirow{4.5}{*}{$d_{e,\mathrm{Kite}}^{NC}$}\\
		& \hspace*{7mm}
		\begin{tikzpicture}[thick,baseline=(cy.base)]
			\begin{feynman}
				\vertex (a0);
				\vertex [right=1.8cm of a0](a1);
				\vertex at ($(a0)!0.5!(a1)+(0,0.9cm)$) (bx);
				\vertex at ($(a0)!0.2!(bx)$) (ax);
				\vertex at ($(bx)!0.8!(a1)$) (ay);
				\vertex at ($(ax)!0.4!(ay)$) (cx);
				\vertex at ($(bx)!0.4!(ay)$) (cy);
				\vertex [above=0.5cm of bx] (by);
				\diagram*{(a0)--[small,fermion](bx)--[small,fermion](a1),
					(ax)--[scalar,edge label'={\scriptsize$\bm{H_j^{}}$}](cx)--[boson,edge label'={\scriptsize$\bm{Z}$}](ay),
					(cx)--[boson](cy),
					(bx) -- [boson](by)};
			\end{feynman}
		\end{tikzpicture} + ... &&&\\
		\hline\hline
	\end{tabular}
	\caption{Dominant two-loop contributions to the eEDM in the A2HDM. The classification shown here is adopted from Ref.~\cite{Altmannshofer:2020shb}. The dots in the ``Kite'' class imply the diagrams containing permutations of the gauge boson and scalar propagators. While the contribution $d_{e,f}^{CC}$ is given by Eqs.~\eqref{eq:tb_loop}, \eqref{eq:up_type_loop} and \eqref{eq:cb_ub_loop}, the rest of the contributions are presented in full generality in Appendix~\ref{sec:eEDM_detail}.}
	\label{tab:two-loop}
\end{table}
 
Nevertheless, it is a well-known fact that the two-loop contribution to the eEDM is much larger than the one-loop result
\cite{Barr:1990vd}. Two-loop diagrams with only gauge and their Goldstone bosons attached to the electron line do not contribute to the eEDM as they are not proportional to any CP-violating parameter. The dominant contribution at the two-loop order emerges from diagrams having one scalar and one or two gauge bosons
(apart from the external photon) attached to the electron line. Since these diagrams contain a lower number of electron-scalar interaction vertices, their contributions to the eEDM are significantly larger than the one-loop diagrams. It is important to mention that while discussing the two-loop effects we neglect the diagrams attaching two or more scalars to the electron line.

Depending on the number of gauge bosons connected to the electron line, the dominant two-loop diagrams contributing to the eEDM in the A2HDM can be classified in two classes:\footnote{This separation is not necessarily gauge invariant, as it will shortly be discussed.} a) Barr-Zee (one gauge boson) and b) Kite (two gauge bosons), as shown in Table~\ref{tab:two-loop}. Moreover, based on the internal gauge boson connected to the electron line, these diagrams can further be divided into three categories: i) electromagnetic-current mediated ($\gamma$), ii) neutral-current mediated ($Z$) and iii) charged-current mediated ($W^\pm$), each of which are induced by three types of loops: fermion loop, charged Higgs loop and gauge boson loop. We will refer to i)-iii) as EM, NC, and CC contributions, respectively. This classification method is adopted from Ref. \cite{Altmannshofer:2020shb}. While the Barr-Zee diagrams come from all three types of contributions, the kite diagrams originate from the neutral and charged currents only.

The eEDM has been discussed in Ref.~\cite{Altmannshofer:2020shb} within the C2HDM scenario, which is a special case of the A2HDM. In the A2HDM there are additional contributions to the eEDM, arising from the charged-current fermion-loop Barr-Zee diagrams, that are completely absent in the C2HDM. They are generated by the complex coupling of the $\bar e \nu H^+$ vertex, which is real valued in the C2HDM.
 We calculate this extra contribution below; the rest of contributions that are common to both A2HDM and C2HDM are delegated to Appendix~\ref{sec:eEDM_detail}. Some technical details about the calculation of the charged-current contribution from fermion loops are found in Appendix~\ref{sec:details_defCC}.
There are no (sub)divergences in the calculation of the Barr-Zee diagrams with fermion loops; accordingly, we perform the Feynman integrals in four dimensions, and in particular $\gamma_5$ anti-commutes.

\begin{figure}[h!]
	\centering
	\scalebox{1.0}{\begin{tikzpicture}[thick, scale=1.5]
		\begin{feynman}
        \vertex (i1) at (0.8, -0.5);
        \vertex (a) at (1.3, -0.5);
        \vertex (b) at (2, 0.45);
        \vertex (c) at (2.5, 1.32);
        \vertex (f1) at (2.5, 2);
        \vertex (d) at (3, 0.45);
        \vertex (e) at (3.7, -0.5);
        \vertex (f2) at (4.2, -0.5); 
        
        \diagram* {
            (i1) -- [small,fermion, edge label= $\bm e$] (a) -- [small,fermion, edge label=\(\bm\nu\)] (e) -- [small,fermion, edge label= $\bm e$] (f2),
            (a) -- [boson, edge label=\(\bm{W^-}\), momentum'={ }, /tikzfeynman/momentum/arrow shorten=0.3] (b),
            (b) -- [anti fermion, looseness=1.1, edge label=\(\bm{q_u}\), out=120, in=180, relative=false] (c) -- [anti fermion, quarter
            left, looseness=1.1, edge label=\(\bm {q_u}\), out=0, in=60, relative=false] (d) -- [anti fermion, half left, looseness=1.1, edge label'=\(\bm{q_d}\), out=240, in=300, relative=false] (b),
            (d) -- [very thick,scalar, edge label=\(\bm{H^-}\), momentum'={ }, /tikzfeynman/momentum/arrow shorten=0.3] (e),
            (c) -- [boson, edge label=$\bm \gamma$] (f1),
        };
    \end{feynman}
	\end{tikzpicture}
    \hspace*{5mm}
		\begin{tikzpicture}[thick, scale=1.5]
		\begin{feynman}
        \vertex (i1) at (0.8, -0.5);
        \vertex (a) at (1.3, -0.5);
        \vertex (b) at (2, 0.45);
        \vertex (c) at (2.5, 1.32);
        \vertex (f1) at (2.5, 2);
        \vertex (d) at (3, 0.45);
        \vertex (e) at (3.7, -0.5);
        \vertex (f2) at (4.2, -0.5); 
        
        \diagram* {
            (i1) -- [small,fermion, edge label= $\bm e$] (a) -- [small,fermion, edge label=\(\bm\nu\)] (e) -- [small,fermion, edge label= $\bm e$] (f2),
            (a) -- [very thick,scalar, edge label=\(\bm{H^-}\), momentum'={ }, /tikzfeynman/momentum/arrow shorten=0.3] (b),
            (b) -- [anti fermion, looseness=1.1, edge label=\(\bm{q_u}\), out=120, in=180, relative=false] (c) -- [anti fermion, quarter
            left, looseness=1.1, edge label=\(\bm {q_u}\), out=0, in=60, relative=false] (d) -- [anti fermion, half left, looseness=1.1, edge label'=\(\bm{q_d}\), out=240, in=300, relative=false] (b),
            (d) -- [boson, edge label=\(\bm{W^-}\), momentum'={ }, /tikzfeynman/momentum/arrow shorten=0.3] (e),
            (c) -- [boson, edge label=$\bm \gamma$] (f1),
        };
    \end{feynman}
	\end{tikzpicture}}

\vspace*{5mm}
   \scalebox{1.0}{ \begin{tikzpicture}[thick, scale=1.5]
		\begin{feynman}
        \vertex (i1) at (0.8, -0.5);
        \vertex (a) at (1.3, -0.5);
        \vertex (b) at (2, 0.45);
        \vertex (c) at (2.5, 1.32);
        \vertex (f1) at (2.5, 2);
        \vertex (d) at (3, 0.45);
        \vertex (e) at (3.7, -0.5);
        \vertex (f2) at (4.2, -0.5); 
        
        \diagram* {
            (i1) -- [small,fermion, edge label= $\bm e$] (a) -- [small,fermion, edge label=\(\bm\nu\)] (e) -- [small,fermion, edge label= $\bm e$] (f2),
            (a) -- [boson, edge label=\(\bm{W^-}\), momentum'={ }, /tikzfeynman/momentum/arrow shorten=0.3] (b),
            (b) -- [fermion, looseness=1.1, edge label=\(\bm{q_d}\), out=120, in=180, relative=false] (c) -- [fermion, quarter
            left, looseness=1.1, edge label=\(\bm {q_d}\), out=0, in=60, relative=false] (d) -- [fermion, half left, looseness=1.1, edge label'=\(\bm{q_u}\), out=240, in=300, relative=false] (b),
            (d) -- [very thick,scalar, edge label=\(\bm{H^-}\), momentum'={ }, /tikzfeynman/momentum/arrow shorten=0.3] (e),
            (c) -- [boson, edge label=$\bm \gamma$] (f1),
        };
    \end{feynman}
	\end{tikzpicture}
    \hspace*{5mm}
		\begin{tikzpicture}[thick, scale=1.5]
		\begin{feynman}
        \vertex (i1) at (0.8, -0.5);
        \vertex (a) at (1.3, -0.5);
        \vertex (b) at (2, 0.45);
        \vertex (c) at (2.5, 1.32);
        \vertex (f1) at (2.5, 2);
        \vertex (d) at (3, 0.45);
        \vertex (e) at (3.7, -0.5);
        \vertex (f2) at (4.2, -0.5); 
        
        \diagram* {
            (i1) -- [small,fermion, edge label= $\bm e$] (a) -- [small,fermion, edge label=\(\bm\nu\)] (e) -- [small,fermion, edge label= $\bm e$] (f2),
            (a) -- [very thick,scalar, edge label=\(\bm{H^-}\), momentum'={ }, /tikzfeynman/momentum/arrow shorten=0.3] (b),
            (b) -- [fermion, looseness=1.1, edge label=\(\bm{q_d}\), out=120, in=180, relative=false] (c) -- [fermion, quarter
            left, looseness=1.1, edge label=\(\bm {q_d}\), out=0, in=60, relative=false] (d) -- [fermion, half left, looseness=1.1, edge label'=\(\bm{q_u}\), out=240, in=300, relative=false] (b),
            (d) -- [boson, edge label=\(\bm{W^-}\), momentum'={ }, /tikzfeynman/momentum/arrow shorten=0.3] (e),
            (c) -- [boson, edge label=$\bm \gamma$] (f1),
        };
    \end{feynman}
	\end{tikzpicture}}
	\caption{Charged-current Barr-Zee diagrams with a fermion loop. Here, $q_u$ and $q_d$ are up-type and down-type quarks.}
	\label{fig:CC_f}
\end{figure}
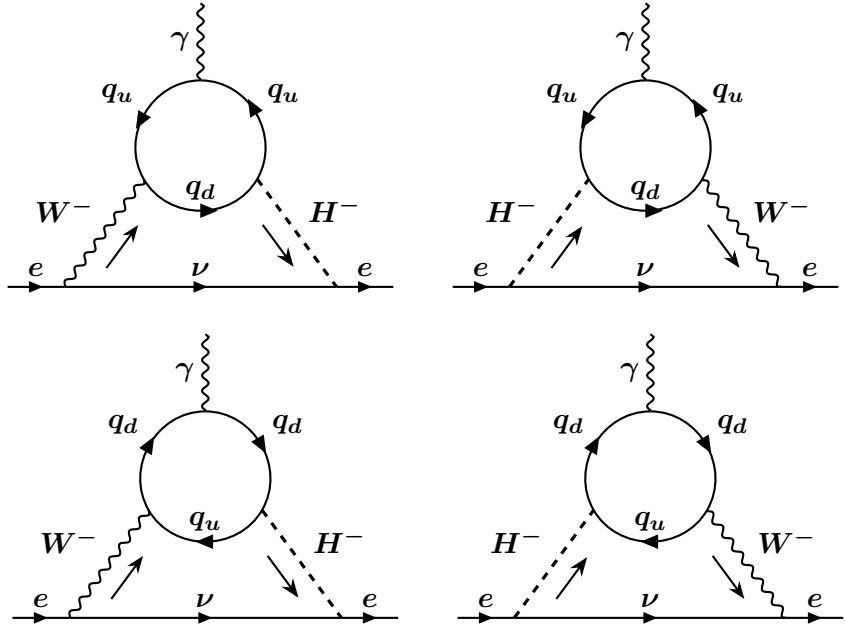

The Feynman diagrams responsible for the charged-current fermion-loop Barr-Zee process are presented in Fig.~\ref{fig:CC_f}. The dominant contribution for this class of diagrams comes from the top-bottom Barr-Zee loop, whose effect in the eEDM is given by:
\begin{equation}\label{eq:tb_loop}
	\frac{d_{e,f}^{CC}}{e}\bigg
	|_{tb-\text{loop}} \, = \, m_e\, \frac{\sqrt 2\, \alpha\, G_F}{(4 \pi)^3\, s_w^2}\, N_C |V_{tb}|^2 \Big\{ \text{Im}(\varsigma_u^* \varsigma_l) (Q_t F_1 + Q_b F_2) + \text{Im}(\varsigma_d^* \varsigma_l) \frac{m_b^2}{m_t^2}  (Q_t G_1 + Q_b G_2) \Big\},
\end{equation}
where $Q_q$ is the electromagnetic charge of the corresponding quark in units of $e$, $N_C$ is the number of colours for the quarks, and the constants $\{\alpha, G_F, s_w, V_{tb}\}$ carry their usual definitions, with $s_w = \sin ( \theta_W )$ and $\theta_W$ being the weak mixing angle. Introducing $z_{H} = (M_{H^{\pm}}^2/m_t^2)$ and $z_W = (m_{W}^2/m_t^2)$, the functions $F_i$ and $G_i$ are defined as: 
\begin{align}
	F_i \, &= \, \frac{T_i(z_H) - T_i(z_W)}{z_H - z_W}, \qquad 
	G_i \, = \, \frac{E_i(z_H) - E_i(z_W)}{z_H - z_W},  \label{eqn: FGq} \\
	T_1(z) \, &= \, \frac{1 - 3z}{z^2} \frac{\pi^2}{6} + \left( \frac{1}{z} - \frac52 \right) \log{z} - \frac{1}{z} - \left( 2 - \frac{1}{z} \right) \left( 1 - \frac{1}{z} \right) \text{Li}_2 (1 - z), \label{eqn: Tt}\\
	T_2(z) \, &= \, \frac{2z - 1}{z^2} \frac{\pi^2}{6} + \left( \frac32 - \frac{1}{z} \right) \log{z} + \frac{1}{z} - \frac{1}{z} \left( 2 - \frac{1}{z} \right) \text{Li}_2 (1 - z), \label{eqn: Tb}\\
	E_1(z) \, &= \, \frac{z-1}{z^2} \frac{\pi^2}{6} + \left( \frac12 - \frac{1}{z} \right) \log{z} + \frac{1}{z} - \frac{1}{z} \left( 1 - \frac{1}{z} \right) \text{Li}_2 (1 - z), \label{eqn: Et}\\
	E_2(z) \, &= \, \frac{1}{z^2} \frac{\pi^2}{6} + \left( \frac{1}{z} + \frac12 \right) \log{z} - \frac{1}{z} - \frac{1}{z^2} \text{Li}_2 (1 - z). \label{eqn: Eb}
\end{align}
This contribution was first calculated in Ref.~\cite{Bowser-Chao:1997kjp} neglecting the mass of the bottom quark. The formula was rectified in Ref.~\cite{Jung:2013hka} by correcting the expression of $T_1(z)$. However, none of the latter two references clearly specifies the sign conventions they use, which determine the overall sign for the eEDM.
Our sign conventions are detailed in Appendix~\ref{sec:Conventions}.
There is a relative sign in Ref.~\cite{Jung:2013hka} between their neutral- and charged-current contributions that we correct.
Our result agrees with the recent Ref.~\cite{Altmannshofer:2025nsl}, where the eEDM has been discussed within the G2HDM.

Apart from the leading top-bottom Barr-Zee loop diagram, several contributions emerge when the three quark families are considered, such as top-strange, top-down, charm-bottom and up-bottom (we only consider contributions depending on either $m_b$ or $m_t$, which introduce dependencies on $\varsigma_d$ and $\varsigma_u$, respectively, while the analogous diagram depending on $m_c$ does not bring any novel coupling in the A2HDM).
These contributions will also be important for our discussion, since they introduce further logarithmically-enhanced terms, although suppressed by off-diagonal elements of the CKM matrix.
Since the functions $F_i$ do not depend on the mass of the down-type quark (provided that such mass is neglected with respect to the mass of the top quark) and the electromagnetic charge is the same for all down-type quarks, we can add together the contributions with $tb$, $ts$ and $td$ loops:
\begin{equation}\label{eq:up_type_loop}
	\frac{d_{e,f}^{CC}}{e}\Bigg
	|_{\varsigma_u} \, = \, m_e\, \frac{\sqrt 2\, \alpha\, G_F}{(4 \pi)^3\, s_w^2}\, N_C \text{Im}(\varsigma_u^* \varsigma_l) (Q_t F_1 + Q_b F_2),
\end{equation}
thus getting rid of the CKM factors (since $\sum_i |V_{ti}|^2=1$). This will later be of importance, since results from the SMEFT framework do not depend on such constants.
Qualitatively, it means that CP violation in charged-current fermion-loop Barr-Zee diagrams can only come from the alignment parameters.

On the other hand, we computed the sub-leading contributions with $cb$ and $ub$ loops and found that the functions that appear in the final result coincide with the functions that already appeared in the top-bottom contribution, but with a dependence on the mass of the bottom quark instead of the top one. In particular, we find:
\begin{equation}\label{eq:cb_ub_loop}
	\frac{d_{e,f}^{CC}}{e}\Bigg
	|_{cb/ub-\text{loop}} \, = \, m_e\, \frac{\sqrt 2\, \alpha\, G_F}{(4 \pi)^3\, s_w^2}\, N_C \text{Im}(\varsigma_d^* \varsigma_l) \sum_{i=u,c} |V_{ib}|^2 \Big\{ Q_i G_2 (m_t \rightarrow m_b) + Q_b G_1 (m_t \rightarrow m_b) \Big\},
\end{equation}
where the mass of the lighter quark (that is, $m_u$ or $m_c$) was neglected with respect to $m_b$; the replacement rules shown as arguments of the $G_i$ functions mean that $z_{H}$ and $z_{W}$ are replaced accordingly. Although we write $Q_i$ with $i=u,c$ to refer to the electromagnetic charges of the up and charm quarks, we will take advantage of the fact that they share the same electromagnetic charge as the top quark, and in the following we will simply write $Q_t$ for every up-type quark.
It is interesting to mention that Feynman diagrams of the same kind as the ones just discussed, but with an internal lepton loop, do not contribute to the eEDM in the A2HDM: these diagrams are proportional to $|\varsigma_l|^2$, which obviously does not have an imaginary part to contribute to the eEDM.

In the decoupling limit $z_H\approx z_M=(M^2/m_{t}^2) \gg 1$ or $z_H\approx z_M=(M^2/m_{b}^2) \gg 1$ and, therefore, ignoring the effects of $z_W$ one can approximate $F_i\approx [T_i(z_M)/z_M]$ and $G_i\approx[E_i(z_M)/z_M]$. Moreover, in the limit $z_H\approx z_M\gg 1$, the dilogarithm function behaves as\footnote{Here, we have used the identities: $\text{Li}_2 (1 - z)+\text{Li}_2 (1 - 1/z)=-\log^2(z)/2$ and $\text{Li}_2 (1)=\pi^2/6$.} $\text{Li}_2 (1 - z_M)\approx -\log^2(z_M)/2$. Thus, in the decoupling limit, the dominant contribution in $d_{e,f}^{CC}$ coming from $\varsigma_u$ is proportional to the \textit{double logarithm} $\log^2(z_M)/z_M$ (where $z_M = (M^2/m_t^2)$), whereas the dominant contributions coming from $\varsigma_d$ are proportional to the \textit{single logarithm} $\log(z_M)/z_M$ (where $z_M = (M^2/m_t^2)$ for the $tb$ loop contribution and $(M^2/m_b^2)$ for the $cb$ and $ub$ loop contributions); this results in the following expressions:
\begin{align}
	\frac{d_{e,f}^{CC}}{e} \bigg|_{\mathlarger{\varsigma_u}}\, \aprx \;\; &m_e\, \frac{\sqrt 2\, \alpha\, G_F}{(4 \pi)^3\, s_w^2}\, N_C \, Q_t \,\text{Im}(\varsigma_u^* \varsigma_l)  \frac{m_t^2}{M^2} \log^2 \left( \frac{M^2}{m_t^2}\right),
	\label{eqn:CC_dec_u}\\
    \frac{d_{e,f}^{CC}}{e} \bigg|_{\mathlarger{\varsigma_d}}\, \aprx \;\; &m_e\, \frac{\sqrt 2\, \alpha\, G_F}{(4 \pi)^3\, s_w^2}\, N_C \left(\frac{Q_t+Q_b}{2}\right)\text{Im}(\varsigma_d^* \varsigma_l) \Bigg\{ |V_{tb}|^2\,\frac{m_b^2}{m_t^2} \frac{m_t^2}{M^2} \log \left( \frac{M^2}{m_t^2}\right) \nonumber \\
    &+ \sum_{i=u,c} |V_{ib}|^2\,\frac{m_b^2}{M^2} \log \left( \frac{M^2}{m_b^2}\right) \Bigg\} \nonumber \\
    \approx \;\; & m_e\, \frac{\sqrt 2\, \alpha\, G_F}{(4 \pi)^3\, s_w^2}\, N_C \left(\frac{Q_t+Q_b}{2}\right)\;\text{Im}(\varsigma_d^* \varsigma_l) \; \frac{m_b^2}{M^2} \; \log \left( \frac{M^2}{m_t^2}\right),
    \label{eqn:CC_dec_d}
\end{align}
where we separated the single logarithmic terms coming from $cb$ and $ub$ loops using

\begin{equation}\label{eq:separate_log}
    \log(M^2/m_b^2) = \log(M^2/m_t^2) + \log(m_t^2/m_b^2) \,.
\end{equation}
(The logarithm $\log(m_t^2/m_b^2)$
is resummed via mixing effects below the EW scale, and we thus omit it in our comparison to the SMEFT.) Finally, adding the contributions from $tb$, $cb$ and $ub$ loops, one gets rid of the $|V_{ib}|^2$ factors using the orthogonality of the CKM matrix, i.e. $\sum_i |V_{ib}|^2 = 1$.
As discussed later, in the SMEFT approach the $\log^2(z_M)/z_M$ dependence arises from three effective operators through a chain of operator mixing, each link of the chain being determined at one-loop order, while the $\log(z_M)/z_M$ dependence emerges from the mixing of two effective operators at the two-loop order.

The total contribution of the A2HDM to the eEDM is given by:
\begin{equation}
    d_{e}^{}= \sum_{L,X}d_{e,L}^{X}+ (d_{e,\rm{Kite}}^{CC}+d_{e,\rm{Kite}}^{NC})\, ,
\end{equation}
where $L \in \{f,\, H^\pm,\, W \}$ and $X \in $ \{CC,\, NC,\, EM\} (see Appendix~\ref{sec:eEDM_detail}). After providing the charged-current contributions, we now give the leading terms for the remaining fermion-loop contributions in the decoupling limit.
On one hand, the expressions for the neutral-current contributions are:
\begin{align}
    \frac{d_{e, f}^{NC}}{e} \bigg|_{\mathlarger{\varsigma_u}}\, \aprx \;\; & \frac{\sqrt{2} \alpha G_F m_e}{(4 \pi)^3} \frac{N_C Q_t (1 - 4 Q_t s_w^2) (1 + 4 Q_e s_w^2)}{2 s_w^2 c_w^2} \text{Im}(\varsigma_u^* \varsigma_l) \frac{m_t^2}{M^2} \log^2\left( \frac{M^2}{m_t^2} \right), 
    \label{eqn:NC_dec_u}\\
    \frac{d_{e, f}^{NC}}{e} \bigg|_{\mathlarger{\varsigma_d}}\, \aprx \;\; & \frac{\sqrt{2} \alpha G_F m_e}{(4 \pi)^3} \frac{N_C Q_b (1 + 4 Q_b s_w^2) (1 + 4 Q_e s_w^2)}{2 s_w^2 c_w^2}\text{Im}(\varsigma_d^* \varsigma_l) \frac{m_b^2}{M^2} \log \left( \frac{M^2}{m_Z^2} \right) \nonumber \\
    \approx \;\; & \frac{\sqrt{2} \alpha G_F m_e}{(4 \pi)^3} \frac{N_C Q_b (1 + 4 Q_b s_w^2) (1 + 4 Q_e s_w^2)}{2 s_w^2 c_w^2}\text{Im}(\varsigma_d^* \varsigma_l) \frac{m_b^2}{M^2} \log \left( \frac{M^2}{m_t^2} \right),
    \label{eqn:NC_dec_d}
\end{align}
where we separated the logarithm in the bottom-loop contribution into two terms using an expression analogous to Eq.~\eqref{eq:separate_log}. On the other hand, for the EM contributions we have: 
\begin{align}
    \frac{d_{e, f}^{EM}}{e} \bigg|_{\mathlarger{\varsigma_u}}\, \aprx \;\; &- \frac{\sqrt{2} \alpha G_F m_e}{(4 \pi)^3} 8 N_C Q_t^2 Q_e \text{Im}(\varsigma_u^* \varsigma_l) \frac{m_t^2}{M^2} \log^2\left( \frac{M^2}{m_t^2} \right), 
    \label{eqn:EM_dec_u}\\
    \frac{d_{e, f}^{EM}}{e} \bigg|_{\mathlarger{\varsigma_d}}\, \aprx \;\; & \frac{\sqrt{2} \alpha G_F m_e}{(4 \pi)^3} 8 N_C Q_b^2 Q_e \text{Im}(\varsigma_d^* \varsigma_l) \frac{m_b^2}{M^2} \log \left( \frac{M^2}{m_b^2} \right) \nonumber \\
    \approx \;\; & \frac{\sqrt{2} \alpha G_F m_e}{(4 \pi)^3} 8 N_C Q_b^2 Q_e \text{Im}(\varsigma_d^* \varsigma_l) \frac{m_b^2}{M^2} \log \left( \frac{M^2}{m_t^2} \right).
    \label{eqn:EM_dec_d}
\end{align}

Adding Eqs.~\eqref{eqn:CC_dec_u}, \eqref{eqn:NC_dec_u} and \eqref{eqn:EM_dec_u} and substituting the values $N_C=3$, $Q_t=2/3$ and $Q_e=-1$, we get the leading squared logarithm coming from fermion-loop Barr-Zee contributions with a dependence on $\varsigma_u$:
\begin{align}
    \frac{d_{e, f}}{e} \bigg|_{\mathlarger{\varsigma_u}} \, = \;\; & \sum_{X} \frac{d_{e, f}^X}{e} \bigg|_{\mathlarger{\varsigma_u}} \, \aprx \;\; \frac{\sqrt{2} \alpha G_F m_e}{(4 \pi)^3} \frac{3+5t_w^2}{s_w^2} \text{Im}(\varsigma_u^* \varsigma_l) \frac{m_t^2}{M^2} \log^2\left( \frac{M^2}{m_t^2} \right) \nonumber \\
    = \;\; & m_e \frac{g^2}{(4 \pi)^4 v^2} (3+5t_w^2) \text{Im}(\varsigma_u^* \varsigma_l) \frac{m_t^2}{M^2} \log^2\left( \frac{M^2}{m_t^2} \right),
    \label{eqn:BZf_dec_u}
\end{align}
where we used the relations $\alpha=e^2/(4\pi)$, $e=g s_w$ and $\sqrt{2}G_F=v^{-2}$. In a similar way, if we add Eqs.~\eqref{eqn:CC_dec_d}, \eqref{eqn:NC_dec_d} and \eqref{eqn:EM_dec_d} and substitute the value $Q_b=-1/3$, we get the leading single logarithm coming from fermion-loop Barr-Zee contributions with a dependence on $\varsigma_d$:
\begin{align}
    \frac{d_{e, f}}{e} \bigg|_{\mathlarger{\varsigma_d}} \, = \;\; & \sum_{X} \frac{d_{e, f}^X}{e} \bigg|_{\mathlarger{\varsigma_d}} \, \aprx \;\; -\frac{\sqrt{2} \alpha G_F m_e}{(4 \pi)^3} \frac{1}{2 c_w^2} \text{Im}(\varsigma_d^* \varsigma_l) \frac{m_b^2}{M^2} \log\left( \frac{M^2}{m_t^2} \right) \nonumber \\
    = \;\; & - m_e \frac{g^2}{(4 \pi)^4 v^2}\frac{t_w^2}{2} \text{Im}(\varsigma_d^* \varsigma_l) \frac{m_b^2}{M^2} \log\left( \frac{M^2}{m_t^2} \right).
    \label{eqn:BZf_dec_d}
\end{align}

Finally, from the sum of the gauge boson loop Barr-Zee contributions and charged-current kite diagrams, we get an additional logarithmic term in the decoupling limit \cite{Altmannshofer:2020shb}:

\begin{align}\label{eqn:HWk_dec}
    \sum_{X} \frac{d_{e, W}^X}{e} + \frac{d_{e,\rm{Kite}}^{CC}}{e} &\aprx \;\; \frac{\sqrt{2} \alpha G_F m_e}{(4 \pi)^3} \frac{3}{4 c_w^2} \sum_{i = 2, 3} \text{Im} \Big( y_l^{H_i}  \Big) \mathcal{R}_{i 1} \log \left( \frac{M^2}{m_W^2} \right) \nonumber \\
    & = \,\, m_e \frac{g^2}{(4 \pi)^4 v^2} \frac{3}{4} t_w^2 \sum_{i = 2, 3} \text{Im} \Big( \varsigma_l (\mathcal{R}_{i 2} + i \mathcal{R}_{i 3}) \Big) \mathcal{R}_{i 1} \log \left( \frac{M^2}{m_W^2} \right) \nonumber \\
    & = \,\, m_e \frac{g^2}{(4 \pi)^4} \frac{3}{4} t_w^2 \, \frac{\text{Im} (\lambda_6^* \varsigma_l)}{M^2} \log \left( \frac{M^2}{m_W^2} \right)
\end{align}
where the decoupling limit expression of the rotation matrix elements $\mathcal{R}_{ij}$ from Eq. (\ref{eqn: R_dec}) has been used in the last step.

For clarity, in Table~\ref{tab: CPV}
we show the vertices whose couplings are a source of CP violation in the A2HDM \cite{Jung:2013hka} and in the C2HDM \cite{Altmannshofer:2020shb}, and a comment about the cancellation (or not) of large logarithms in each category, in the decoupling limit. It is interesting to observe from Eqs.~\eqref{eqn:BZf_dec_u}-\eqref{eqn:HWk_dec} that, in the decoupling limit, the eEDM does not depend on the complex parameter $\lambda_5$.\footnote{This should not be confused with the $\widetilde{\lambda}_5$ in the C2HDM scenario, which is typically defined in the $\mathcal{Z}_2$-basis.}

Before concluding this section, a comment about gauge invariance is in order.
In studying the Barr-Zee diagrams one can first consider the one-loop sub-diagrams not carrying the lepton line. The corresponding one-loop functions for one external on-shell photon, one off-shell gauge boson and one off-shell scalar lines have been analyzed in great detail, and are not gauge invariant in general, see e.g. Ref.~\cite{Abe:2013qla}.
Gauge invariance is obtained when combining the Barr-Zee diagrams with some portions of the non-Barr-Zee diagrams, in a way termed as \textit{pinch technique} \cite{Papavassiliou:1989zd,Degrassi:1992ue,Papavassiliou:1994pr,Cornwall:2010upa}; see Appendix~\ref{sec:details_defCC} for further discussion.
Ref.~\cite{Altmannshofer:2020shb} shows the gauge invariance of the full contribution to the eEDM from the C2HDM using instead the \textit{background field gauge}
as well as the \textit{'t~Hooft $R_\xi$ gauge} explicitly \cite{Weinberg:1996kr}.
The loop functions that we reproduce from Refs.~\cite{Altmannshofer:2020shb,Altmannshofer:2025nsl} are independent of the choice of the gauge when added together. Barr-Zee diagrams built from internal fermion loops are gauge invariant by themselves.

\begin{table}[ht!]
\centering
{\renewcommand{\arraystretch}{1.7}
\begin{tabular}{c m{7em} m{6em} m{15em}} 
 \hline \hline
  & \text{(a)~CPV~in} A2HDM & \text{(c)~CPV~in} C2HDM & Cancellation of large logarithms in the decoupling limit \\ 
 \hline
 \hline
 Barr-Zee diagrams & & &  \\
 \hline
 \hline
 EM Fermion Loop & $H_i e e$, $H_i f f$ & $H_i e e$, $H_i f f$ & \multirow{2}{18em}{(c) $\mathcal{R}_{i1} \mathcal{R}_{i3} = \mathcal{O} \left( \frac{v^2}{M^2} \right)$ for $i=2,3$, $\sum\limits_{i=2,3}\mathcal{R}_{i2} \mathcal{R}_{i3}= \mathcal{O} \left( \frac{v^4}{M^4} \right)$} \\ 
 NC Fermion Loop & $H_i e e$, $H_i f f$ & $H_i e e$, $H_i f f$ & \\
 \hline
 CC Fermion Loop & $H^{\pm} e \nu$, $H^{\pm} f f'$ & \xmark & (c) \xmark \\
 \hline
 EM $W$ Loop & $H_i e e$ & $H_i e e$ & \multirow{4}{15em}{(a, c) Large logarithms do not cancel completely} \\ 
 NC $W$ Loop & $H_i e e$ & $H_i e e$ & \\
 CC $W$ Loop & $H^{\pm} e \nu$, $H^{\pm} W^{\mp} H_i$, $H^{\pm} W^{\mp} H_i \gamma$ & $H^{\pm} W^{\mp} H_i$, $H^{\pm} W^{\mp} H_i \gamma$ & \\
 \hline
 EM $H^{\pm}$ Loop & $H_i e e$ & $H_i e e$ & \multirow{4}{15em}{(a, c) $\mathcal{R}_{1i} \lambda_{h H^+ H^-} = \mathcal{O} \left( \frac{v^2}{M^2} \right)$ for $i=2,3$, and cancellation of identical terms} \\ 
 NC $H^{\pm}$ Loop & $H_i e e$ & $H_i e e$ & \\
 CC $H^{\pm}$ Loop & $H^{\pm} e \nu$, $H^{\pm} W^{\mp} H_i$, $H^{\pm} W^{\mp} H_i \gamma$ & $H^{\pm} W^{\mp} H_i$, $H^{\pm} W^{\mp} H_i \gamma$ & \\
 \hline
 \hline
 Kite diagrams & & &  \\
 \hline
 \hline
 Neutral Current & $H_i e e$ & $H_i e e$ & \multirow{2}{15em}{(a, c) $\mathcal{R}_{i1} \mathcal{R}_{i3} = \mathcal{O} \left( \frac{v^2}{M^2} \right)$ for $i=2,3$, and cancellation of identical terms} \\ 
 Charged Current & $H_i e e$ & $H_i e e$ & \\ 
 \hline \hline
\end{tabular}}
\caption{Vertices introducing CP violation (CPV) in each category of diagram contributing to the eEDM in the ``(a)'' A2HDM and in the ``(c)'' C2HDM, together with the origin of cancellations of possible large logarithms. In the vertices, ``$ff$'' represents fermions of the same flavours, and ``$ff'$'' represents fermions of different flavours. The cross (\xmark) indicates that no CP-violating coupling is present.
}\label{tab: CPV}
\end{table}

\section{Model-independent contributions to the eEDM}\label{sec:SMEFT}

In order to better comprehend the features of the A2HDM in the decoupling limit, such as the presence of double logarithms in some cases and only single logarithms in others, we are interested in this section in using the EFT approach in order to compute the leading contributions to the eEDM in the A2HDM. We will explicitly compare their expressions to the ones provided in the previous section. This will only be valid when the scale of NP (i.e. the masses of the heavy scalars) is much larger than the EW scale. This way, we can characterize their effects via Wilson coefficients $C_i$ accompanying SMEFT operators $Q_i$ of dimension higher than 4:
\begin{equation}
    \mathcal{L} \, = \, \mathcal{L}_{SM} + \sum_i C_i (\mu) Q_i.
\end{equation}

\noindent
Both the set of fundamental interactions and the scalar spectrum are distinct with respect to the SM, and these differences are all encoded at low energies in the sum on the right-hand side of this equation, starting from operators of dimension 6.
We remind the reader that ``pure SM contributions'', defined when $M \to + \infty$, are only relevant to the eEDM starting at the four-loop level, and are very suppressed. Therefore, at two loops we can entirely focus on those contributions to the eEDM accompanied by inverse powers of $M^2$.

Once generated, the coefficients of interest will run from the NP scale down to the EW scale, mixing with the coefficients of the EW dipole operators
\begin{align}
    Q_{e W} \, &= \, (\bar{l} \sigma^{\mu \nu} e) \tau^I H W_{\mu \nu}^I, \\
    Q_{e B} \, &= \, (\bar{l} \sigma^{\mu \nu} e) H B_{\mu \nu},
\end{align}

\noindent
of Wilson coefficients $C_{e W}$ and $C_{e B}$, respectively.
Then, a linear combination of both Wilson coefficients will give the electromagnetic dipole coefficient
\begin{equation}
    \mathscr{C}_{e \gamma} \, = \, c_w C_{e B} - s_w C_{e W},
\end{equation}
whose imaginary part is proportional to the EDM:
\begin{equation}
    d_e \, = \, - \sqrt{2} v \text{Im}(\mathscr{C}_{e \gamma}).
\end{equation}

The evolution of the Wilson coefficients is determined by the RGEs:
\begin{equation}
    \mu \frac{d}{d \mu} C_i \, = \, \left( \frac{1}{(4 \pi)^2} \gamma_{ij}^{(1)} + \frac{1}{(4 \pi)^4} \gamma_{ij}^{(2)} \right) C_j,
    \label{eqn: RGE}
\end{equation}
where $\gamma_{ij}^{(1)}$ and $\gamma_{ij}^{(2)}$ are the one-loop and two-loop elements of the anomalous-dimension matrix $\gamma$ that mix the coefficient $C_j$ into the coefficient $C_i$.

The leading contribution to the eEDM will come from dimension-6 operators.
In order to reproduce the leading logarithmic contributions that appear in the decoupling limit of the A2HDM, we only need to take into account four independent effective non-dipole operators, that are conveniently found in the Warsaw basis \cite{Grzadkowski:2010es}. In the following expressions $p$, $r$, $m$ and $n$ are flavour indices, and $y_q$, $y_u$, $y_d$, $y_l$ and $y_e$ are the hypercharges of left-handed quarks, right-handed up-type quarks, right-handed down-type quarks, left-handed leptons and right-handed charged leptons, respectively.
\begin{itemize}
    \item First, the dimension-6 Yukawa operator $Q_{eH}$
    \begin{equation}
        Q_{eH}^{pr} \, = \, (H^{\dagger}H)(\bar{l}_p e_r H),
    \end{equation}
    which mixes at the two-loop level into the electromagnetic dipole operator
    \begin{equation}\label{eq:mixing_QeH_dipole}
        \mu \frac{d}{d \mu} \mathscr{C}_{e \gamma}^{pr} \, = \, - \frac{g^3}{(4 \pi)^4} \frac34 s_w \Big( y_h - \frac12 + t_w^2 (y_l + y_e) y_h \left( 4 y_h - \frac23 \right) \Big) C_{eH}^{pr},
    \end{equation}
    thus generating a single logarithmic contribution
    \begin{equation}
        \mathscr{C}_{e \gamma}^{pr} \, = \, - \frac{e g^2}{(4 \pi)^4} \frac38 t_w^2 C_{eH}^{pr} \log \left( \frac{M^2}{m_{EW}^2} \right).
    \end{equation}
    A comment about the overall sign of the contribution to $\mathscr{C}_{e \gamma}^{pr}$ induced by $ Q_{eH} $ is in order. Ref.~\cite{Altmannshofer:2020shb} obtains a result consistent with the overall sign displayed above, while Refs.~\cite{Panico:2018hal,EliasMiro:2020tdv} obtain the opposite sign. The overall sign of Eq.~\eqref{eq:mixing_QeH_dipole} has been checked by one of the authors in Ref.~\cite{Jager:2019wkc}.\footnote{In Ref.~\cite{Jager:2019wkc}, one employs the convention $ D_\mu = \partial_\mu - i g_3 T^A A^A_\mu - i g_2 t^I W^I_\mu - i g_1 \text{y} B_\mu $ for the covariant derivative, opposite to the convention in use in Ref.~\cite{Alonso:2013hga}, which means that gauge couplings carry a relative minus sign between these two references.}
    \item Then, the four-fermion scalar operator $Q_{ledq}$
    \begin{equation}
        Q_{ledq}^{prmn} \, = \, (\bar{l}_p^j e_r) (\bar{d}_m q_n^j),
    \end{equation}
    which also mixes at the two-loop level into the electromagnetic dipole operator \cite{Panico:2018hal}\footnote{An independent verification is currently ongoing \cite{Silva:2020vyx,ValeSilva:2022nzm}, and will be the subject of a separate publication.}
    \begin{equation}
        \mu \frac{d}{d \mu} \mathscr{C}_{e \gamma}^{pr} \, = \, \sum_{m,n} \frac{g^3}{(4 \pi)^4} \frac{N_C}{4} s_w \Big( 3 y_q - \frac12 + 2 t_w^2 (y_l + y_e) (2 y_q^2 - y_q + 2 y_d^2) \Big) C_{ledq}^{prmn} [Y_d]_{nm},
    \end{equation}
    and also generates a single logarithmic contribution
    \begin{equation}
        \mathscr{C}_{e \gamma}^{pr} \, = \, \sum_{m,n} \frac{e g^2}{(4 \pi)^4} [Y_d]_{nm} \frac18 t_w^2 C_{ledq}^{prmn} \log \left( \frac{M^2}{m_{EW}^2} \right).
    \end{equation}
Here, $Y_f = \frac{\sqrt{2}}{v} M_f$ denotes the SM Yukawa matrix of the fermion type $f$, which is diagonal in the fermion mass-eigenstate basis.
    
    \item Finally, the four-fermion scalar operator $Q_{lequ}^{(1)}$
    \begin{equation}
        Q_{lequ}^{(1), prmn} \, = \, (\bar{l}_p^j e_r) \epsilon_{jk} (\bar{q}_m^k u_n),
    \end{equation}
    mixes into the four-fermion tensor operator $Q_{lequ}^{(3)}$
    \begin{equation}
        Q_{lequ}^{(3), prmn} \, = \, (\bar{l}_p^j \sigma_{\mu \nu} e_r) \epsilon_{jk} (\bar{q}_m^k \sigma^{\mu \nu} u_n)
    \end{equation}
    at the one-loop level via the RG equation \cite{Alonso:2013hga}
    \begin{equation}
        \mu \frac{d}{d \mu} C_{lequ}^{(3), prmn} \, = \, \frac{g^2}{(4 \pi)^2} \frac18 \left( -4(y_q + y_u) (2 y_e - y_q + y_u) t_w^2 + 3 \right) C_{lequ}^{(1), prmn},
    \end{equation}
    which then gets mixed into the electromagnetic dipole \cite{Jenkins:2013zja,Alonso:2013hga}
    \begin{equation}
        \mu \frac{d}{d \mu} \mathscr{C}_{e \gamma}^{pr} \, = \, \sum_{m,n} \frac{e}{(4 \pi)^2} \Big( 4 N_C (y_u + y_q) + 2 N_C \Big) [Y_u]_{nm} C_{lequ}^{(3), prmn},
    \end{equation}
    thus generating a squared logarithmic contribution to the eEDM:
    \begin{equation}
        \mathscr{C}_{e \gamma}^{pr} \, = \, \sum_{m,n} \frac{e g^2}{(4 \pi)^4} [Y_u]_{nm} \frac14 (5 t_w^2 + 3) C_{lequ}^{(1), prmn} \log^2\left( \frac{M^2}{m_{EW}^2} \right).
    \end{equation}
\end{itemize}


The expressions given above consist of the logarithmic contributions to the eEDM coming from SMEFT operators, which can be compared to the leading contributions from the full-model computation in the decoupling limit. In order to do this, we now determine the Wilson coefficients at the tree level, by integrating out the heavy scalars from the A2HDM. More details are found in Appendix~\ref{app:decoupling_limit}.

\begin{itemize}
    \item For the scalar-fermionic operator $Q_{eH}$, we select the indices $p,r = 1$, since we are computing the eEDM. Thus, we get the following Wilson coefficient:
    \begin{equation}
        C_{eH}^{11} \, = \, \frac{\sqrt{2} m_e}{v} \frac{\lambda_6^* \, \varsigma_l}{M^2},
    \end{equation}
    so the leading logarithmic contribution to the eEDM coming from $Q_{eH}$ is:
    \begin{equation}
        \frac{d_{e, \, eH}^{\text{SMEFT}}}{e} \, = \, m_e \frac{g^2}{(4 \pi)^4} \frac34 t_w^2 \, \frac{\text{Im} (\lambda_6^* \, \varsigma_l)}{M^2} \log \left( \frac{M^2}{m_{EW}^2} \right).
    \end{equation}
    
    We find that this contribution coming from SMEFT is the same as the one already found in Eq.~\eqref{eqn:HWk_dec}, which was obtained from the sum of gauge boson loop Barr-Zee and charged-current kite diagrams in A2HDM. Note that this contribution only depends on $ \lambda_6 $, which is the coefficient that accompanies the operator $(\Phi_1^\dagger \Phi_1) (\Phi_1^\dagger \Phi_2)$ in Eq.~\eqref{eq:pot}. Aside from its complex conjugate, this is the only dimension-4 operator carrying three light fields ($\Phi_1$), and a heavy one ($\Phi_2$), which when integrated out in combination with the Yukawa interaction produces the dimension-6 $Q_{eH}$ operator; see Appendix~\ref{app:decoupling_limit} for further discussion (from where it is also transparent the effect at low energies produced by the dimension-2 $\Phi_1^\dagger \Phi_2$ operator). Consistently, this means that no dependence on $\lambda_5$ or $\lambda_7$ (which was chosen real, by a rephasing of $\Phi_2$) in the decoupling limit was to be expected. Note that by choosing $\lambda_6$ real (instead of $\lambda_7$) by a rephasing of $\Phi_2$, the expression of $\text{Im} (\lambda_6^* \, \varsigma_l)$ above simplifies to $\lambda_6 \, \text{Im} (\varsigma_l)$, but we cannot choose $ \varsigma_l $ real simultaneously.
    \item For the four-fermion scalar operator $Q_{ledq}$, we select again the indices $p,r = 1$, and $m = 3$, since the main contributions come from diagrams involving a bottom quark in the loop. Thus, we get the following Wilson coefficient:
    \begin{equation}
        C_{ledq}^{113n} \, = \, 
        \frac{\varsigma_l [Y_e]_{11} \varsigma_d^* [Y_d^{\dagger}]_{3n}}{M^2} 
       \, = \, \frac{2}{v^2} \varsigma_d^* \varsigma_l \frac{m_e m_b}{M^2}\,\delta_{3n},
    \end{equation}
    so the leading logarithmic contribution to the eEDM coming from $Q_{ledq}$ is:
    \begin{align}\label{eq:ledq_SMEFT}
        \frac{d_{e,b}^{\text{SMEFT}}}{e} \, = \, - m_e \frac{g^2}{(4 \pi)^4 v^2} \frac{t_w^2}{2} \text{Im}(\varsigma_d^* \varsigma_l) \frac{m_b^2}{M^2} \log \left( \frac{M^2}{m_{EW}^2} \right).
    \end{align}
    
    We find that this contribution from SMEFT is the same as the one found in Eq.~\eqref{eqn:BZf_dec_d}, which was obtained from fermion-loop Barr-Zee diagrams in the A2HDM.
    \item Finally, for the four-fermion scalar operator $Q_{lequ}^{(1)}$, we select again $p,r = 1$, but now $n = 3$, since the main contributions come from diagrams involving a top quark. Thus, we get the Wilson coefficient:
    \begin{equation}
        C_{lequ}^{(1),11m3} \, = \, 
     -\frac{\varsigma_l [Y_e]_{11} \varsigma_u^* [Y_u^{\dagger}]_{m3}}{M^2} 
        \, = \, -\frac{2}{v^2} \varsigma_u^* \varsigma_l \frac{m_e m_t}{M^2}\,\delta_{m3},
    \end{equation}
    so the leading logarithmic contribution to the eEDM coming from $Q_{lequ}$ is:
    \begin{align}\label{eq:lequ_SMEFT}
        \frac{d_{e,t}^{\text{SMEFT}}}{e} \, = \, m_e \frac{g^2}{(4 \pi)^4 v^2} (5 t_w^2 + 3) \text{Im}(\varsigma_u^* \varsigma_l) \frac{m_t^2}{M^2} \log^2 \left( \frac{M^2}{m_{EW}^2} \right).
    \end{align}
    In this last case we find that this contribution coming from SMEFT is the same as the one found in Eq.~\eqref{eqn:BZf_dec_u}, which was obtained from fermion-loop Barr-Zee diagrams in the A2HDM.
\end{itemize}

As previously stressed, the A2HDM carries extra logarithmically-enhanced contributions with respect to more studied versions of 2HDMs where a $\mathcal{Z}_2$ symmetry is enforced, which consist of the later two sets of contributions, Eqs.~\eqref{eq:ledq_SMEFT} and \eqref{eq:lequ_SMEFT}.
It is clear that due to the $ SU(2)_L \otimes U(1)_Y $ gauge symmetry a particular SMEFT contribution encodes in the decoupling limit the charged-current, neutral-current and EM gauge boson contributions found in the discussion of the full model, or equivalently it collects simultaneously the contributions from heavy neutral and charged scalars at the origin of the contact four-fermion interactions.
This fact helps us understanding why large logarithmic contributions to the eEDM associated with fermion loops must be absent in constrained versions of 2HDMs, where the charged-current category does not contribute to the eEDM since the $ H^\pm e \nu $ and $ H^\pm f f' $ couplings, together with the coupling $ W^\mp f' f $, lead to real contributions:
in this case, neutral-current and EM gauge boson contributions alone cannot match a gauge invariant counter-part in SMEFT, and cannot thus carry large logarithms in the decoupling limit.

Similarly, we have already pointed out that due to the structure of the A2HDM in study charged-current lepton-loop Barr-Zee diagrams do not contribute to the eEDM, since the CP-violating parameter $ \varsigma_l $ is the same across lepton generations. Here as well, no large logarithms induced by SMEFT operators of dimension 6 can be present.\footnote{Large logarithms could still appear in association with the dimension-8 SMEFT operator $ (\bar{L} e H) (\bar{L} e H) $ \cite{Davidson:2016utf}.}
The A2HDM can be trivially generalized with diagonal alignment matrices instead of the 3 alignment parameters $\varsigma_f$ (i.e. the generalized alignment framework \cite{Penuelas:2017ikk}). In that case there would be lepton-loop contributions to the eEDM proportional to $\text{Im} (\varsigma^*_l \varsigma_e)$.

On the other hand, charged
scalar loops do not introduce large logarithmic contributions to the eEDM in the decoupling limit. This is due to the structure of the effective operators produced in the decoupling limit, which are not of the kind $ H^2 X^2 $ with a dual field strength tensor, or which are of mass dimension higher than six.

At this point, it is worth mentioning that there are some logarithms important for the phenomenological analysis of the next section that we are not able to deal with based on the RGEs provided above for the SMEFT. Namely, resumming the sub-leading single logarithms $ m_t^2 / M^2 \times \log (m_t / M) $ (that we have not attempted to identify in our discussion of the full model) based on the SMEFT would require identifying the anomalous-dimension matrix element at two loops describing the direct mixing of the four-fermion operator $ Q_{\ell e q u}^{(1)} $ into dipole operators, as well as finite one-loop contributions (i.e., one-loop matching effects) in the chain of RGEs involving $ Q_{\ell e q u}^{(1)} $ and $ Q_{\ell e q u}^{(3)} $, which are both sub-leading contributions to the RGEs.
In contrast to the leading double logarithm, the sub-leading single logarithm involves in principle arbitrary powers of the square of the unsuppressed top Yukawa coupling, i.e., $y_t^2$, $y_t^4$, etc.
Therefore, based on currently known ingredients needed in the RGEs of the SMEFT Wilson coefficients we only discuss the leading powers of the logarithms $ \log ( M^2 / m_{EW}^2 ) $ in the decoupling limit.

As previously pointed out, other phenomenologically important logarithms also appear, such as $ \log ( m_{EW}^2 / m_b^2 ) $. Their discussion, however, goes beyond the scope of this paper, since their analysis would require considering a low-energy EFT able to resum such logarithms systematically. We however depict next the main features of this EFT.

The presence of the dimension-6 Yukawa SMEFT operator $Q_{eH}$ will lead at the tree level to four-fermion low-energy EFT operators below the EW scale, with a Wilson coefficient proportional to the combination of scalar potential parameters $\lambda_6^\ast / \mu_2$.
A similar comment holds for the analogous quark operators, $Q_{uH}$ and $Q_{dH}$.
Below the EW scale, the four-fermion SMEFT operator $Q_{l e d q}$ will also match at the tree level onto four-fermion operators of the low-energy EFT, this time involving only the parameter $\mu_2$ from the scalar potential.
These effects must also be combined with other dimension-6 operators, suppressed by inverse powers of the EW scale instead, namely $1/v^2$, resulting from the interchange at the tree level of heavy EW gauge bosons, which must be combined with the latter dimension-6 operators suppressed by $1/M^2$.
A complete discussion, e.g. in the case of the operator $Q_{l e q u}^{(1)}$, singles out the top flavour, when compared to the bottom or lighter flavours, which must also be integrated out at the EW scale.
In the matching to the low-energy EFT, the previously discussed dipole operators produced via RGE effects must also be taken into account, but in their case only their tree-level contributions matter to our discussion at the two-loop level.
Altogether, this long basis of effective operators must provide an account of the dependence at two loops on the arbitrary scale at which one moves to the low-energy EFT starting from the SMEFT framework.
Of course,
the EFT below the EW scale also includes the renormalizable interactions of the EM and QCD gauge boson fields.


\section{Phenomenology}\label{sec:phenomenology}

We now explore how the incorporation of the novel charged-current fermion-loop Barr-Zee contributions affects the prediction of the eEDM. We use the following set of inputs for the SM parameters:
\begin{align}
m_{\tau} &= 1.777 \, \text{GeV} \,,   &  m_{W} &= 80.34 \, \text{GeV} \,, \nonumber \\
m_{b} &= 2.88 \, \text{GeV} \,,   &  m_{Z} &= 91.19 \, \text{GeV} \,, \nonumber \\
m_{t} &= 163 \, \text{GeV} \,,   &  m_{h} &= 125 \, \text{GeV} \,, \nonumber \\
\alpha &= 1/129 \,,   &  v &= 246 \, \text{GeV},
\end{align}
with $c_w = m_W/m_Z$, and the running bottom and top quark masses are taken at the EW scale. Although not producing enhanced logarithmic contributions reproduced by the SMEFT as previously pointed out, tau loops can still provide sub-leading contributions to the eEDM.

The first steps towards a global fit analysis of the A2HDM parameter space have been taken, but the results will be the subject of a separate work.
We will therefore only discuss a benchmark point, which displays the main features we would like to illustrate.
We fix the A2HDM parameters to the following benchmark values consistent with Ref.~\cite{Karan:2023kyj}:
\begin{align}\label{eq:benchmark_values}
\lambda_3 &= 0.02 \,,  &  \lambda_4 &= 0.04 \,, \nonumber \\
\lambda_7 &= 0.03 \,,  &  \text{Re}(\lambda_5) &= 0.05 \,, \nonumber \\
\text{Re}(\lambda_6) &= -0.05 \,,   &  \text{Im}(\lambda_6) &= 0.01 \,, \nonumber \\
\alpha_3 &= \pi/6.
\end{align}
The remaining parameters of the scalar potential are either determined by this set, or not relevant to our analysis. For instance, the parameters $\mu_1$ and $\mu_3$ are obtained from the minimization conditions of the
scalar potential (see Eq.~\eqref{eq:min} in Appendix~\ref{sec:pot}). In the decoupling limit, the parameters $\lambda_1$ and Im$(\lambda_5)$ are fixed by the mass of the SM-like Higgs $m_h$ and the mixing angle $\alpha_3$ between the second and third neutral scalars, respectively (see Eqs.~\eqref{eq:mass_eigenvalues} and \eqref{eq:lm5}), while the parameter $\lambda_2$ that accompanies the operator $  (\Phi_2^\dagger \Phi_2)^2 $ remains irrelevant to our discussion.

\begin{figure}[t]
    \centering
    \includegraphics[scale=0.3]{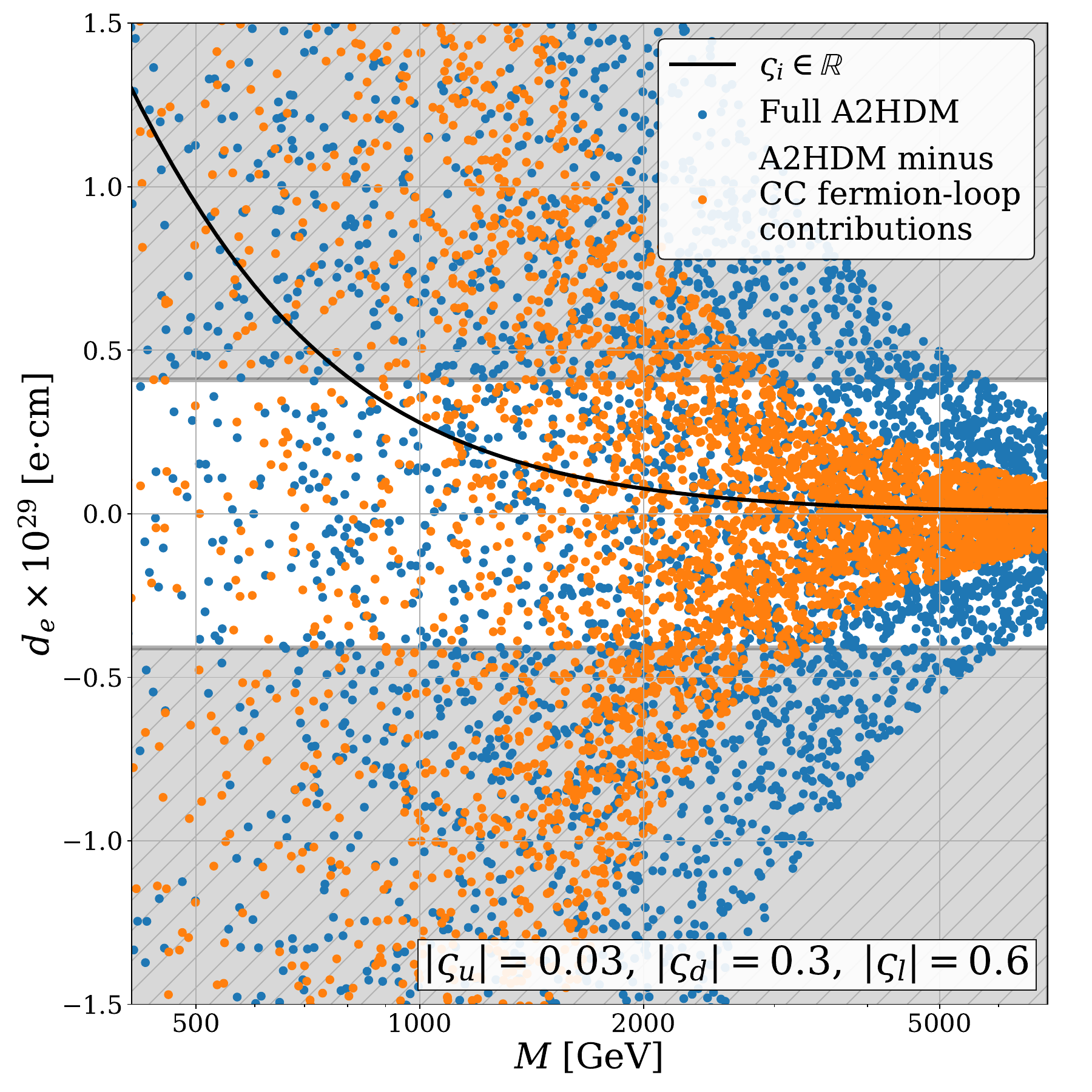}
    \caption{Scatter plot of the eEDM in the A2HDM as a function of $M$, at the benchmark point Eq.~\eqref{eq:benchmark_values}. The absolute values of the alignment parameters are taken as: $|\varsigma_u|=0.03$, $|\varsigma_d|=0.3$ and $|\varsigma_l|=0.6$. The blue dots correspond to the full A2HDM, while the expression at the origin of the orange dots does not include charged-current (CC) fermion-loop Barr-Zee contributions. The black line assumes real alignment parameters $\varsigma_i$. The gray bands show the upper bound on the modulus of the eEDM from Eq.~\eqref{eq:current_exp_value}.
    \label{fig:pheno_scatter}}
\end{figure}

By varying the parameter $\mu_2$, we change the mass scale $M$ and study the dependence of the eEDM on it.
In order to work in the decoupling limit (i.e. $\mu_2\gg \lambda_i v^2$) for the whole range of masses considered in the plot, i.e. from 400~GeV to 6~TeV, all the dimensionless parameters ($\lambda_i$, except for $\lambda_1$) from the scalar potential are considered to be $\mathcal O(10^{-2})$. In order to understand this choice, we need to take a look at Appendix \ref{sec:pot}, in particular Eqs. (\ref{eq:mu2l1l6}) and (\ref{eq:mass_eigenvalues}), where the masses of the neutral and charged scalars are defined in terms of the scalar potential parameters. If a value of $\sqrt{\mu_2} \sim \mathcal O (500 \text{ GeV})$ is considered, the ratio $v^2/\mu_2$ takes a value of approximately $0.25$. Consequently, by choosing the parameters $\lambda_i\sim\mathcal O(10^{-2})$, the masses of the new scalars remain within $1\%$ of $\sqrt\mu_2$, thereby validating the decoupling limit framework even at 400~GeV.

In Fig.~\ref{fig:pheno_scatter}, we show a scatter plot of the eEDM, where random values are given to the parameter $\sqrt{\mu_2} \approx M_{H_2}\approx M_{H_3}\approx M_{H^{\pm}}=M$, and the phases of the flavour alignment parameters $\varsigma_i$ (with $i=u,d,l$), while keeping their moduli fixed at $|\varsigma_u|=0.03$, $|\varsigma_d|=0.3$ and $|\varsigma_l|=0.6$, which is consistent with Ref.~\cite{Karan:2023kyj}. The blue dots are predictions from the full A2HDM, while the orange ones are computed by subtracting the new charged-current fermion-loop Barr-Zee contributions from the full result. The black line shows the prediction for a scenario with real flavour alignment parameters\footnote{While the parameters $\varsigma_u$ and $\varsigma_d$ were chosen to be positive in this case, the parameter $\varsigma_l$ was chosen to be negative so that the total contribution to the eEDM has a positive value.} (similar to the C2HDM scenario) in which the only sources of CP violation come from the parameters $\text{Im}(\lambda_5)$ and $\text{Im}
(\lambda_6)$, although the contribution from $\text{Im}(\lambda_5)$ in the decoupling limit is negligible. Finally, we show in gray the upper bound on the modulus of the eEDM from Eq.~\eqref{eq:current_exp_value}, which was obtained neglecting the electron-nucleon coupling $C_S$ (whose detailed analysis will be the subject of a future work). The separation between the clusters of blue and orange points indicates that the charged-current fermion loop, which is absent in the C2HDM, can significantly contribute to the eEDM in the A2HDM scenario. Furthermore, the deviation of the orange or blue points from the black curve highlights the crucial role played by the phases of the alignment parameters in determining the eEDM in the A2HDM. Notably, the current experimental limit on the eEDM can be easily satisfied with complex $\varsigma_i$ even at a relatively low mass of 400~GeV, which is challenging to achieve with real $\varsigma_i$ parameters, as in the case of the C2HDM.

In the A2HDM, the main contributions to the eEDM are very sensitive to the flavour alignment parameters $\varsigma_i$ (both to their moduli and phases). Since different contributions are dependent on different $\varsigma_i$, this can lead to cancellations that lower the value of the eEDM, causing it to drop below the current constraints for the whole mass range assumed in the previous plot. It is also worth mentioning the cancellation that can happen between fermionic and non-fermionic contributions depending on the value of the alignment parameters. In particular, we find strong cancellations between fermion-loop Barr-Zee contributions (which are proportional to the coefficients $\text{Im}(\varsigma_{u, d}^* \varsigma_l)$ in the decoupling limit) and several $W$ boson loop contributions (which are dependent on the coefficient $\text{Im}(y_e^i) = \mathcal{R}_{i 2} \text{Im}(\varsigma_l) + \mathcal{R}_{i 3} \text{Re}(\varsigma_l)$) for the particular values of Eq.~\eqref{eq:benchmark_values}.
Further discussion about accidental
cancellation mechanisms operating in 2HDMs is found in Refs.~\cite{Bian:2014zka,Fuyuto_2020,Kanemura:2020ibp,Enomoto:2021dkl}.
Other than accidental suppression mechanisms, we have detailed in previous sections the cancellation mechanism stemming from the presence of a $\mathcal{Z}_2$ symmetry, that would limit the contributions of large logarithms in the decoupling limit.

\begin{figure}[t]
    \centering
    \includegraphics[scale=0.3]{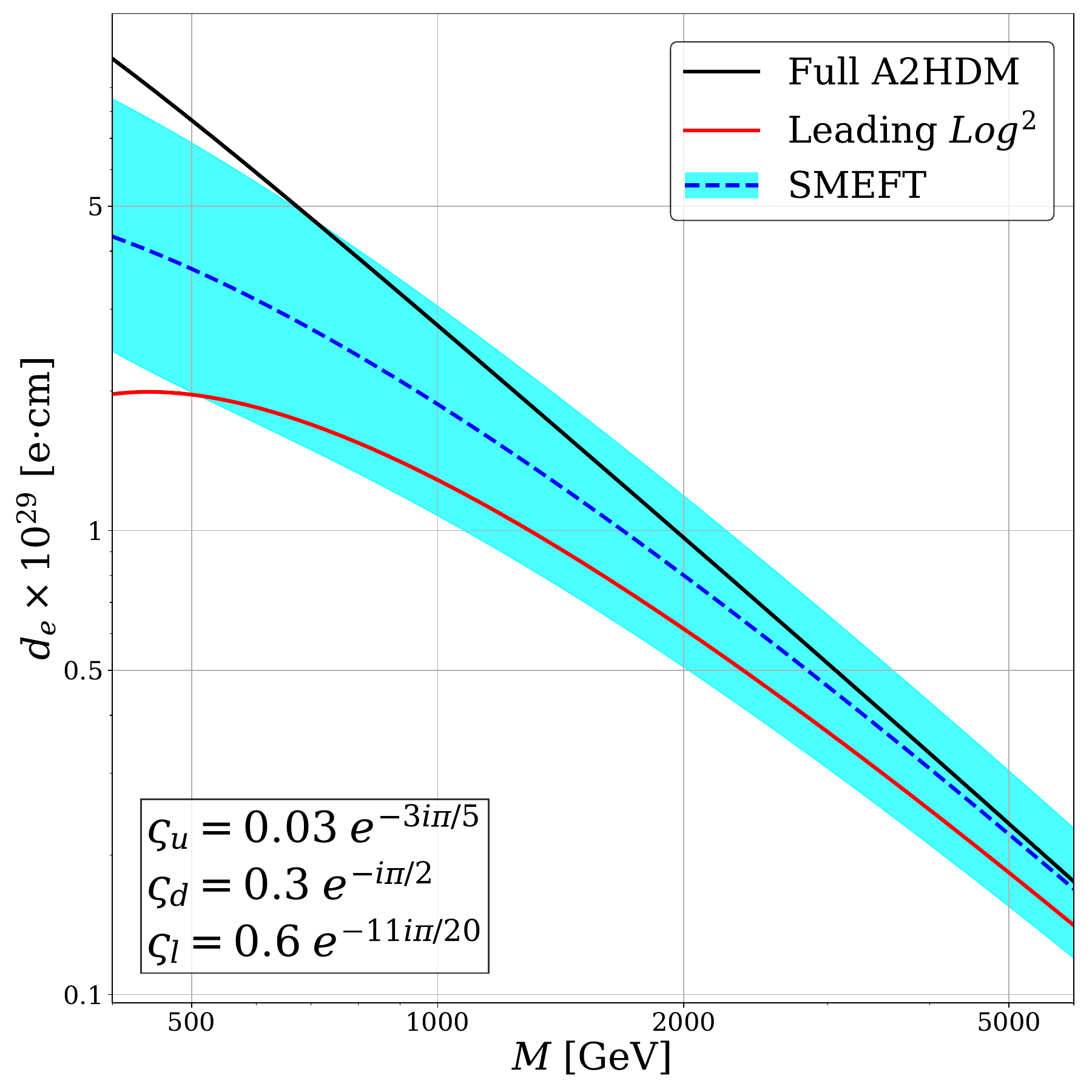}
    \caption{Approximations to predictions of the eEDM in the A2HDM as a function of $M$, at the benchmark point in Eq.~\eqref{eq:benchmark_values}. The black line is the full two-loop result in the A2HDM. The solid red curve is the leading squared logarithmic approximation, and the dashed blue curve includes some sub-leading logarithms from SMEFT. The shaded blue region is obtained by varying the UV scale from $M/2$ up to $2M$.
    The $\mathcal{Z}_2$-symmetric case carries no large squared-logarithm in the decoupling limit.
    }
    \label{fig:pheno_num}
\end{figure}

In Fig.~\ref{fig:pheno_num}, we numerically compare several approximations to the eEDM as a function of the mass scale $M$ of the new scalars, for particular values of the alignment parameters: $\varsigma_u=0.03 \,e^{-3i\pi/5}$, $\varsigma_d=0.3 \,e^{-i\pi/2}$ and $\varsigma_l=0.6 \,e^{-11i\pi/20}$. All other parameters are fixed according to the benchmark point in Eq.~\eqref{eq:benchmark_values}.
With these values, the pre-factor that multiplies the bottom-loop Barr-Zee diagrams are much below than the ones for the top-loop and the EW-loop Barr-Zee diagrams, i.e., $ ( \text{Im}(\lambda_6 \varsigma_l^\ast) \, v ) / ( \sqrt{2} \, \text{Im}(\varsigma_u \varsigma_l^\ast) \, m_t ) \sim 10 $ while $ ( \text{Im}(\lambda_6 \varsigma_l^\ast) \, v ) / ( \sqrt{2} \, \text{Im}(\varsigma_d \varsigma_l^\ast) \, m_b ) \sim -70 $. However, the top-loop contribution is accompanied by an additional large logarithm in the decoupling limit, which also impacts the hierarchy of contributions.
The black line in Fig.~\ref{fig:pheno_num} shows the result of the full two-loop calculation in the A2HDM. The solid red curve shows the leading squared logarithmic approximation. For the values of $M$ displayed in the plot, it manages to give a rather accurate prediction for the eEDM. Finally, the dashed blue line shows the SMEFT result (also including the sub-leading single logarithms discussed above, namely, related to the SMEFT operators $Q_{ledq}$ and $Q_{eH}$), and the shaded band is obtained by varying the Ultra-Violet (UV) scale $\mu$ at which the SMEFT logarithmic terms are evaluated, between $\mu=M/2$ and $\mu=2M$. For large values of $M$, the SMEFT and the full model descriptions match quite well, though by construction some sub-leading single-logarithmic terms are not contained in
the SMEFT curve (namely, the single logarithm associated to the top-quark mass squared).
All contributions to the eEDM being suppressed by the scale of NP squared $M^2$, logarithmically enhanced contributions gain in importance as we go deeper into the decoupling limit.

\section{Conclusions}\label{sec:conclusions}

In this article we present a discussion about the contributions stemming from the A2HDM to the electron EDM. In contrast to more constrained versions of the 2HDM, the A2HDM carries extra sources of CP violation in the fermionic sector while still avoiding large FCNCs.
In particular, this leads to new contributions to EDMs under the form of Barr-Zee diagrams with fermion loops and the exchange of charged bosons, which are absent in $\mathcal Z_2$-symmetric 2HDMs.
Another contribution to CP violation in EDMs comes from effective couplings of electrons to nucleons.
Due to hadronic uncertainties and the need to extend our discussion to a low-energy EFT, we do not discuss this category of contributions in details, focusing on the electron dipole generated by the top- and bottom-loop contributions, which at the order analyzed do not carry sizable hadronic uncertainties.

Our main focus consists of the contributions from the A2HDM in the decoupling limit. The features observed in this limit
can be explained by relying on the analysis of dimension-6 operators of SMEFT.
We then explicitly show that the leading logarithmic contributions proportional to the inverse of the decoupling-scale squared are described by a few effective operators of dimension 6, including semileptonic four-fermion operators, that mix into the electron dipole due to one- and two-loop anomalous-dimension matrix elements.
The comparison was only possible after reviewing relative and overall signs in the literature, for A2HDM and SMEFT expressions.
The latter renormalization-group mixing produces a pattern of double and single logarithmic enhanced contributions.
These contributions are absent in the more studied 2HDMs, where a discrete $\mathcal Z_2$ symmetry controls the basis of effective operators necessary in the decoupling limit. In other words, the $\mathcal Z_2$ symmetry controls the ways in which logarithmically-enhanced contributions can appear, therefore providing a suppression mechanism. Additionally, the $\mathcal Z_2$-symmetric 2HDM and the A2HDM frameworks receive contributions from the mixing of Yukawa interactions of dimension 6 into dipole operators, more studied in the literature.
We stress that the discussion of logarithmic contributions concerns solely CP violation, while CP-conserving contributions to the dipole structure would not suffer from the same suppression.

In our phenomenological study, we find that the contributions from charged-current fermion-loop Barr-Zee diagrams can be dominant with respect to other contributions to the eEDM, depending on the values assumed by the parameters of the scalar potential and the alignment parameters.
Apart from the cancellation mechanism operating when a $\mathcal{Z}_2$ symmetry is 
imposed, thus controlling the presence of some logarithmically-enhanced contributions, accidental cancellation mechanisms can also take place,
for instance between different fermion-loop Barr-Zee diagrams, or between fermionic and non-fermionic Barr-Zee diagrams. Therefore, with respect to the C2HDM, with the same values of the parameters in both cases, we can have lower masses of the NP scalars and still satisfy the experimental upper bound, just by varying the phases of the alignment parameters.

It would be interesting to study more systematically effects that appear in the A2HDM, or even in more general 2HDM realizations,
while being absent in more constrained versions of 2HDMs.

\section*{Acknowledgements}

We are happy to thank Wolfgang Altmannshofer, Joachim Brod, Sebastian J\"ager, Kevin Monsalvez Pozo, Verónica Sanz and Mustafa Tabet for useful discussions.
This work is supported by the Spanish Government (Agencia Estatal de Investigación MCIN/AEI/10.13039/501100011033) Grants No.  PID2020–114473GB-I00 and No. PID2023-146220NB-I00, and CEX2023-001292-S (Agencia Estatal de Investigación MCIU/AEI (Spain) under grant IFIC Centro de Excelencia Severo Ochoa).
We receive support from the Generalitat Valenciana (Spain) through the plan GenT program CIDEGENT/2021/037.
The work of JMD is also supported by Generalitat Valenciana, grant CIACIF/2023/409.
The work of EP is also supported by the U.S. National Science Foundation under grant PHY-2310149. The work of AK has partially been supported by the research grant number 20227S3M3B “Bubble Dynamics in Cosmological Phase Transitions” under the program PRIN 2022 of the Italian Ministero dell’Università e Ricerca (MUR).

\appendix

\section{Scalar potential and spectrum}
\label{sec:pot}

The scalar potential of the A2HDM is given by Eq.~\eqref{eq:pot}. Applying the minimization conditions of the scalar potential with respect to the two scalar fields at their corresponding VEVs, one obtains:
\begin{equation}
\label{eq:min}
v^2=-\frac{2\mu_1}{\lambda_1}=-\frac{2\mu_3}{\lambda_6}\, ,
\end{equation}
which removes the dependency of the model on the parameters $\mu_1$ and $\mu_3$ if $\lambda_1$ and $\lambda_6$ are known. Furthermore, by redefining the phase of $\Phi_2$, one of the three parameters $\lambda_{\{5,6,7\}}$ can be made real; for our convenience we assume $\lambda_7$ to be real. Therefore, given that the value of $v$ is known, the scalar potential depends on ten parameters: $\mu_2$, $\lambda_{\{1,2,3,4\}}$, $|\lambda_{\{5,6,7\}}|$, and two relative phases between $\lambda_{\{5,6,7\}}$.

Denoting the neutral scalars in the Higgs-basis as $\mathbf{S}\equiv(S_1,S_2,S_3)^T$, the mass terms from the scalar potential in Eq.~\eqref{eq:pot} can easily be identified as:
\begin{equation}
    V_M = M_{H^\pm}^2 H^+H^-+\frac{1}{2}\,\mathbf{S}^T \mathcal{M}\,\mathbf{S}\,, \quad \text {with}\quad M_{H^\pm}^2= \mu_2+\frac{1}{2}\lambda_3 v^2 \,,
\end{equation}
and 
\begin{equation}
\mathcal{M}=\begin{pmatrix} v^2\, \lambda_1  && v^2\, \text{Re}(\lambda_6) && -v^2\, \text{Im}(\lambda_6) \\
    v^2\, \text{Re}(\lambda_6) && \mu_2+\frac{v^2}{2}\{\lambda_3+\lambda_4+\text{Re}(\lambda_5)\} && -\frac{1}{2}v^2\, \text{Im}(\lambda_5) \\
    -v^2\, \text{Im}(\lambda_6) && -\frac{1}{2}v^2\, \text{Im}(\lambda_5) && \mu_2+\frac{v^2}{2}\{\lambda_3 + \lambda_4 - \text{Re}(\lambda_5) \} \end{pmatrix}.
    \label{eq:massmatrix}
\end{equation}
This mass matrix $\mathcal{M}$ can be diagonalized by an orthogonal transformation $\R$ as:
\begin{equation}
  \mathcal{M}=\R^T \mathcal{M}_D \R\quad \text{and} \quad \bm{H}=\R\, \mathbf{S},
\end{equation}
where $\bm H \equiv(H_1,H_2,H_3)^T$ denotes the neutral scalars in the mass basis and the diagonal matrix $\mathcal{M}_D=\text{diag} (M_{H_1}^2, M_{H_2}^2, M_{H_3}^2)$ indicates the corresponding squared masses. Thus, identifying $H_1$ as the SM-like Higgs $h$, the scalar potential can be described with nine parameters: $\mu_2$ and $\lambda_{\{2,3,4,5,6,7\}}$ (considering $\lambda_7$ to be real, and $\lambda_5$ and $\lambda_6$ to be complex).

It is interesting to mention that Refs.~\cite{Karan:2023kyj,Karan:2023xze,Karan:2024kgr,Eberhardt:2020dat,Coutinho:2024vzm,Coutinho:2024zyp} use the masses and the mixing angles of the new scalars as independent parameters. The scalar potential parameters are related to the masses and the mixing angles by the following equations:
\begin{align}
 &\lambda_4  =  \frac{1}{v^2} \Big\{- 2 M_{H^{\pm}}^2 + \sum_j (\mathcal{R}_{j2}^2 + \mathcal{R}_{j3}^2) M_{H_j}^2 \Big\}, \;\;
    \lambda_5  =  \frac{1}{v^2} \sum_j \Big\{(\mathcal{R}_{j2}^2 - \mathcal{R}_{j3}^2)-2i\,\mathcal{R}_{j2} \mathcal{R}_{j3}\Big\} M_{H_j}^2,\\
  &\mu_2 \,= \, M_{H^{\pm}}^2 - \frac{1}{2}\lambda_3 v^2 , \quad 
  \lambda_1 \, = \, \frac{1}{v^2} \sum_j \mathcal{R}_{j1}^2 M_{H_j}^2, \quad  \lambda_6 \, =\, \frac{1}{v^2} \sum_j \mathcal{R}_{j1} (\mathcal{R}_{j2}-i\,\mathcal{R}_{j3}) M_{H_j}^2.
  \label{eq:mu2l1l6}
\end{align}
Assuming the SM-like Higgs to be the lightest scalar and in the CP-conserving scenario, the global A2HDM fit with heavy scalars under this parametrization indicates the following ranges \cite{Karan:2023kyj}: $M\gtrsim 400$ GeV (at 95\% C.L.), $\alpha_1=(0.05\pm 21.0) \times 10^{-3}$, $\lambda_2=3.2 \pm 1.9$, $\lambda_3=5.9\pm 3.5$ and $\lambda_7= 0.0 \pm 1.1$, where both the mean value and one standard deviation (i.e. at 68\% C.L.) are mentioned. Here, $M$ denotes the mass of the BSM scalars and $\alpha_1$ is the mixing angle between the two CP-even states $S_1$ and $S_2$.

However, the advantage of using the current parametrization is that in the decoupling limit, i.e. $M_{\{H^\pm,\,H_2,\,H_3\}}\gg m_h$, two of the mixing angles, $\alpha_1$ and $\alpha_2$ (that mix $S_1$ with $S_2$ and $S_3$, respectively), tend to zero automatically.

The couplings of the physical neutral scalars with the electroweak gauge bosons become:
\begin{equation}
    g_{H_jVV}=\R_{j1}\;g_{hVV}^{\text{SM}} \qquad \text{where $VV\in \{W^+W^-, ZZ\}$}.
\end{equation}

\input{ScalarMassesDecoupling}

\section{Effective Lagrangian for the eEDM}
\label{sec:LeEDM}

The electromagnetic interaction of a fermion ($\psi_f$) with a photon ($A_\mu$) is given by the interaction Hamiltonian:
\begin{equation}
	\mathcal H_I=J^\mu A_\mu \qquad \text{with} \qquad J^\mu=\bar \psi_f \,\widetilde O^\mu \,\psi_f\, ,
\end{equation}
where $\widetilde O^\mu$ denotes the effective vertex\footnote{Here, we are using the convention $D_\mu=\partial_\mu+ieQA_\mu$, where $e$ is the positron charge and $Q$ is the charge operator.} in position space. Considering all possible Lorentz structures, the effective fermion-photon interaction vertex in momentum space can in general be expressed in terms of four independent real-valued form factors ($\mathcal F_j$) as \cite{Nowakowski:2004cv}:
\begin{eqnarray}
    &\bar \psi_f \,\widetilde O^\mu \,\psi_f\rightarrow -i\,\bar u(p_f)\,\Gamma^\mu \,u(p_i) \qquad\text{with}\nonumber\\
	&\Gamma^\mu= \mathcal F_1 (q^2)\, \gamma^\mu+\frac{i\sigma^{\mu \nu}}{2m_f}\, q_\nu \,\mathcal F_2 (q^2) + i\epsilon^{\mu\nu\alpha\beta} \,\frac{\sigma_{\alpha\beta}}{4m_f}\, q_\nu \,\mathcal F_3 (q^2)+\frac{1}{2m_f} \Big(q^\mu-\frac{q^2}{2m_f}\gamma^\mu\Big)\,\gamma^5 \,\mathcal F_4 (q^2),
\end{eqnarray}
where $q_\nu$ is the four-momentum of the photon (towards the vertex). At the lowest order (tree-level interaction) and in the non-relativistic limit ($q^2\to 0$), $\mathcal F_1$ is equal to the electric charge of the fermion while the rest of form factors vanish. However, higher-order loop corrections make these form factors non-vanishing. In the non-relativistic limit, while a combination of the form factors $\mathcal{F}_{1,2}$ determines the \textit{magnetic moment} and $\mathcal{F}_4$ indicates the \textit{anapole moment} of the fermion, the EDM ($d_f$) of the fermion is estimated by $\mathcal{F}_3$ as:
\begin{equation}
	d_f=-\,\big[\mathcal{F}_3(q^2)/(2m_f)\big]_{q^2\to 0}.
\end{equation}
Therefore, in the EFT approach, the eEDM $d_e$ is defined as the coefficient of the following dimension five operator in the effective Lagrangian at a very low-energy scale ($\mu\sim m_e$):
\begin{equation}
	\mathcal L \supset -\frac{i}{2} d_e (\mu)\, \bar \psi_e\, \sigma^{\mu \nu}\gamma^5 \, \psi_e\, F_{\mu\nu}
\end{equation}
where $F^{\mu\nu}$ is the field strength tensor of the photon.
We have also used the identity: $ \sigma^{\mu \nu}\gamma^5=\frac{i}{2}\epsilon^{\mu\nu\alpha\beta} \,\sigma_{\alpha\beta}$
with $\epsilon^{0123}=+1$. In momentum space, the Feynman rule corresponding to the eEDM is given by:
\begin{equation}
	\begin{tikzpicture}[baseline=(a0.base),thick]
		\begin{feynman}
			\vertex  [small,blob](a0) {};
			\vertex [below left = 1 cm of a0] (a1);
			\vertex [below right = 1 cm of a0] (a2);
			\vertex [above = 1 cm of a0] (a3);
			
			\diagram* { (a1) -- [small,fermion, edge label = $e$](a0),
				(a0) -- [small,fermion, edge label = {$e$}](a2),
				(a3)--[boson, edge label' = {$\gamma(\epsilon_\mu)$}, momentum = {$q^\mu$}](a0)};
		\end{feynman}
	\end{tikzpicture}	
	\;\equiv \;-i\,\Gamma^\mu \;\supset \;i\,d_e(q^2)\, \sigma^{\mu \nu} q_\nu \gamma^5.
\end{equation}

\section{Contributions to the eEDM common in both A2HDM and C2HDM}
\label{sec:eEDM_detail}

\subsection{Fermion-loop contributions}

\begin{figure}[ht!]
	\centering
	\begin{tikzpicture}[thick, scale=1.5]
		\begin{feynman}
        \vertex (i1) at (0.8, -0.5);
        \vertex (a) at (1.3, -0.5);
        \vertex (b) at (2, 0.4);
        \vertex (c) at (2.5, 1.27);
        \vertex (f1) at (2.5, 2);
        \vertex (d) at (3, 0.4);
        \vertex (e) at (3.7, -0.5);
        \vertex (f2) at (4.2, -0.5); 
        
        \diagram* {
            (i1) -- [small,fermion, edge label= $\bm e$] (a) -- [small,fermion, edge label=\(\bm e\)] (e) -- [small,fermion, edge label= $\bm e$] (f2),
            (a) -- [boson, edge label=\(\bm{\gamma/Z}\), /tikzfeynman/momentum/arrow shorten=0.3] (b),
            (b) -- [fermion, looseness=1.1, out=120, in=180, relative=false] (c) -- [fermion, quarter
            left, looseness=1.1, edge label=\(\bm f\), out=0, in=60, relative=false] (d) -- [fermion, half left, looseness=1.1, out=240, in=300, relative=false] (b),
            (d) -- [very thick, scalar, edge label=\(\bm{H_i}\), /tikzfeynman/momentum/arrow shorten=0.3] (e),
            (c) -- [boson, edge label=$\bm \gamma$] (f1),
        };
    \end{feynman}
	\end{tikzpicture}
	\caption{EM and NC Barr-Zee diagrams with a fermion loop.
	}
	\label{fig:EM_NC_f}
\end{figure}
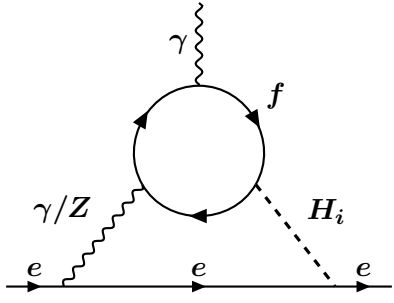

We first discuss the contributions displayed in Fig.~\ref{fig:EM_NC_f}.

\subsubsection{EM fermion-loop contribution}

The computation of the contribution of EM fermion-loop diagrams to the eEDM was carried out in Refs.~\cite{Altmannshofer:2020shb,Jung:2013hka}; it is explicitly shown in Ref.~\cite{Altmannshofer:2020shb}, while in Ref.~\cite{Jung:2013hka} the result is left in integral form, and an agreement between the two is found. The expression for such contribution (which is dominated by top and bottom quarks) is the following:
\begin{align}
	\frac{d_{e, f}^{EM}}{e} \, = \, &- \frac{\sqrt{2} \alpha G_F m_e}{(4 \pi)^3} 4 N_C Q_f^2 Q_e \sum_i \Big( \text{Im}(y_f^i) \text{Re}(y_e^i) r_i \Phi(r_i) \nonumber \\
	&+ \text{Re}(y_f^i) \text{Im}(y_e^i) r_i (4 + 2 \log{r_i} + (1 - 2 r_i) \Phi(r_i)) \Big),
	\label{eqn: de_EMloop}
\end{align}
where $N_C$ is the number of colors, $Q_f$ and $Q_e$ are respectively the EM charges of the fermion in the loop and the electron, $y_f^i$ is the Yukawa coupling of the fermion $f$ to the scalar $H_i$, $r_i = m_f^2 / M_{H_i}^2$ and $\Phi(x)$ is the Davydychev-Tausk function, whose small-argument expansion is the following:
\begin{equation}
	\Phi(x) \, = \, \left( \log^2(x) + \frac{\pi^2}{3} \right) + 2x \left( \log^2(x) + 2 \log(x) + \frac{\pi^2}{3} - 2 \right) + \mathcal{O}(x^2).
\end{equation}

In the decoupling limit, and as opposed to the C2HDM case \cite{Altmannshofer:2020shb}, some $\log^2 ( m_f^2/M^2 )$ terms that come from $\Phi ( m_f^2/M^2 )$ (and also some $\log ( m_f^2 / M^2 )$ terms) are not suppressed, when we consider complex values for the parameters $\varsigma_f$. Due to the different Yukawa couplings of fermions with different weak-isospin quantum numbers, the contributions in the decoupling limit of a top quark and a bottom quark in the loop will not be the same:
\begin{align}
	\frac{d_{e, t}^{EM}}{e} \, =& \, - \frac{\sqrt{2} \alpha G_F m_e}{(4 \pi)^3} 8 N_C Q_t^2 Q_e \text{Im}(\varsigma_u^* \varsigma_l) \frac{m_t^2}{M^2} \Big\{ \log \left( \frac{m_t^2}{M^2} \right) + \log^2\left( \frac{m_t^2}{M^2} \right) \Big\} + \mathcal{O} \left( \frac{m_t^4}{M^4} \right) \nonumber \\
	\frac{d_{e, b}^{EM}}{e} \, =& \, - \frac{\sqrt{2} \alpha G_F m_e}{(4 \pi)^3} 8 N_C Q_b^2 Q_e \text{Im}(\varsigma_d^* \varsigma_l) \frac{m_b^2}{M^2} \log \left( \frac{m_b^2}{M^2} \right) + \mathcal{O} \left( \frac{m_b^4}{M^4} \right).
	\label{eqn: de_EMloop_dec}
\end{align}

\subsubsection{NC fermion-loop contribution}

The computation of the NC fermion-loop diagrams was carried out in Ref.~\cite{Altmannshofer:2020shb}, and is similar to the EM one, having different overall factors:
\begin{align}
	\frac{d_{e, f}^{NC}}{e} \, = \, &- \frac{\sqrt{2} \alpha G_F m_e}{(4 \pi)^3} \frac{N_C Q_f Q_f^w Q_e^w}{4 c_w^2 s_w^2} \sum_i \Bigg\{ \text{Im}(y_f^i) \text{Re}(y_e^i) \frac{r_i}{1 - z_i} \left( \Phi(r_i) - \Phi \left( \frac{r_i}{z_i} \right) \right) \nonumber \\
	&+ \text{Re}(y_f^i) \text{Im}(y_e^i) \frac{r_i}{1 - z_i} \left( 2 \log{z_i} + (1 - 2 r_i) \Phi(r_i) - \left( 1 - \frac{2 r_i}{z_i} \right) \Phi \left( \frac{r_i}{z_i} \right) \right) \Bigg\},
	\label{eqn: de_NCloop}
\end{align}
where $Q_f^w = 2 T_f^3 - 4 Q_f s_w^2$ is the weak charge of the fermion $f$, $s_w$ and $c_w$ are the sine and cosine of the weak angle and $z_i = m_Z^2 / M_{H_i}^2$.\\

In the decoupling limit, we have:
\begin{align}
	\frac{d_{e, t}^{NC}}{e} \, =& \, - \frac{\sqrt{2} \alpha G_F m_e}{(4 \pi)^3} \frac{N_C Q_t (1 - 4 Q_t s_w^2) (-1 - 4 Q_e s_w^2)}{2 s_w^2 c_w^2} \nonumber \\
	&\times \text{Im}(\varsigma_u^* \varsigma_l) \frac{m_t^2}{M^2} \Big\{ \log \left( \frac{m_Z^2}{M^2} \right) + \log^2\left( \frac{m_t^2}{M^2} \right) \Big\} + \mathcal{O} \left( \frac{m_t^4}{M^4} \right) \nonumber \\
	\frac{d_{e, b}^{NC}}{e} \, =& \, - \frac{\sqrt{2} \alpha G_F m_e}{(4 \pi)^3} \frac{N_C Q_b (1 + 4 Q_b s_w^2) (1 + 4 Q_e s_w^2)}{2 s_w^2 c_w^2}\text{Im}(\varsigma_d^* \varsigma_l) \frac{m_b^2}{M^2} \log \left( \frac{m_Z^2}{M^2} \right) + \mathcal{O} \left( \frac{m_b^4}{M^4} \right).
	\label{eqn: de_NCloop_dec}
\end{align}

\subsection{Charged Higgs loop contributions}

The contributions of charged Higgs loop Barr-Zee diagrams to the eEDM were computed in \cite{Altmannshofer:2020shb}.

\subsubsection{EM and NC charged Higgs loop contributions}

\begin{figure}[ht!]
	\centering
	\begin{tikzpicture}[thick, scale=1.5]
		\begin{feynman}
        \vertex (i1) at (0.8, -0.5);
        \vertex (a) at (1.3, -0.5);
        \vertex (b) at (2, 0.4);
        \vertex (c) at (2.5, 1.27);
        \vertex (f1) at (2.5, 2);
        \vertex (d) at (3, 0.4);
        \vertex (e) at (3.7, -0.5);
        \vertex (f2) at (4.2, -0.5); 
        
        \diagram* {
            (i1) -- [small,fermion, edge label= $\bm e$] (a) -- [small,fermion, edge label=\(\bm e\)] (e) -- [small,fermion, edge label= $\bm e$] (f2),
            (a) -- [boson, edge label=\(\bm{\gamma/Z}\), /tikzfeynman/momentum/arrow shorten=0.3] (b),
            (b) -- [very thick, scalar, looseness=1.1, out=120, in=180, relative=false] (c) -- [very thick, scalar, quarter
            left, looseness=1.1, edge label=\(\bm{H^{\pm}}\), out=0, in=60, relative=false] (d) -- [very thick, scalar, half left, looseness=1.1, out=240, in=300, relative=false] (b),
            (d) -- [very thick, scalar, edge label=\(\bm{H_i}\), /tikzfeynman/momentum/arrow shorten=0.3] (e),
            (c) -- [boson, edge label=$\bm \gamma$] (f1),
        };
    \end{feynman}
	\end{tikzpicture}
	\caption{EM and NC charged Higgs loop Barr-Zee diagrams.
	}
	\label{fig:EM_H}
\end{figure}
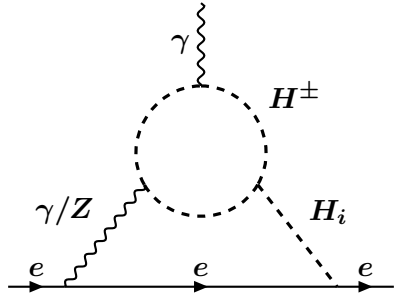

The expressions for the contributions coming from EM and NC charged Higgs loop Barr-Zee diagrams displayed in Fig.~\ref{fig:EM_H} are the following:
\begin{align}
	\frac{d_{e, H^+}^{EM}}{e} \, =& \, \frac{\sqrt{2} \alpha G_F m_e}{(4 \pi)^3} \frac{2 Q_e s_w^2}{\pi \alpha} \sum_i \text{Im} (y_l^i) \lambda_{i H^+ H^-} w_i [ 2 + \log(w_i) - h_i \Phi(h_i) ], \label{eqn: de_EM_H}  \\
	\frac{d_{e, H^+}^{NC}}{e} \, =& \, \frac{\sqrt{2} \alpha G_F m_e}{(4 \pi)^3} \frac{Q_e^w c_{2w}}{4 \pi \alpha} \sum_i \text{Im} (y_l^i) \lambda_{i H^+ H^-} \frac{z_i}{1 - z_i} \left[ \log(z_i) - h_i \Phi(h_i) + \frac{h_i}{z_i} \Phi \left( \frac{h_i}{z_i} \right) \right],
	\label{eqn: de_NC_H} 
\end{align}
where $w_i = m_W^2/M_{H_i}^2$, $h_i = M_{H^{\pm}}^2 / M_{H_i}^2$, $c_{2 w} = \cos (2 \theta_W)$ and $\lambda_{i H^+ H^-}$ is the triple Higgs coupling, which is given by:
\begin{equation}
	\lambda_{i H^+ H^-} \, = \, \mathcal{R}_{i 1} \lambda_3 + \text{Re} (\lambda_7 (\mathcal{R}_{i 2} + i \mathcal{R}_{i 3})).
\end{equation}

\subsubsection{CC charged Higgs loop contribution}

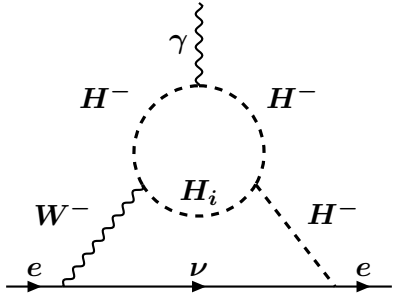
\begin{figure}[ht!]
	\centering
	\begin{tikzpicture}[thick, scale=1.5]
		\begin{feynman}
        \vertex (i1) at (0.8, -0.5);
        \vertex (a) at (1.3, -0.5);
        \vertex (b) at (2, 0.4);
        \vertex (c) at (2.5, 1.27);
        \vertex (f1) at (2.5, 2);
        \vertex (d) at (3, 0.4);
        \vertex (e) at (3.7, -0.5);
        \vertex (f2) at (4.2, -0.5); 
        
        \diagram* {
            (i1) -- [small,fermion, edge label= $\bm e$] (a) -- [small,fermion, edge label=\(\bm\nu\)] (e) -- [small,fermion, edge label= $\bm e$] (f2),
            (a) -- [boson, edge label=\(\bm{W^-}\), /tikzfeynman/momentum/arrow shorten=0.3] (b),
            (b) -- [very thick, scalar, edge label=\(\bm{H^-}\), looseness=1.1, out=120, in=180, relative=false] (c) -- [very thick, scalar, quarter
            left, looseness=1.1, edge label=\(\bm{H^-}\), out=0, in=60, relative=false] (d) -- [very thick, scalar, edge label'=\(\bm{H_i}\), half left, looseness=1.1, out=240, in=300, relative=false] (b),
            (d) -- [very thick, scalar, edge label=\(\bm{H^-}\), /tikzfeynman/momentum/arrow shorten=0.3] (e),
            (c) -- [boson, edge label=$\bm \gamma$] (f1),
        };
    \end{feynman}
	\end{tikzpicture}
	\caption{CC charged Higgs loop Barr-Zee diagram.
	}
	\label{fig:CC_H}
\end{figure}

The result for the CC charged Higgs loop Barr-Zee diagrams shown in Fig.~\ref{fig:CC_H} is the following:
\begin{align}
	\frac{d_{e, H^+}^{CC}}{e} \, =& \, \frac{\sqrt{2} \alpha G_F m_e}{(4 \pi)^3} \frac{(- 2 T_e^3)}{4 \pi \alpha} \sum_i \text{Im} (y_l^i) \lambda_{i H^+ H^-} \Big[ 2 - \frac{2}{h_i} + \frac{2}{h_i} \log (h_i) \nonumber \\
	&- \frac{2 - 2 h_i + w_i}{h_i - w_i} \log \left( \frac{h_i}{w_i} \right) - \frac{1 + h_i^2 - h_i (2 + w_i)}{w_i (h_i - w_i)} \log (h_i) \log \left( \frac{h_i}{w_i} \right) \nonumber \\
	&- \frac{2 (h_i - 2 h_i^2 + h_i^3 + w_i - 2 h_i w_i)}{h_i^2 w_i} \text{Li}_2 \left( 1 - \frac{1}{h_i} \right) + \frac{w_i (1 - 4 h_i + 2 h_i^2)}{h_i^2 (h_i - w_i)} \Phi(h_i) \nonumber \\
	&- \frac{1 - h_i^3 - w_i + h_i^2 (3 + 2 w_i) - h_i (3 + w_i + w_i^2)}{w_i (h_i - w_i)} \Phi(h_i, w_i) \Big].
	\label{eqn: de_CC_H} 
\end{align}

\subsection{$W$ boson loop contributions}

The contributions of $W$ boson loop Barr-Zee diagrams to the eEDM were computed in Ref.~\cite{Altmannshofer:2020shb}. They present the full gauge-dependent expressions for each kind of diagram. Here we show only the gauge independent parts of such expressions, since the sum of all contributions is independent of the choice of gauge.

\subsubsection{EM and NC $W$ boson loop contributions}

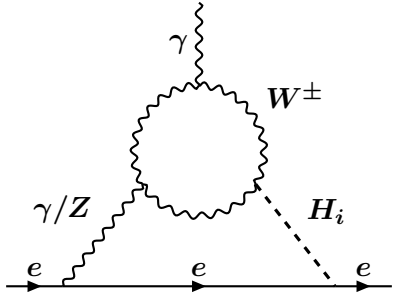
\begin{figure}[ht!]
	\centering
	\begin{tikzpicture}[thick, scale=1.5]
		\begin{feynman}
        \vertex (i1) at (0.8, -0.5);
        \vertex (a) at (1.3, -0.5);
        \vertex (b) at (2, 0.4);
        \vertex (c) at (2.5, 1.27);
        \vertex (f1) at (2.5, 2);
        \vertex (d) at (3, 0.4);
        \vertex (e) at (3.7, -0.5);
        \vertex (f2) at (4.2, -0.5); 
        
        \diagram* {
            (i1) -- [small,fermion, edge label= $\bm e$] (a) -- [small,fermion, edge label=\(\bm e\)] (e) -- [small,fermion, edge label= $\bm e$] (f2),
            (a) -- [boson, edge label=\(\bm{\gamma/Z}\), /tikzfeynman/momentum/arrow shorten=0.3] (b),
            (b) -- [boson, looseness=1.1, out=120, in=180, relative=false] (c) -- [boson, quarter
            left, looseness=1.1, edge label=\(\bm{W^{\pm}}\), out=0, in=60, relative=false] (d) -- [boson, half left, looseness=1.1, out=240, in=300, relative=false] (b),
            (d) -- [very thick, scalar, edge label=\(\bm{H_i}\), /tikzfeynman/momentum/arrow shorten=0.3] (e),
            (c) -- [boson, edge label=$\bm \gamma$] (f1),
        };
    \end{feynman}
	\end{tikzpicture}
	\caption{EM and NC $W$ boson loop Barr-Zee diagrams.
	}
	\label{fig:EM_W}
\end{figure}

The expressions for the contributions coming from EM and NC $W$ boson loop Barr-Zee diagrams found in Fig.~\ref{fig:EM_W} are the following:
\begin{align}
	\frac{d_{e, W}^{EM}}{e} \, =& \, \frac{\sqrt{2} \alpha G_F m_e}{(4 \pi)^3} Q_e \sum_i \text{Im} (y_l^i) \mathcal{R}_{i 1} [ 4 (1 + 6 w_i) + 2 (1 + 6 w_i) \log(w_i) \nonumber \\
	&- (3 - 16 w_i + 12 w_i^2) \Phi(w_i) ], \label{eqn: de_EM_W}  \\
	\frac{d_{e, W}^{NC}}{e} \, =& \, \frac{\sqrt{2} \alpha G_F m_e}{(4 \pi)^3} \frac{Q_e^w}{4 s_w^2} \sum_i \text{Im} (y_l^i) \mathcal{R}_{i 1} \Big[ - \frac{3 - 16 w_i + 12 w_i^2}{1 - z_i} \Phi(w_i) \nonumber \\
	&+ \frac{1 - 2 s_w^2 + 2 (5 - 6 s_w^2) w_i}{c_w^2 (1 - z_i)} \log(z_i) - \frac{(1 + 8 s_w^2 - 12 s_w^2) z_i}{1 - z_i} \Phi(c_w^2) \Big].
	\label{eqn: de_NC_W} 
\end{align}

\subsubsection{CC $W$ boson loop contribution}

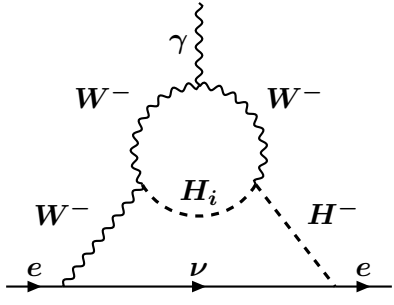
\begin{figure}[ht!]
	\centering
	\begin{tikzpicture}[thick, scale=1.5]
		\begin{feynman}
        \vertex (i1) at (0.8, -0.5);
        \vertex (a) at (1.3, -0.5);
        \vertex (b) at (2, 0.4);
        \vertex (c) at (2.5, 1.27);
        \vertex (f1) at (2.5, 2);
        \vertex (d) at (3, 0.4);
        \vertex (e) at (3.7, -0.5);
        \vertex (f2) at (4.2, -0.5); 
        
        \diagram* {
            (i1) -- [small,fermion, edge label= $\bm e$] (a) -- [small,fermion, edge label=\(\bm \nu \)] (e) -- [small,fermion, edge label= $\bm e$] (f2),
            (a) -- [boson, edge label=\(\bm{W^-}\), /tikzfeynman/momentum/arrow shorten=0.3] (b),
            (b) -- [boson, edge label=\(\bm{W^-}\), looseness=1.1, out=120, in=180, relative=false] (c) -- [boson, quarter
            left, looseness=1.1, edge label=\(\bm{W^-}\), out=0, in=60, relative=false] (d) -- [very thick, scalar, edge label'=\(\bm{H_i}\), half left, looseness=1.1, out=240, in=300, relative=false] (b),
            (d) -- [very thick, scalar, edge label=\(\bm{H^-}\), /tikzfeynman/momentum/arrow shorten=0.3] (e),
            (c) -- [boson, edge label=$\bm \gamma$] (f1),
        };
    \end{feynman}
	\end{tikzpicture}
	\caption{CC $W$ boson loop Barr-Zee diagram.
	}
	\label{fig:CC_W}
\end{figure}

The result for the CC $W$ boson loop Barr-Zee diagrams displayed in Fig.~\ref{fig:CC_W}, more complicated due to the presence of another mass scale from the charged Higgs, is the following:
\begin{align}
	\frac{d_{e, W}^{CC}}{e} \, =& \, \frac{\sqrt{2} \alpha G_F m_e}{(4 \pi)^3} \frac{(- 2 T_e^3)}{4 s_w^2} \sum_i \text{Im} \Big( \varsigma_l (\mathcal{R}_{i 2} + i \mathcal{R}_{i 3}) \Big) \mathcal{R}_{i 1} \Big[ \frac{2}{w_i} - \frac{2 (1 - w_i)^2}{h_i w_i} \nonumber \\
	&- \frac{2 (1 - w_i) w_i^2 + (2 + w_i) h_i^2 - h_i (2 - w_i - 7 w_i^2)}{h_i w_i (h_i - w_i)} \log(w_i) \nonumber \\
	&+ \frac{h_i^2 - 2 (1 - w_i^2) + h_i (1 + 7 w_i)}{h_i (h_i - w_i)} \log (h_i) \nonumber \\
	&- \frac{(1 - w_i)^3 - 3 h_i^2 w_i - h_i (1 + 3 w_i - 4 w_i^2)}{h_i^2 (h_i - w_i)} \log \left( \frac{h_i}{w_i} \right) \log (w_i) \nonumber \\
	&- \frac{2 w_i (1 - w_i)^3 + h_i (2 - 8 w_i + 6 w_i^3)}{h_i^2 w_i^2} \text{Li}_2 \left( 1 - \frac{1}{w_i} \right) - \frac{1 - 6 w_i + 6 w_i^2 + 4 w_i^3}{(h_i - w_i) w_i^2} \Phi(w_i) \nonumber \\
	&+ \frac{(1 - w_i)^4 - 3 h_i^3 w_i - h_i (2 + 5 w_i) (1 - w_i)^2 + h_i^2 (1 + 7 w_i^2)}{h_i^2 (h_i - w_i)} \Phi(h_i, w_i) \Big].
	\label{eqn: de_CC_W} 
\end{align}

\subsection{Kite contributions}

The contributions coming from kite diagrams to the eEDM were computed in Ref.~\cite{Altmannshofer:2020shb}.

\subsubsection{NC kite contribution}

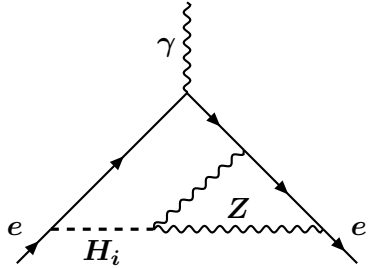
\begin{figure}[ht!]
	\centering
	\begin{tikzpicture}[thick, scale=1.5]
		\begin{feynman}
			\vertex (i1) at (1, -0.8);
            \vertex (a) at (1.3, -0.5);
            \vertex (b) at (2.2, -0.5);
            \vertex (c) at (2.5, 0.7);
            \vertex (f1) at (2.5, 1.5);
            \vertex (d) at (3, 0.2);
            \vertex (e) at (3.7, -0.5);
            \vertex (f2) at (4, -0.8); 
			
			\diagram* {
				(i1) -- [small, fermion, , edge label= $\bm e$] (a) -- [small, fermion] (c) -- [small, fermion] (d) -- [small, fermion] (e) -- [small, fermion, edge label= $\bm e$] (f2),
				(a) -- [very thick, scalar, edge label'=\(\bm{H_i}\)] (b) -- [boson, edge label=\(\bm Z\)] (e),
				(b) -- [boson] (d),
				(c) -- [boson, edge label=$\bm \gamma$] (f1),
			};
		\end{feynman}
	\end{tikzpicture}
	\caption{NC kite diagram.
	}
	\label{fig:NC_K}
\end{figure}

The NC kite contribution of Fig.~\ref{fig:NC_K} does not depend on the choice of gauge:
\begin{align}
	\frac{d_{e, \text{kite}}^{NC}}{e} \, =& \, - \frac{\sqrt{2} \alpha G_F m_e}{(4 \pi)^3} Q_e \frac{(Q_e^w)^2 - 1}{8 s_w^2 c_w^2} \sum_i \text{Im} (y_e^i) \mathcal{R}_{i 1} \frac{1}{z_i^3} \Big[ z_i^2 + \frac{\pi^2}{6} (1 - 4 z_i) - 2 z_i^2 \log (z_i) \nonumber \\
	&- \frac{1 - 4 z_i}{2} \log^2 (z_i) - 2 (1 - 4 z_i + z_i^2) \text{Li}_2 \left( 1 - \frac{1}{z_i} \right) - \frac{1 - 6 z_i + 8 z_i^2}{2} \Phi(z_i) \Big] \nonumber \\
	&- \frac{\sqrt{2} \alpha G_F m_e}{(4 \pi)^3} Q_e \frac{(Q_e^w)^2 + 1}{24 s_w^2 c_w^2} \sum_i \text{Im} (y_e^i) \mathcal{R}_{i 1} \frac{1}{z_i} \Big[ 2 z_i (1 - 4 z_i) + \frac{\pi^2}{3} (3 z_i^2 + z_i^3) \nonumber \\
	&- 2 z_i (1 + 4 z_i) \log (z_i) + 2 (1 - 3 z_i^2 - 4 z_i^3) \text{Li}_2 \left( 1 - \frac{1}{z_i} \right) - (1 - 2 z_i - 8 z_i^2) \Phi(z_i) \Big].
	\label{eqn: de_NC_K} 
\end{align}

\subsubsection{CC kite contribution}

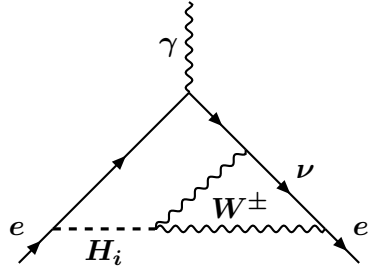
\begin{figure}[ht!]
	\centering
	\begin{tikzpicture}[thick, scale=1.5]
		\begin{feynman}
			\vertex (i1) at (1, -0.8);
            \vertex (a) at (1.3, -0.5);
            \vertex (b) at (2.2, -0.5);
            \vertex (c) at (2.5, 0.7);
            \vertex (f1) at (2.5, 1.5);
            \vertex (d) at (3, 0.2);
            \vertex (e) at (3.7, -0.5);
            \vertex (f2) at (4, -0.8); 
			
			\diagram* {
				(i1) -- [small, fermion, , edge label= $\bm e$] (a) -- [small, fermion] (c) -- [small, fermion] (d) -- [small, fermion, edge label= $\bm\nu$] (e) -- [small, fermion, edge label= $\bm e$] (f2),
				(a) -- [very thick, scalar, edge label'=\(\bm{H_i}\)] (b) -- [boson, edge label=\(\bm{W^{\pm}}\)] (e),
				(b) -- [boson] (d),
				(c) -- [boson, edge label=$\bm \gamma$] (f1),
			};
		\end{feynman}
	\end{tikzpicture}
	\caption{CC kite diagram.
	}
	\label{fig:CC_K}
\end{figure}

The CC kite contribution of Fig.~\ref{fig:CC_K} does depend on the choice of gauge. Again we only show the gauge-independent part of the contribution:
\begin{align}
	\frac{d_{e, \text{kite}}^{CC}}{e} \, =& \, \frac{\sqrt{2} \alpha G_F m_e}{(4 \pi)^3} \frac{(- 2 T_e^3)}{4 s_w^2} \sum_i \text{Im} (y_e^i) \mathcal{R}_{i 1} \Big[ \frac{2 \pi^2}{9} w_i (3 + 4 w_i) + \frac23 (5 - 8 w_i) - \frac{16}{3} (1 + w_i) \log (w_i) \nonumber \\
	& + \frac{2 (3 + 2 w_i - 6 w_i^3 - 8 w_i^4)}{3 w_i^2} \text{Li}_2 \left( 1 - \frac{1}{w_i} \right) + \frac{(1 + 2 w_i) (3 - 10 w_i + w_i^2)}{3 w_i^2} \Phi(w_i) \Big].
	\label{eqn: de_CC_K}
\end{align}\\

In the decoupling limit $w_{2,3} = m_W^2/M^2$, $z_{2,3} = m_Z^2/M^2$ and $h_1 = M^2/m_h^2$. Adding the contributions from $W$ loop Barr-Zee diagrams and the CC kite in this limit, we find a logarithmically enhanced contribution, such as was already carried out in Ref.~\cite{Altmannshofer:2020shb}:
\begin{equation}
	\frac{d_{e, W, \text{kite}}}{e} \Bigg|_{\log} \, = \, \frac{\sqrt{2} \alpha G_F m_e}{(4 \pi)^3} \sum_{i = 2, 3} \frac{3}{4 c_w^2} \text{Im} (y_e^i) \mathcal{R}_{i 1} \log \left( \frac{M^2}{m_W^2} \right),
	\label{eqn: de_Wkite_dec}
\end{equation}
where it is important to recall that the factor $\sum_{i = 2, 3} \text{Im} (y_e^i) \mathcal{R}_{i 1}$ involves the desired mass suppression of order $v^2/M^2$.

\section{Details of calculating $d_{e,f}^{CC}$}\label{sec:details_defCC}

In order to calculate the Barr-Zee diagrams, we first consider the one-loop effective $\phi\gamma V$ vertex ($\widetilde\Gamma^{\mu\nu}_{\phi V\gamma}$). Considering all possible Lorentz structures, this vertex can generally be expressed in terms of five form factors as:
\begin{eqnarray}
&\begin{tikzpicture}[baseline=(a0.base)]
		\begin{feynman}
			\vertex  [small,blob,thick](a0) {};
			\vertex [below left = 1.2 cm of a0] (a1);
			\vertex [below right = 1.2 cm of a0] (a2);
			\vertex [above = 1 cm of a0] (a3);
			
			\diagram* { (a1) -- [very thick,scalar, edge label = $\bm{\phi}$](a0),
				(a0) -- [thick,boson, edge label = {$\bm{V}(\epsilon_\nu)$},momentum' = {$k^\nu$}](a2),
				(a3)--[boson,thick, edge label' = {$\bm{\gamma}(\epsilon_\mu)$}, momentum = {$q^\mu$}](a0)};
		\end{feynman}
	\end{tikzpicture}
    \equiv  \,\Gamma^{\mu\nu}_{\phi V\gamma}&\nonumber\\
    \text{where,} \qquad &\,\Gamma^{\mu\nu}_{\phi V\gamma}=S_0\, g^{\mu\nu}+S_1\, q^\mu q^\nu+ S_2\, k^\mu k^\nu + S_3\, q^\mu k^\nu +S_4\, k^\mu q^\nu + i\,P \epsilon^{\mu\nu\rho\sigma} q_\rho k_\sigma.\hskip 1.8cm &
\end{eqnarray}
Since we are interested in the EDM, we have to evaluate the form factor $\mathcal F_3$ at $q^2\to 0$, i.e. the photon must be on-shell (OS). Therefore, this effective vertex must obey the Ward identity:  $q_\mu \,\Gamma^{\mu\nu}_{\phi V\gamma_{\text{OS}}^{}}=0$ along with $q^2=0$, which enforces: $S_2=0$ and $S_4=-S_0/ (q \cdot k)\equiv -S$. Moreover, the amplitude of the Barr-Zee diagram would contain the factor:
$\,\epsilon_\mu\,\Gamma^{\mu\nu}_{\phi V\gamma_{\text{OS}}^{}}$, and since we are dealing with a real photon we must have $\epsilon \cdot q=0$. Therefore, dropping the unnecessary terms proportional to $q^\mu$ in the effective $\phi\gamma V$ vertex for a real photon, we can write:
\begin{equation}
    \,\Gamma^{\mu\nu}_{\phi V\gamma_{\text{OS}}^{}}=S (g^{\mu\nu} q \cdot k - k^\mu q^\nu) + i\,P \epsilon^{\mu\nu\rho\sigma} q_\rho k_\sigma~.
\end{equation}
This tensorial structure ensures the gauge symmetry of the $\phi\gamma V$ vertex with an on-shell photon. However, while performing the actual one-loop calculation, one finds the $\phi\gamma V$ vertex factor to be:
\begin{equation}
    \widetilde\Gamma^{\mu\nu}_{\phi V\gamma_{\text{OS}}^{}}=\Gamma^{\mu\nu}_{\phi V\gamma_{\text{OS}}^{}} + \Gamma_P\,g^{\mu\nu}+\Gamma_D\,k^\mu k^\nu,
\end{equation}
which apparently breaks the gauge symmetry.\footnote{The extra terms vanish only if all the three particles $\phi$, $V$ and $\gamma$ are on-shell.} However, the $k^\mu k^\nu$ term ($\Gamma_D$) cannot contribute to the EDM at the two-loop level as it does not generate the required $\sigma^{\mu\nu} \gamma^5$ structure for the effective electron-photon vertex \cite{Abe:2013qla}. On the other hand, the appearance of the extra $g^{\mu\nu}$ term ($\Gamma_P$) is not anomalous since the gauge invariance is guaranteed for the $S$-matrix element, not for any effective vertex. Nevertheless, the effect of $\Gamma_P$ can be canceled by borrowing some terms from the Kite diagrams or further non-Barr-Zee diagrams through the \textit{pinch technique} and thus the Barr-Zee diagrams (understood now to include these extra pinch terms) are made gauge invariant \cite{Abe:2013qla}. Therefore, in the one-loop calculation of the effective $\phi\gamma V$ vertex with a real photon, one should identify the form factors $S$ and $P$ from the coefficients of $k^\mu q^\nu$ and $\epsilon^{\mu\nu\rho\sigma} q_\rho k_\sigma$ respectively.

In our case, we need to calculate the one-loop effective $H^-W^+\gamma$ vertex, for which the coefficients $S$ and $P$ are given by:
\begin{align}
    S&=i\,\frac{\alpha N_C |V_{tb}|^2}{2\pi v s_w}\int_0^1 \frac{[Q_{t}x+Q_b (1-x)][\varsigma_u m_t^2x^2-\varsigma_d m_b^2(1-x)^2]}{k^2 x (1-x)-m_b^2(1-x)-m_t^2 x}, \\
    P&=i\,\frac{\alpha N_C |V_{tb}|^2}{2\pi v s_w}\int_0^1 \frac{[Q_{t}x+Q_b (1-x)][\varsigma_u m_t^2x+\varsigma_d m_b^2(1-x)]}{k^2 x (1-x)-m_b^2(1-x)-m_t^2 x}~.
\end{align}
The final contributions to the eEDM are found in the main text, Section~\ref{sec:calculation_A2HDM}. See also the discussion of Ref.~\cite{Kanemura:2024ezz}.

\section{Conventions}\label{sec:Conventions}

It is worth addressing the convention that was chosen in this work for some relevant terms from the SM Lagrangian. In particular, we define the covariant derivative of a left-handed fermion field $SU(2)_L$-doublet $\psi_L$ as:
\begin{align}
    D_{\mu} \psi_L \, = \, \Big[ \partial_{\mu} + i \frac{g}{\sqrt{2}} (\tau_+ W_{\mu}^+ + \tau_- W_{\mu}^-) + i e Q A_{\mu} + i\frac{g}{\cos \theta_W} \left( \frac{\tau_3}{2} - Q \sin^2 \theta_W \right) Z_{\mu} \Big] \, \psi_L,
\end{align}
where $Q = T_3 + Y$ ($T_3$ is the third component of the weak isospin and $Y$ is the hypercharge) is the charge operator, and
\begin{align}
    \tau_{\pm} \, = \, \frac{\tau_1 \pm i\tau_2}{2}\, ,
\end{align}
where $\tau_a$ are the Pauli matrices. On the other hand, we define the covariant derivative of a right-handed ($SU(2)_L$-singlet) fermion field $SU(2)_L$-singlet $\psi_R$ with electromagnetic charge $Q$ as:
\begin{align}
    D_{\mu} \psi_R \, = \, \Big[ \partial_{\mu} + i e Q A_{\mu} - i\frac{g}{\cos \theta_W} Q \sin^2 \theta_W Z_{\mu} \Big] \, \psi_R.
\end{align}

We checked that this convention is in agreement with the one chosen in Ref.~\cite{Altmannshofer:2020shb}.

\bibliography{mybib_draft}
\bibliographystyle{utphys}

\end{document}

%% file: ScalarMassesDecoupling.tex
\subsection{\boldmath Diagonalization of the neutral scalar mass matrix when $\mu_2\gg v^2$}

In the decoupling limit $\mu_2\gg v^2$ the three eigenvalues of the neutral scalar mass matrix (\ref{eq:massmatrix}) can be easily obtained through an expansion in powers of $v^2/\mu_2$:
\begin{eqnarray}
M_{H_1}^2 &=& \lambda_1 v^2 + \cO(v^4/\mu_2)\, ,
\nonumber\\
M_{H_2}^2 &=& \mu_2 + \left(\lambda_3 + \lambda_4 + |\lambda_5|\right)\,\frac{v^2}{2} + \cO(v^4/\mu_2)\, ,
\nonumber\\
M_{H_3}^2 &=& \mu_2 + \left(\lambda_3 + \lambda_4 - |\lambda_5|\right)\,\frac{v^2}{2} + \cO(v^4/\mu_2)\, .
\label{eq:mass_eigenvalues}
\end{eqnarray}
Since there are two degenerate eigenvalues in the limit $v^2\to 0^+$, the diagonalization of the corresponding $2\times 2$ submatrix involves a rotation of $\cO(v^0)$:
\begin{equation}
\mathcal{R}^{(23)}(\alpha_3) = \left( \begin{array}{ccc}
1 & 0 & 0 \\ 0 & \cos{\alpha_3} & - \sin{\alpha_3} \\ 0 & \sin{\alpha_3} & \cos{\alpha_3}
\end{array} \right)
\qquad\qquad\text{with}\qquad\qquad
\tan{(2\alpha_3)} = \frac{\lambda_5^I}{\lambda_5^R}\, ,
\label{eq:lm5}
\end{equation}
where $ \text{Re}(\lambda_i) = \lambda_i^R $ and $ \text{Im}(\lambda_i) = \lambda_i^I $. This rotation becomes a trivial identity in the CP-conserving limit.

The full $3\times 3$ matrix can be diagonalized through three consecutive rotations
$\mathcal{R}^{(23)}(\alpha_3)$, $\mathcal{R}^{(12)}(\alpha_1)$ and $\mathcal{R}^{(13)}(\alpha_2)$,
where $\alpha_1$ and $\alpha_2$ are of $\cO(v^2)$. Keeping only terms up to $\cO(v^2)$,
\begin{eqnarray} \label{eqn: R_dec}
\mathcal{R} &=& \mathcal{R}^{(13)}(\alpha_2)\,\mathcal{R}^{(12)}(\alpha_1)\,\mathcal{R}^{(23)}(\alpha_3) \\
&=& \left(\begin{array}{ccc}
1 & -\lambda_6^R\,\frac{v^2}{\mu_2} & \lambda_6^I\,\frac{v^2}{\mu_2} \\
\left( \lambda_6^R \cos{\alpha_3} + \lambda_6^I \sin{\alpha_3}\right) \frac{v^2}{\mu_2} &
\cos{\alpha_3} & - \sin{\alpha_3} \\
\left( \lambda_6^R \sin{\alpha_3} - \lambda_6^I \cos{\alpha_3}\right) \frac{v^2}{\mu_2} &
\sin{\alpha_3} & \cos{\alpha_3}
\end{array}\right)
\quad + \quad \cO\!\left(\frac{v^4}{\mu_2^2}\right) . \nonumber
\end{eqnarray}

\section{Yukawa sector}
\label{sec:Yuk}

In the Higgs basis, the interactions of fermions with the scalar fields read:
\begin{equation}
    -\mathcal L_Y\, =\,\frac{\sqrt 2}{v}\Big[\overline Q_L'(M_d' \Phi_1+Y_d' \Phi_2)d_R'+\overline Q_L'(M_u' \widetilde\Phi_1+Y_u' \widetilde\Phi_2)u_R'+\overline L_L'(M_l' \Phi_1+Y_l' \Phi_2)l_R'\Big] + \rm{h.c.},
\end{equation}
where $Q_L'$ and $L_L'$ are the quark and lepton doublets and $\widetilde\Phi_a=i\tau_2 \Phi_a^*$ is the charge-conjugate scalar doublet. The complex matrices $M_f'$ and $Y_f'$ (with $f\in\{u,d,l\}$) set
the Yukawa interactions, and in general they cannot be diagonalized simultaneously. In the mass basis of fermions, the Yukawa interaction with the doublet $\Phi_1$ is governed by the diagonal matrices $M_f$ while the interaction with $\Phi_2$ is controlled by the non-diagonal matrices $Y_f$. Thus, in the fermionic mass basis and scalar Higgs basis their interactions take the following form:
\begin{align}
-\mathcal L_Y = &\Big(1+\frac{S_1}{v}\Big)\big\{\bar u_L\,M_u\,u_R+\bar d_L\,M_d\,d_R+\bar\nl_L\, M_\nl\,\nl_R\big\}+\Big(\frac{S_2+i S_3}{v}\Big)\big\{\bar u_R\,Y_u^\dagger\,u_L+\bar d_L\,Y_d\,d_R+\bar\nl_L\, Y_\nl\,\nl_R\big\}\nonumber\\
&+\frac{\sqrt 2}{v}\, H^+ \left\{\bar u_L\,V\,Y_d\,d_R-\bar u_R \,Y_u^\dagger\,V\,d_L+\bar\nu_L\, Y_\nl\,\nl_R\right\} + \mathrm{h.c.}\, ,
\end{align}
where $V$ is the CKM matrix. The non-diagonal nature of the matrices $Y_f$ creates undesirable FCNCs at tree level. These FCNCs can easily be avoided by demanding the alignment condition of $M_f$ and $Y_f$ in the flavour-space:
\begin{equation}\label{eq:Yalignment}
Y_u=\varsigma^*_u\,M_u \qquad\qquad \text{and} \qquad\qquad Y_{d,\nl}=\varsigma_{d,\nl}\,M_{d,\nl}\, ,
\end{equation}
where the complex-valued couplings $\varsigma_f$ are termed as alignment parameters, and we obtain the interaction part of the Yukawa Lagrangian in Eq.~\eqref{eq:Yukawa}. In the CP-conserving scenario, the global A2HDM fit with heavy scalars indicates the following ranges of the alignment parameters with mean value and one standard deviation (at 68\% C.L.) \cite{Karan:2023kyj}: $\varsigma_u=0.006\pm 0.257$, $\varsigma_d=0.12\pm 4.12$ and $\varsigma_l=-0.39\pm 11.69$.

\section{Matching to the SMEFT in the decoupling limit}\label{app:decoupling_limit}

In the Higgs basis, the $\Phi_2$-dependent part of the A2HDM Lagrangian can be formally written as
\begin{equation}
\mathcal{L}_{\Phi_2} = (D_\mu\Phi_2)^\dagger D^\mu\Phi_2 - V_{\Phi_2} -
\frac{\sqrt{2}}{v}\left[ (\Phi_2^a \mathcal{F}^a) + (\Phi_2^a \mathcal{F}^a)^\dagger \right] ,
\end{equation}
where a sum over the $SU(2)_L$ superindex `a' is implied,
\begin{eqnarray}
V_{\Phi_2} &=& \mu_2\, (\Phi_2^\dagger\Phi_2) +\left[\mu_3\, (\Phi_1^\dagger\Phi_2) +\mathrm{h.c.}\right] + \frac{\lambda_2}{2}\, (\Phi_2^\dagger\Phi_2)^2 +\lambda_3\, (\Phi_1^\dagger\Phi_1) (\Phi_2^\dagger\Phi_2) + \lambda_4\, (\Phi_1^\dagger\Phi_2) (\Phi_2^\dagger\Phi_1)
\nonumber\\ &+&
\left\{ \left[ \frac{\lambda_5}{2}\, (\Phi_1^\dagger\Phi_2)
+\lambda_6\, (\Phi_1^\dagger\Phi_1) + \lambda_7\, (\Phi_2^\dagger\Phi_2)
\right] (\Phi_1^\dagger\Phi_2) +\mathrm{h.c.}\right\} ,
\end{eqnarray}
with $\mu_3 =-\frac{1}{2}\lambda_6 v^2$,
and the Yukawa interactions are parametrized by
\begin{equation}
\mathcal{F}^a = \varsigma_d\,\bar Q_L^a\mathcal{M}_d d_R 
-\varsigma_u\epsilon_{ab}\, \bar u_R\mathcal{M}_u Q_L^b
+ \varsigma_l\, \bar L_L^a\mathcal{M}_\ell \ell_R\, .
\end{equation}

The equations of motion of $\Phi_2$ take the form
\begin{equation}
(D_\mu D^\mu\Phi_2)^a + \frac{\partial V_{\Phi_2}}{\partial\Phi_2^{a*}} + \frac{\sqrt{2}}{v}\,
\mathcal{F}^{a\dagger} = 0\, .
\end{equation}
In the limit $\mu_2\gg v^2$, these equations of motion can be solved as an expansion in inverse powers of $\mu_2$ that determines the classical field $(\Phi_2)^{\mathrm{cl}}$ in terms of the light degrees of freedom:
\begin{equation}
(\Phi_2)^{\mathrm{cl}} = \sum_{n=1}\frac{\omega_n}{\mu_2^n} \, .
\end{equation}
To determine the dimension-6 contributions to the SMEFT Lagrangian we only need the leading term, 
\begin{equation}
\omega_1^a = \lambda_6^*\left[\frac{1}{2}\, v^2 - (\Phi_1^\dagger\Phi_1)\right] \Phi_1^a -\frac{\sqrt{2}}{v}\, \mathcal{F}^{a\dagger}\, .
\end{equation}
Inserting back this expression in $\mathcal{L}_{\Phi_2}$, one gets the wanted low-energy effective Lagrangian:
\begin{eqnarray}
\mathcal{L}_{\Phi_2}^{\mathrm{eff}} &=& \frac{1}{\mu_2}\Bigg\{
|\lambda_6|^2 \left[ (\Phi_1^\dagger\Phi_1)- \frac{1}{2}\, v^2 \right]^2 (\Phi_1^\dagger\Phi_1)
\nonumber\\ &&\hskip .6cm
+ \frac{\sqrt{2}}{v}\left[ (\Phi_1^\dagger\Phi_1)- \frac{1}{2}\, v^2 \right]
\left( \lambda_6^*\left[
\varsigma_d\,\bar Q_L\mathcal{M}_d \Phi_1 d_R + \varsigma_u\, \bar u_R\mathcal{M}_u \tilde\Phi_1^\dagger Q_L + \varsigma_l\,\bar L_L\mathcal{M}_\ell \Phi_1 \ell_R\right]
+ \mathrm{h.c.}\right)
\nonumber\\ &&\hskip .6cm
+ \frac{2}{v^2} \left[ \varsigma_d\,\bar Q_L^a\mathcal{M}_d d_R 
-\varsigma_u\epsilon_{ab}\,  u_R\mathcal{M}_u Q_L^b
+ \varsigma_l\,\bar L_L^a\mathcal{M}_\ell \ell_R\right]
\nonumber\\ &&\hskip 1.3cm \times
\left[ \varsigma_d^*\,\bar d_R \mathcal{M}_d Q_L^a 
- \varsigma_u^*\epsilon_{ac}\,\bar Q_L^c\mathcal{M}_u  u_R
+ \varsigma_l^*\,\bar \ell_R\mathcal{M}_\ell L_L^a\right]
\Bigg\} \quad + \quad \cO\!\left(\frac{1}{\mu_2^2}\right)\, .
\end{eqnarray}
From this expression, one easily reads the Wilson coefficients of the operators
$Q_{eH}^{pr} = (H^\dagger H) (\bar \ell_p H e_r)$, 
$Q_{\ell e dq}^{prmn} = (\bar\ell_p^a e_r) (\bar d_m q_n^a)$
and $Q_{\ell equ}^{(1),prmn} = (\bar\ell_p^a e_r) \epsilon_{ab} (\bar q_m^b u_n)$:
\begin{eqnarray}
C_{eH}^{11} &=& \frac{\sqrt{2}}{\mu_2 v}\,\lambda_6^*\varsigma_l m_e
\, =\, -\frac{\sqrt{2} m_e}{v^3}\,\varsigma_l\, (\mathcal{R}_{12} + i \mathcal{R}_{13})
\,\dot=\, \frac{\sqrt{2} m_e}{v^3}\,\sum_{i=2}^3 y_l^{H_i} \mathcal{R}_{i1} \, ,
\nonumber\\
C_{\ell e dq}^{113n}&=& \frac{2}{\mu_2 v^2}\,\varsigma_l\,\varsigma_d^* m_e m_b\,\delta_{3n}\, ,
\nonumber\\
C_{\ell e q u}^{(1),11m3}&=& -\frac{2}{\mu_2 v^2}\,\varsigma_l\,\varsigma_u^* m_e m_t\,\delta_{m3}\, .
\end{eqnarray}
%

%% file: draft.bbl
\providecommand{\href}[2]{#2}\begingroup\begin{thebibliography}{100}

\bibitem{Bernreuther:1990jx}
W.~Bernreuther and M.~Suzuki, ``{The electric dipole moment of the electron},''
  \href{http://dx.doi.org/10.1103/RevModPhys.63.313}{{\em Rev. Mod. Phys.}
  {\bfseries 63} (1991) 313--340}. [Erratum: Rev.Mod.Phys. 64, 633 (1992)].

\bibitem{Pospelov:2005pr}
M.~Pospelov and A.~Ritz, ``{Electric dipole moments as probes of new
  physics},'' \href{http://dx.doi.org/10.1016/j.aop.2005.04.002}{{\em Annals
  Phys.} {\bfseries 318} (2005) 119--169},
  \href{http://arxiv.org/abs/hep-ph/0504231}{{\ttfamily arXiv:hep-ph/0504231}}.

\bibitem{Engel:2013lsa}
J.~Engel, M.~J. Ramsey-Musolf, and U.~van Kolck, ``{Electric Dipole Moments of
  Nucleons, Nuclei, and Atoms: The Standard Model and Beyond},''
  \href{http://dx.doi.org/10.1016/j.ppnp.2013.03.003}{{\em Prog. Part. Nucl.
  Phys.} {\bfseries 71} (2013) 21--74},
  \href{http://arxiv.org/abs/1303.2371}{{\ttfamily arXiv:1303.2371 [nucl-th]}}.

\bibitem{Dwivedi:2015nta}
S.~Dwivedi, D.~K. Ghosh, B.~Mukhopadhyaya, and A.~Shivaji, ``{Constraints on
  CP-violating gauge-Higgs operators},''
  \href{http://dx.doi.org/10.1103/PhysRevD.92.095015}{{\em Phys. Rev. D}
  {\bfseries 92} no.~9, (2015) 095015},
  \href{http://arxiv.org/abs/1505.05844}{{\ttfamily arXiv:1505.05844
  [hep-ph]}}.

\bibitem{Ferreira:2016jea}
F.~Ferreira, B.~Fuks, V.~Sanz, and D.~Sengupta, ``{Probing ${CP}$-violating
  Higgs and gauge-boson couplings in the Standard Model effective field
  theory},'' \href{http://dx.doi.org/10.1140/epjc/s10052-017-5226-6}{{\em Eur.
  Phys. J. C} {\bfseries 77} no.~10, (2017) 675},
  \href{http://arxiv.org/abs/1612.01808}{{\ttfamily arXiv:1612.01808
  [hep-ph]}}.

\bibitem{Cirigliano:2019vfc}
V.~Cirigliano, A.~Crivellin, W.~Dekens, J.~de~Vries, M.~Hoferichter, and
  E.~Mereghetti, ``{CP Violation in Higgs-Gauge Interactions: From Tabletop
  Experiments to the LHC},''
  \href{http://dx.doi.org/10.1103/PhysRevLett.123.051801}{{\em Phys. Rev.
  Lett.} {\bfseries 123} no.~5, (2019) 051801},
  \href{http://arxiv.org/abs/1903.03625}{{\ttfamily arXiv:1903.03625
  [hep-ph]}}.

\bibitem{Haisch:2019xyi}
U.~Haisch and A.~Hala, ``{Bounds on CP-violating Higgs-gluon interactions: the
  case of vanishing light-quark Yukawa couplings},''
  \href{http://dx.doi.org/10.1007/JHEP11(2019)117}{{\em JHEP} {\bfseries 11}
  (2019) 117}, \href{http://arxiv.org/abs/1909.09373}{{\ttfamily
  arXiv:1909.09373 [hep-ph]}}.

\bibitem{Rossia:2024rfo}
A.~N. Rossia and E.~Vryonidou, ``{CP-odd effects at NLO in SMEFT WH and ZH
  production},'' \href{http://dx.doi.org/10.1007/JHEP11(2024)142}{{\em JHEP}
  {\bfseries 11} (2024) 142}, \href{http://arxiv.org/abs/2409.00168}{{\ttfamily
  arXiv:2409.00168 [hep-ph]}}.

\bibitem{Asteriadis:2024xts}
K.~Asteriadis, S.~Dawson, P.~P. Giardino, and R.~Szafron, ``{e$^{+}$e$^{-}$
  \textrightarrow{} ZH process in the SMEFT beyond leading order},''
  \href{http://dx.doi.org/10.1007/JHEP02(2025)162}{{\em JHEP} {\bfseries 02}
  (2025) 162}, \href{http://arxiv.org/abs/2409.11466}{{\ttfamily
  arXiv:2409.11466 [hep-ph]}}.

\bibitem{Thomas:2024dwd}
M.~O.~A. Thomas and E.~Vryonidou, ``{CP violation in loop-induced diboson
  production},'' \href{http://dx.doi.org/10.1007/JHEP03(2025)038}{{\em JHEP}
  {\bfseries 03} (2025) 038}, \href{http://arxiv.org/abs/2411.00959}{{\ttfamily
  arXiv:2411.00959 [hep-ph]}}.

\bibitem{Grzadkowski:2010es}
B.~Grzadkowski, M.~Iskrzynski, M.~Misiak, and J.~Rosiek, ``{Dimension-Six Terms
  in the Standard Model Lagrangian},''
  \href{http://dx.doi.org/10.1007/JHEP10(2010)085}{{\em JHEP} {\bfseries 10}
  (2010) 085},
\href{http://arxiv.org/abs/1008.4884}{{\ttfamily arXiv:1008.4884 [hep-ph]}}.

\bibitem{Hisano:2012cc}
J.~Hisano, K.~Tsumura, and M.~J.~S. Yang, ``{QCD Corrections to Neutron
  Electric Dipole Moment from Dimension-six Four-Quark Operators},''
  \href{http://dx.doi.org/10.1016/j.physletb.2012.06.038}{{\em Phys. Lett. B}
  {\bfseries 713} (2012) 473--480},
  \href{http://arxiv.org/abs/1205.2212}{{\ttfamily arXiv:1205.2212 [hep-ph]}}.

\bibitem{Alonso:2013hga}
R.~Alonso, E.~E. Jenkins, A.~V. Manohar, and M.~Trott, ``{Renormalization Group
  Evolution of the Standard Model Dimension Six Operators III: Gauge Coupling
  Dependence and Phenomenology},''
  \href{http://dx.doi.org/10.1007/JHEP04(2014)159}{{\em JHEP} {\bfseries 04}
  (2014) 159}, \href{http://arxiv.org/abs/1312.2014}{{\ttfamily arXiv:1312.2014
  [hep-ph]}}.

\bibitem{Chien:2015xha}
Y.~T. Chien, V.~Cirigliano, W.~Dekens, J.~de~Vries, and E.~Mereghetti,
  ``{Direct and indirect constraints on CP-violating Higgs-quark and
  Higgs-gluon interactions},''
  \href{http://dx.doi.org/10.1007/JHEP02(2016)011}{{\em JHEP} {\bfseries 02}
  (2016) 011}, \href{http://arxiv.org/abs/1510.00725}{{\ttfamily
  arXiv:1510.00725 [hep-ph]}}.

\bibitem{Cirigliano:2016njn}
V.~Cirigliano, W.~Dekens, J.~de~Vries, and E.~Mereghetti, ``{Is there room for
  CP violation in the top-Higgs sector?},''
  \href{http://dx.doi.org/10.1103/PhysRevD.94.016002}{{\em Phys. Rev. D}
  {\bfseries 94} no.~1, (2016) 016002},
  \href{http://arxiv.org/abs/1603.03049}{{\ttfamily arXiv:1603.03049
  [hep-ph]}}.

\bibitem{Cirigliano:2016nyn}
V.~Cirigliano, W.~Dekens, J.~de~Vries, and E.~Mereghetti, ``{Constraining the
  top-Higgs sector of the Standard Model Effective Field Theory},''
  \href{http://dx.doi.org/10.1103/PhysRevD.94.034031}{{\em Phys. Rev. D}
  {\bfseries 94} no.~3, (2016) 034031},
  \href{http://arxiv.org/abs/1605.04311}{{\ttfamily arXiv:1605.04311
  [hep-ph]}}.

\bibitem{Eeg:2016fsy}
J.~O. Eeg, ``{Electric dipole moment of the neutron from a flavor changing
  Higgs boson},'' \href{http://dx.doi.org/10.1140/epjc/s10052-018-6477-6}{{\em
  Eur. Phys. J. C} {\bfseries 78} no.~12, (2018) 998},
  \href{http://arxiv.org/abs/1611.07778}{{\ttfamily arXiv:1611.07778
  [hep-ph]}}.

\bibitem{Panico:2018hal}
G.~Panico, A.~Pomarol, and M.~Riembau, ``{EFT approach to the electron Electric
  Dipole Moment at the two-loop level},''
  \href{http://dx.doi.org/10.1007/JHEP04(2019)090}{{\em JHEP} {\bfseries 04}
  (2019) 090}, \href{http://arxiv.org/abs/1810.09413}{{\ttfamily
  arXiv:1810.09413 [hep-ph]}}.

\bibitem{Jager:2019wkc}
S.~J\"ager, K.~Leslie, and L.~Vale~Silva, ``{Probing the flavor of New Physics
  with dipoles},'' \href{http://dx.doi.org/10.22323/1.364.0248}{{\em PoS}
  {\bfseries EPS-HEP2019} (2020) 248}.

\bibitem{Silva:2020vyx}
L.~V. Silva, S.~J\"ager, and K.~Leslie, ``{Using dipole processes to constrain
  the flavour of four-fermion effective interactions},''
  \href{http://dx.doi.org/10.22323/1.390.0434}{{\em PoS} {\bfseries ICHEP2020}
  (2021) 434}, \href{http://arxiv.org/abs/2012.05630}{{\ttfamily
  arXiv:2012.05630 [hep-ph]}}.

\bibitem{Aebischer:2021uvt}
J.~Aebischer, W.~Dekens, E.~E. Jenkins, A.~V. Manohar, D.~Sengupta, and
  P.~Stoffer, ``{Effective field theory interpretation of lepton magnetic and
  electric dipole moments},''
  \href{http://dx.doi.org/10.1007/JHEP07(2021)107}{{\em JHEP} {\bfseries 07}
  (2021) 107}, \href{http://arxiv.org/abs/2102.08954}{{\ttfamily
  arXiv:2102.08954 [hep-ph]}}.

\bibitem{Haisch:2021hcg}
U.~Haisch and G.~Koole, ``{Beautiful and charming chromodipole moments},''
  \href{http://dx.doi.org/10.1007/JHEP09(2021)133}{{\em JHEP} {\bfseries 09}
  (2021) 133}, \href{http://arxiv.org/abs/2106.01289}{{\ttfamily
  arXiv:2106.01289 [hep-ph]}}.

\bibitem{Kley:2021yhn}
J.~Kley, T.~Theil, E.~Venturini, and A.~Weiler, ``{Electric dipole moments at
  one-loop in the dimension-6 SMEFT},''
  \href{http://dx.doi.org/10.1140/epjc/s10052-022-10861-5}{{\em Eur. Phys. J.
  C} {\bfseries 82} no.~10, (2022) 926},
  \href{http://arxiv.org/abs/2109.15085}{{\ttfamily arXiv:2109.15085
  [hep-ph]}}.

\bibitem{Brod:2022bww}
J.~Brod, J.~M. Cornell, D.~Skodras, and E.~Stamou, ``{Global constraints on
  Yukawa operators in the standard model effective theory},''
  \href{http://dx.doi.org/10.1007/JHEP08(2022)294}{{\em JHEP} {\bfseries 08}
  (2022) 294}, \href{http://arxiv.org/abs/2203.03736}{{\ttfamily
  arXiv:2203.03736 [hep-ph]}}.

\bibitem{Dermisek:2023nhe}
R.~Dermisek, K.~Hermanek, N.~McGinnis, and S.~Yoon, ``{Effective field theory
  of chirally enhanced muon mass and dipole operators},''
  \href{http://dx.doi.org/10.1103/PhysRevD.107.095043}{{\em Phys. Rev. D}
  {\bfseries 107} no.~9, (2023) 095043},
  \href{http://arxiv.org/abs/2302.14144}{{\ttfamily arXiv:2302.14144
  [hep-ph]}}.

\bibitem{Fajfer:2023gie}
S.~Fajfer, J.~F. Kamenik, N.~Ko{\v{s}}nik, A.~Smolkovi{\v{c}}, and M.~Tammaro,
  ``{New Physics in CP violating and flavour changing quark dipole
  transitions},'' \href{http://dx.doi.org/10.1007/JHEP10(2023)133}{{\em JHEP}
  {\bfseries 10} (2023) 133}, \href{http://arxiv.org/abs/2306.16471}{{\ttfamily
  arXiv:2306.16471 [hep-ph]}}.

\bibitem{Miralles:2024huv}
V.~Miralles, Y.~Peters, E.~Vryonidou, and J.~K. Winter, ``{Sensitivity to $
  \mathcal{CP} $-violating effective couplings in the top-Higgs sector},''
  \href{http://dx.doi.org/10.1007/JHEP06(2025)052}{{\em JHEP} {\bfseries 06}
  (2025) 052}, \href{http://arxiv.org/abs/2412.10309}{{\ttfamily
  arXiv:2412.10309 [hep-ph]}}.

\bibitem{Brod:2013cka}
J.~Brod, U.~Haisch, and J.~Zupan, ``{Constraints on CP-violating Higgs
  couplings to the third generation},''
  \href{http://dx.doi.org/10.1007/JHEP11(2013)180}{{\em JHEP} {\bfseries 11}
  (2013) 180}, \href{http://arxiv.org/abs/1310.1385}{{\ttfamily arXiv:1310.1385
  [hep-ph]}}.

\bibitem{Brod:2018pli}
J.~Brod and E.~Stamou, ``{Electric dipole moment constraints on CP-violating
  heavy-quark Yukawas at next-to-leading order},''
  \href{http://dx.doi.org/10.1007/JHEP07(2021)080}{{\em JHEP} {\bfseries 07}
  (2021) 080}, \href{http://arxiv.org/abs/1810.12303}{{\ttfamily
  arXiv:1810.12303 [hep-ph]}}.

\bibitem{Brod:2018lbf}
J.~Brod and D.~Skodras, ``{Electric dipole moment constraints on CP-violating
  light-quark Yukawas},'' \href{http://dx.doi.org/10.1007/JHEP01(2019)233}{{\em
  JHEP} {\bfseries 01} (2019) 233},
  \href{http://arxiv.org/abs/1811.05480}{{\ttfamily arXiv:1811.05480
  [hep-ph]}}.

\bibitem{Brod:2023wsh}
J.~Brod, Z.~Polonsky, and E.~Stamou, ``{A precise electron EDM constraint on
  CP-odd heavy-quark Yukawas},''
  \href{http://dx.doi.org/10.1007/JHEP06(2024)091}{{\em JHEP} {\bfseries 06}
  (2024) 091}, \href{http://arxiv.org/abs/2306.12478}{{\ttfamily
  arXiv:2306.12478 [hep-ph]}}.

\bibitem{Nierste:2019fbx}
U.~Nierste, M.~Tabet, and R.~Ziegler, ``{Cornering Spontaneous $CP$ Violation
  with Charged-Higgs-Boson Searches},''
  \href{http://dx.doi.org/10.1103/PhysRevLett.125.031801}{{\em Phys. Rev.
  Lett.} {\bfseries 125} no.~3, (2020) 031801},
  \href{http://arxiv.org/abs/1912.11501}{{\ttfamily arXiv:1912.11501
  [hep-ph]}}.

\bibitem{Weinberg:1989dx}
S.~Weinberg, ``{Larger Higgs Exchange Terms in the Neutron Electric Dipole
  Moment},'' \href{http://dx.doi.org/10.1103/PhysRevLett.63.2333}{{\em Phys.
  Rev. Lett.} {\bfseries 63} (1989) 2333}.

\bibitem{Barr:1990vd}
S.~M. Barr and A.~Zee, ``{Electric Dipole Moment of the Electron and of the
  Neutron},'' \href{http://dx.doi.org/10.1103/PhysRevLett.65.21}{{\em Phys.
  Rev. Lett.} {\bfseries 65} (1990) 21--24}. [Erratum: Phys.Rev.Lett. 65, 2920
  (1990)].

\bibitem{Pilaftsis:2002fe}
A.~Pilaftsis, ``{Higgs mediated electric dipole moments in the MSSM: An
  application to baryogenesis and Higgs searches},''
  \href{http://dx.doi.org/10.1016/S0550-3213(02)00826-X}{{\em Nucl. Phys. B}
  {\bfseries 644} (2002) 263--289},
  \href{http://arxiv.org/abs/hep-ph/0207277}{{\ttfamily arXiv:hep-ph/0207277}}.

\bibitem{Demir:2003js}
D.~A. Demir, O.~Lebedev, K.~A. Olive, M.~Pospelov, and A.~Ritz, ``{Electric
  dipole moments in the MSSM at large tan beta},''
  \href{http://dx.doi.org/10.1016/j.nuclphysb.2003.12.026}{{\em Nucl. Phys. B}
  {\bfseries 680} (2004) 339--374},
  \href{http://arxiv.org/abs/hep-ph/0311314}{{\ttfamily arXiv:hep-ph/0311314}}.

\bibitem{Ning:2025zfh}
K.~Ning and M.~Ramsey-Musolf, ``{Revisiting the Electron EDM in the NMSSM},''
  \href{http://arxiv.org/abs/2507.06320}{{\ttfamily arXiv:2507.06320
  [hep-ph]}}.

\bibitem{Zhang:2007da}
Y.~Zhang, H.~An, X.~Ji, and R.~N. Mohapatra, ``{General CP Violation in Minimal
  Left-Right Symmetric Model and Constraints on the Right-Handed Scale},''
  \href{http://dx.doi.org/10.1016/j.nuclphysb.2008.05.019}{{\em Nucl. Phys. B}
  {\bfseries 802} (2008) 247--279},
  \href{http://arxiv.org/abs/0712.4218}{{\ttfamily arXiv:0712.4218 [hep-ph]}}.

\bibitem{Bernard:2015boz}
V.~Bernard, S.~Descotes-Genon, and L.~Vale~Silva, ``{Short-distance QCD
  corrections to $ {K}^0{\overline{K}}^0 $ mixing at next-to-leading order in
  Left-Right models},'' \href{http://dx.doi.org/10.1007/JHEP08(2016)128}{{\em
  JHEP} {\bfseries 08} (2016) 128},
  \href{http://arxiv.org/abs/1512.00543}{{\ttfamily arXiv:1512.00543
  [hep-ph]}}.

\bibitem{Bertolini:2019out}
S.~Bertolini, A.~Maiezza, and F.~Nesti, ``{Kaon CP violation and neutron EDM in
  the minimal left-right symmetric model},''
  \href{http://dx.doi.org/10.1103/PhysRevD.101.035036}{{\em Phys. Rev. D}
  {\bfseries 101} no.~3, (2020) 035036},
  \href{http://arxiv.org/abs/1911.09472}{{\ttfamily arXiv:1911.09472
  [hep-ph]}}.

\bibitem{Bernard:2020cyi}
V.~Bernard, S.~Descotes-Genon, and L.~Vale~Silva, ``{Constraining the gauge and
  scalar sectors of the doublet left-right symmetric model},''
  \href{http://dx.doi.org/10.1007/JHEP09(2020)088}{{\em JHEP} {\bfseries 09}
  (2020) 088}, \href{http://arxiv.org/abs/2001.00886}{{\ttfamily
  arXiv:2001.00886 [hep-ph]}}.

\bibitem{Dekens:2018bci}
W.~Dekens, J.~de~Vries, M.~Jung, and K.~K. Vos, ``{The phenomenology of
  electric dipole moments in models of scalar leptoquarks},''
  \href{http://dx.doi.org/10.1007/JHEP01(2019)069}{{\em JHEP} {\bfseries 01}
  (2019) 069}, \href{http://arxiv.org/abs/1809.09114}{{\ttfamily
  arXiv:1809.09114 [hep-ph]}}.

\bibitem{Gisbert:2021htg}
H.~Gisbert, V.~Miralles, and J.~Ruiz-Vidal, ``{Electric dipole moments from
  colour-octet scalars},''
  \href{http://dx.doi.org/10.1007/JHEP04(2022)077}{{\em JHEP} {\bfseries 04}
  (2022) 077}, \href{http://arxiv.org/abs/2111.09397}{{\ttfamily
  arXiv:2111.09397 [hep-ph]}}.

\bibitem{Davidson:2016utf}
S.~Davidson, ``{$\mu \rightarrow e \gamma $ in the 2HDM: an exercise in EFT},''
  \href{http://dx.doi.org/10.1140/epjc/s10052-016-4076-y}{{\em Eur. Phys. J. C}
  {\bfseries 76} no.~5, (2016) 258},
  \href{http://arxiv.org/abs/1601.01949}{{\ttfamily arXiv:1601.01949
  [hep-ph]}}.

\bibitem{Altmannshofer:2020shb}
W.~Altmannshofer, S.~Gori, N.~Hamer, and H.~H. Patel, ``{Electron EDM in the
  complex two-Higgs doublet model},''
  \href{http://dx.doi.org/10.1103/PhysRevD.102.115042}{{\em Phys. Rev. D}
  {\bfseries 102} no.~11, (2020) 115042},
  \href{http://arxiv.org/abs/2009.01258}{{\ttfamily arXiv:2009.01258
  [hep-ph]}}.

\bibitem{Branco:2011iw}
G.~C. Branco, P.~M. Ferreira, L.~Lavoura, M.~N. Rebelo, M.~Sher, and J.~P.
  Silva, ``{Theory and phenomenology of two-Higgs-doublet models},''
  \href{http://dx.doi.org/10.1016/j.physrep.2012.02.002}{{\em Phys. Rept.}
  {\bfseries 516} (2012) 1--102},
  \href{http://arxiv.org/abs/1106.0034}{{\ttfamily arXiv:1106.0034 [hep-ph]}}.

\bibitem{Gunion:1989we}
J.~F. Gunion, H.~E. Haber, G.~L. Kane, and S.~Dawson,
  \href{http://dx.doi.org/10.1201/9780429496448}{{\em {The Higgs Hunter's
  Guide}}}, vol.~80.
\newblock 2000.

\bibitem{Ivanov:2017dad}
I.~P. Ivanov, ``{Building and testing models with extended Higgs sectors},''
  \href{http://dx.doi.org/10.1016/j.ppnp.2017.03.001}{{\em Prog. Part. Nucl.
  Phys.} {\bfseries 95} (2017) 160--208},
  \href{http://arxiv.org/abs/1702.03776}{{\ttfamily arXiv:1702.03776
  [hep-ph]}}.

\bibitem{LopezHonorez:2006gr}
L.~Lopez~Honorez, E.~Nezri, J.~F. Oliver, and M.~H.~G. Tytgat, ``{The Inert
  Doublet Model: An Archetype for Dark Matter},''
  \href{http://dx.doi.org/10.1088/1475-7516/2007/02/028}{{\em JCAP} {\bfseries
  02} (2007) 028}, \href{http://arxiv.org/abs/hep-ph/0612275}{{\ttfamily
  arXiv:hep-ph/0612275}}.

\bibitem{Belyaev:2016lok}
A.~Belyaev, G.~Cacciapaglia, I.~P. Ivanov, F.~Rojas-Abatte, and M.~Thomas,
  ``{Anatomy of the Inert Two Higgs Doublet Model in the light of the LHC and
  non-LHC Dark Matter Searches},''
  \href{http://dx.doi.org/10.1103/PhysRevD.97.035011}{{\em Phys. Rev. D}
  {\bfseries 97} no.~3, (2018) 035011},
  \href{http://arxiv.org/abs/1612.00511}{{\ttfamily arXiv:1612.00511
  [hep-ph]}}.

\bibitem{Tsai:2019eqi}
Y.-L.~S. Tsai, V.~Q. Tran, and C.-T. Lu, ``{Confronting dark matter
  co-annihilation of Inert two Higgs Doublet Model with a compressed mass
  spectrum},'' \href{http://dx.doi.org/10.1007/JHEP06(2020)033}{{\em JHEP}
  {\bfseries 06} (2020) 033}, \href{http://arxiv.org/abs/1912.08875}{{\ttfamily
  arXiv:1912.08875 [hep-ph]}}.

\bibitem{Gunion:2005ja}
J.~F. Gunion and H.~E. Haber, ``{Conditions for CP-violation in the general
  two-Higgs-doublet model},''
  \href{http://dx.doi.org/10.1103/PhysRevD.72.095002}{{\em Phys. Rev. D}
  {\bfseries 72} (2005) 095002},
  \href{http://arxiv.org/abs/hep-ph/0506227}{{\ttfamily arXiv:hep-ph/0506227}}.

\bibitem{Wu:1994ja}
Y.~L. Wu and L.~Wolfenstein, ``{Sources of CP violation in the two Higgs
  doublet model},'' \href{http://dx.doi.org/10.1103/PhysRevLett.73.1762}{{\em
  Phys. Rev. Lett.} {\bfseries 73} (1994) 1762--1764},
  \href{http://arxiv.org/abs/hep-ph/9409421}{{\ttfamily arXiv:hep-ph/9409421}}.

\bibitem{Keus:2015hva}
V.~Keus, S.~F. King, S.~Moretti, and K.~Yagyu, ``{CP Violating
  Two-Higgs-Doublet Model: Constraints and LHC Predictions},''
  \href{http://dx.doi.org/10.1007/JHEP04(2016)048}{{\em JHEP} {\bfseries 04}
  (2016) 048}, \href{http://arxiv.org/abs/1510.04028}{{\ttfamily
  arXiv:1510.04028 [hep-ph]}}.

\bibitem{Chen:2017com}
C.-Y. Chen, H.-L. Li, and M.~Ramsey-Musolf, ``{CP-Violation in the Two Higgs
  Doublet Model: from the LHC to EDMs},''
  \href{http://dx.doi.org/10.1103/PhysRevD.97.015020}{{\em Phys. Rev. D}
  {\bfseries 97} no.~1, (2018) 015020},
  \href{http://arxiv.org/abs/1708.00435}{{\ttfamily arXiv:1708.00435
  [hep-ph]}}.

\bibitem{Iguro:2019zlc}
S.~Iguro and Y.~Omura, ``{The direct CP violation in a general two Higgs
  doublet model},'' \href{http://dx.doi.org/10.1007/JHEP08(2019)098}{{\em JHEP}
  {\bfseries 08} (2019) 098}, \href{http://arxiv.org/abs/1905.11778}{{\ttfamily
  arXiv:1905.11778 [hep-ph]}}.

\bibitem{Chun:2019oix}
E.~J. Chun, J.~Kim, and T.~Mondal, ``{Electron EDM and Muon anomalous magnetic
  moment in Two-Higgs-Doublet Models},''
  \href{http://dx.doi.org/10.1007/JHEP12(2019)068}{{\em JHEP} {\bfseries 12}
  (2019) 068}, \href{http://arxiv.org/abs/1906.00612}{{\ttfamily
  arXiv:1906.00612 [hep-ph]}}.

\bibitem{Cheung:2020ugr}
K.~Cheung, A.~Jueid, Y.-N. Mao, and S.~Moretti, ``{Two-Higgs-doublet model with
  soft $CP$ violation confronting electric dipole moments and colliders},''
  \href{http://dx.doi.org/10.1103/PhysRevD.102.075029}{{\em Phys. Rev. D}
  {\bfseries 102} no.~7, (2020) 075029},
  \href{http://arxiv.org/abs/2003.04178}{{\ttfamily arXiv:2003.04178
  [hep-ph]}}.

\bibitem{Darvishi:2023fjh}
N.~Darvishi, A.~Pilaftsis, and J.-H. Yu, ``{Maximising CP Violation in
  naturally aligned Two-Higgs Doublet Models},''
  \href{http://dx.doi.org/10.1007/JHEP05(2024)233}{{\em JHEP} {\bfseries 05}
  (2024) 233}, \href{http://arxiv.org/abs/2312.00882}{{\ttfamily
  arXiv:2312.00882 [hep-ph]}}.

\bibitem{Kim:1986ax}
J.~E. Kim, ``{Light Pseudoscalars, Particle Physics and Cosmology},''
  \href{http://dx.doi.org/10.1016/0370-1573(87)90017-2}{{\em Phys. Rept.}
  {\bfseries 150} (1987) 1--177}.

\bibitem{Espriu:2015mfa}
D.~Espriu, F.~Mescia, and A.~Renau, ``{Axion-Higgs interplay in the two
  Higgs-doublet model},''
  \href{http://dx.doi.org/10.1103/PhysRevD.92.095013}{{\em Phys. Rev. D}
  {\bfseries 92} no.~9, (2015) 095013},
  \href{http://arxiv.org/abs/1503.02953}{{\ttfamily arXiv:1503.02953
  [hep-ph]}}.

\bibitem{Celis:2014zaa}
A.~Celis, J.~Fuentes-Mart\'\i{}n, and H.~Ser\^odio, ``{Effective Aligned 2HDM
  with a DFSZ-like invisible axion},''
  \href{http://dx.doi.org/10.1016/j.physletb.2014.08.032}{{\em Phys. Lett. B}
  {\bfseries 737} (2014) 185--190},
  \href{http://arxiv.org/abs/1407.0971}{{\ttfamily arXiv:1407.0971 [hep-ph]}}.

\bibitem{Ma:2006km}
E.~Ma, ``{Verifiable radiative seesaw mechanism of neutrino mass and dark
  matter},'' \href{http://dx.doi.org/10.1103/PhysRevD.73.077301}{{\em Phys.
  Rev. D} {\bfseries 73} (2006) 077301},
  \href{http://arxiv.org/abs/hep-ph/0601225}{{\ttfamily arXiv:hep-ph/0601225}}.

\bibitem{Hirsch:2013ola}
M.~Hirsch, R.~A. Lineros, S.~Morisi, J.~Palacio, N.~Rojas, and J.~W.~F. Valle,
  ``{WIMP dark matter as radiative neutrino mass messenger},''
  \href{http://dx.doi.org/10.1007/JHEP10(2013)149}{{\em JHEP} {\bfseries 10}
  (2013) 149}, \href{http://arxiv.org/abs/1307.8134}{{\ttfamily arXiv:1307.8134
  [hep-ph]}}.

\bibitem{Turok:1990zg}
N.~Turok and J.~Zadrozny, ``{Electroweak baryogenesis in the two doublet
  model},'' \href{http://dx.doi.org/10.1016/0550-3213(91)90356-3}{{\em Nucl.
  Phys. B} {\bfseries 358} (1991) 471--493}.

\bibitem{Cline:2011mm}
J.~M. Cline, K.~Kainulainen, and M.~Trott, ``{Electroweak Baryogenesis in Two
  Higgs Doublet Models and B meson anomalies},''
  \href{http://dx.doi.org/10.1007/JHEP11(2011)089}{{\em JHEP} {\bfseries 11}
  (2011) 089}, \href{http://arxiv.org/abs/1107.3559}{{\ttfamily arXiv:1107.3559
  [hep-ph]}}.

\bibitem{Fuyuto:2015jha}
K.~Fuyuto and E.~Senaha, ``{Sphaleron and critical bubble in the scale
  invariant two Higgs doublet model},''
  \href{http://dx.doi.org/10.1016/j.physletb.2015.05.061}{{\em Phys. Lett. B}
  {\bfseries 747} (2015) 152--157},
  \href{http://arxiv.org/abs/1504.04291}{{\ttfamily arXiv:1504.04291
  [hep-ph]}}.

\bibitem{Enomoto:2021dkl}
K.~Enomoto, S.~Kanemura, and Y.~Mura, ``{Electroweak baryogenesis in aligned
  two Higgs doublet models},''
  \href{http://dx.doi.org/10.1007/JHEP01(2022)104}{{\em JHEP} {\bfseries 01}
  (2022) 104}, \href{http://arxiv.org/abs/2111.13079}{{\ttfamily
  arXiv:2111.13079 [hep-ph]}}.

\bibitem{Ferreira:2015rha}
P.~Ferreira, H.~E. Haber, and E.~Santos, ``{Preserving the validity of the
  Two-Higgs Doublet Model up to the Planck scale},''
  \href{http://dx.doi.org/10.1103/PhysRevD.92.033003}{{\em Phys. Rev. D}
  {\bfseries 92} (2015) 033003},
  \href{http://arxiv.org/abs/1505.04001}{{\ttfamily arXiv:1505.04001
  [hep-ph]}}. [Erratum: Phys.Rev.D 94, 059903 (2016)].

\bibitem{Das:2015mwa}
D.~Das and I.~Saha, ``{Search for a stable alignment limit in two-Higgs-doublet
  models},'' \href{http://dx.doi.org/10.1103/PhysRevD.91.095024}{{\em Phys.
  Rev. D} {\bfseries 91} no.~9, (2015) 095024},
  \href{http://arxiv.org/abs/1503.02135}{{\ttfamily arXiv:1503.02135
  [hep-ph]}}.

\bibitem{Schuh:2018hig}
P.~Schuh, ``{Vacuum Stability of Asymptotically Safe Two Higgs Doublet
  Models},'' \href{http://dx.doi.org/10.1140/epjc/s10052-019-7426-8}{{\em Eur.
  Phys. J. C} {\bfseries 79} no.~11, (2019) 909},
  \href{http://arxiv.org/abs/1810.07664}{{\ttfamily arXiv:1810.07664
  [hep-ph]}}.

\bibitem{Glashow:1976nt}
S.~L. Glashow and S.~Weinberg, ``{Natural Conservation Laws for Neutral
  Currents},'' \href{http://dx.doi.org/10.1103/PhysRevD.15.1958}{{\em Phys.
  Rev. D} {\bfseries 15} (1977) 1958}.

\bibitem{Paschos:1976ay}
E.~A. Paschos, ``{Diagonal Neutral Currents},''
  \href{http://dx.doi.org/10.1103/PhysRevD.15.1966}{{\em Phys. Rev. D}
  {\bfseries 15} (1977) 1966}.

\bibitem{Pich:2009sp}
A.~Pich and P.~Tuzon, ``{Yukawa Alignment in the Two-Higgs-Doublet Model},''
  \href{http://dx.doi.org/10.1103/PhysRevD.80.091702}{{\em Phys. Rev. D}
  {\bfseries 80} (2009) 091702},
  \href{http://arxiv.org/abs/0908.1554}{{\ttfamily arXiv:0908.1554 [hep-ph]}}.

\bibitem{Pich:2010ic}
A.~Pich, ``{Flavour constraints on multi-Higgs-doublet models: Yukawa
  alignment},'' \href{http://dx.doi.org/10.1016/j.nuclphysbps.2010.12.030}{{\em
  Nucl. Phys. B Proc. Suppl.} {\bfseries 209} (2010) 182--187},
  \href{http://arxiv.org/abs/1010.5217}{{\ttfamily arXiv:1010.5217 [hep-ph]}}.

\bibitem{Ferreira:2010xe}
P.~M. Ferreira, L.~Lavoura, and J.~P. Silva, ``{Renormalization-group
  constraints on Yukawa alignment in multi-Higgs-doublet models},''
  \href{http://dx.doi.org/10.1016/j.physletb.2010.04.033}{{\em Phys. Lett. B}
  {\bfseries 688} (2010) 341--344},
  \href{http://arxiv.org/abs/1001.2561}{{\ttfamily arXiv:1001.2561 [hep-ph]}}.

\bibitem{Braeuninger:2010td}
C.~B. Braeuninger, A.~Ibarra, and C.~Simonetto, ``{Radiatively induced flavour
  violation in the general two-Higgs doublet model with Yukawa alignment},''
  \href{http://dx.doi.org/10.1016/j.physletb.2010.07.039}{{\em Phys. Lett. B}
  {\bfseries 692} (2010) 189--195},
  \href{http://arxiv.org/abs/1005.5706}{{\ttfamily arXiv:1005.5706 [hep-ph]}}.

\bibitem{Bijnens:2011gd}
J.~Bijnens, J.~Lu, and J.~Rathsman, ``{Constraining General Two Higgs Doublet
  Models by the Evolution of Yukawa Couplings},''
  \href{http://dx.doi.org/10.1007/JHEP05(2012)118}{{\em JHEP} {\bfseries 05}
  (2012) 118}, \href{http://arxiv.org/abs/1111.5760}{{\ttfamily arXiv:1111.5760
  [hep-ph]}}.

\bibitem{Botella:2015yfa}
F.~J. Botella, G.~C. Branco, A.~M. Coutinho, M.~N. Rebelo, and J.~I.
  Silva-Marcos, ``{Natural Quasi-Alignment with two Higgs Doublets and RGE
  Stability},'' \href{http://dx.doi.org/10.1140/epjc/s10052-015-3487-5}{{\em
  Eur. Phys. J. C} {\bfseries 75} (2015) 286},
  \href{http://arxiv.org/abs/1501.07435}{{\ttfamily arXiv:1501.07435
  [hep-ph]}}.

\bibitem{Han:2015yys}
T.~Han, S.~K. Kang, and J.~Sayre, ``{Muon $g-2$ in the aligned two Higgs
  doublet model},'' \href{http://dx.doi.org/10.1007/JHEP02(2016)097}{{\em JHEP}
  {\bfseries 02} (2016) 097}, \href{http://arxiv.org/abs/1511.05162}{{\ttfamily
  arXiv:1511.05162 [hep-ph]}}.

\bibitem{Penuelas:2017ikk}
A.~Pe\~nuelas and A.~Pich, ``{Flavour alignment in multi-Higgs-doublet
  models},'' \href{http://dx.doi.org/10.1007/JHEP12(2017)084}{{\em JHEP}
  {\bfseries 12} (2017) 084}, \href{http://arxiv.org/abs/1710.02040}{{\ttfamily
  arXiv:1710.02040 [hep-ph]}}.

\bibitem{Gori:2017qwg}
S.~Gori, H.~E. Haber, and E.~Santos, ``{High scale flavor alignment in
  two-Higgs doublet models and its phenomenology},''
  \href{http://dx.doi.org/10.1007/JHEP06(2017)110}{{\em JHEP} {\bfseries 06}
  (2017) 110}, \href{http://arxiv.org/abs/1703.05873}{{\ttfamily
  arXiv:1703.05873 [hep-ph]}}.

\bibitem{Jung:2010ik}
M.~Jung, A.~Pich, and P.~Tuzon, ``{Charged-Higgs phenomenology in the Aligned
  two-Higgs-doublet model},''
  \href{http://dx.doi.org/10.1007/JHEP11(2010)003}{{\em JHEP} {\bfseries 11}
  (2010) 003}, \href{http://arxiv.org/abs/1006.0470}{{\ttfamily arXiv:1006.0470
  [hep-ph]}}.

\bibitem{Li:2014fea}
X.-Q. Li, J.~Lu, and A.~Pich, ``{$B_{s,d}^0 \to \ell^+\ell^-$ Decays in the
  Aligned Two-Higgs-Doublet Model},''
  \href{http://dx.doi.org/10.1007/JHEP06(2014)022}{{\em JHEP} {\bfseries 06}
  (2014) 022}, \href{http://arxiv.org/abs/1404.5865}{{\ttfamily arXiv:1404.5865
  [hep-ph]}}.

\bibitem{Abbas:2015cua}
G.~Abbas, A.~Celis, X.-Q. Li, J.~Lu, and A.~Pich, ``{Flavour-changing top
  decays in the aligned two-Higgs-doublet model},''
  \href{http://dx.doi.org/10.1007/JHEP06(2015)005}{{\em JHEP} {\bfseries 06}
  (2015) 005}, \href{http://arxiv.org/abs/1503.06423}{{\ttfamily
  arXiv:1503.06423 [hep-ph]}}.

\bibitem{Celis:2013ixa}
A.~Celis, V.~Ilisie, and A.~Pich, ``{Towards a general analysis of LHC data
  within two-Higgs-doublet models},''
  \href{http://dx.doi.org/10.1007/JHEP12(2013)095}{{\em JHEP} {\bfseries 12}
  (2013) 095}, \href{http://arxiv.org/abs/1310.7941}{{\ttfamily arXiv:1310.7941
  [hep-ph]}}.

\bibitem{Celis:2013rcs}
A.~Celis, V.~Ilisie, and A.~Pich, ``{LHC constraints on two-Higgs doublet
  models},'' \href{http://dx.doi.org/10.1007/JHEP07(2013)053}{{\em JHEP}
  {\bfseries 07} (2013) 053}, \href{http://arxiv.org/abs/1302.4022}{{\ttfamily
  arXiv:1302.4022 [hep-ph]}}.

\bibitem{Jung:2010ab}
M.~Jung, A.~Pich, and P.~Tuzon, ``{The $B \to X_s \gamma$ Rate and CP Asymmetry
  within the Aligned Two-Higgs-Doublet Model},''
  \href{http://dx.doi.org/10.1103/PhysRevD.83.074011}{{\em Phys. Rev. D}
  {\bfseries 83} (2011) 074011},
  \href{http://arxiv.org/abs/1011.5154}{{\ttfamily arXiv:1011.5154 [hep-ph]}}.

\bibitem{Jung:2012vu}
M.~Jung, X.-Q. Li, and A.~Pich, ``{Exclusive radiative B-meson decays within
  the aligned two-Higgs-doublet model},''
  \href{http://dx.doi.org/10.1007/JHEP10(2012)063}{{\em JHEP} {\bfseries 10}
  (2012) 063}, \href{http://arxiv.org/abs/1208.1251}{{\ttfamily arXiv:1208.1251
  [hep-ph]}}.

\bibitem{Kanemura:2022cth}
S.~Kanemura, T.~Mondal, and K.~Yagyu, ``{Exploring wrong sign scenarios in the
  Yukawa-Aligned 2HDM},'' \href{http://dx.doi.org/10.1007/JHEP02(2023)237}{{\em
  JHEP} {\bfseries 02} (2023) 237},
  \href{http://arxiv.org/abs/2211.08803}{{\ttfamily arXiv:2211.08803
  [hep-ph]}}.

\bibitem{Kanemura:2020ibp}
S.~Kanemura, M.~Kubota, and K.~Yagyu, ``{Aligned CP-violating Higgs sector
  canceling the electric dipole moment},''
  \href{http://dx.doi.org/10.1007/JHEP08(2020)026}{{\em JHEP} {\bfseries 08}
  (2020) 026}, \href{http://arxiv.org/abs/2004.03943}{{\ttfamily
  arXiv:2004.03943 [hep-ph]}}.

\bibitem{Iguro:2023tbk}
S.~Iguro, T.~Kitahara, M.~S. Lang, and M.~Takeuchi, ``{Current status of the
  muon g-2 interpretations within two-Higgs-doublet models},''
  \href{http://dx.doi.org/10.1103/PhysRevD.108.115012}{{\em Phys. Rev. D}
  {\bfseries 108} no.~11, (2023) 115012},
  \href{http://arxiv.org/abs/2304.09887}{{\ttfamily arXiv:2304.09887
  [hep-ph]}}.

\bibitem{Ilisie:2014hea}
V.~Ilisie and A.~Pich, ``{Low-mass fermiophobic charged Higgs phenomenology in
  two-Higgs-doublet models},''
  \href{http://dx.doi.org/10.1007/JHEP09(2014)089}{{\em JHEP} {\bfseries 09}
  (2014) 089}, \href{http://arxiv.org/abs/1405.6639}{{\ttfamily arXiv:1405.6639
  [hep-ph]}}.

\bibitem{Abbas:2018pfp}
G.~Abbas, D.~Das, and M.~Patra, ``{Loop induced $H^\pm \to W^\pm Z$ decays in
  the aligned two-Higgs-doublet model},''
  \href{http://dx.doi.org/10.1103/PhysRevD.98.115013}{{\em Phys. Rev. D}
  {\bfseries 98} no.~11, (2018) 115013},
  \href{http://arxiv.org/abs/1806.11035}{{\ttfamily arXiv:1806.11035
  [hep-ph]}}.

\bibitem{Cai:2024xjq}
F.-M. Cai, R.-L. Fan, X.-Q. Li, and Y.-D. Yang, ``{CP asymmetries of $t
  \rightarrow c \gamma $ and $t \rightarrow cg$ decays in the aligned
  two-Higgs-doublet model},''
  \href{http://dx.doi.org/10.1140/epjc/s10052-025-13965-w}{{\em Eur. Phys. J.
  C} {\bfseries 85} no.~3, (2025) 285},
  \href{http://arxiv.org/abs/2409.04179}{{\ttfamily arXiv:2409.04179
  [hep-ph]}}.

\bibitem{Connell:2023jqq}
J.~M. Connell, P.~Ferreira, and H.~E. Haber, ``{Accommodating hints of new
  heavy scalars in the framework of the flavor-aligned two-Higgs-doublet
  model},'' \href{http://dx.doi.org/10.1103/PhysRevD.108.055031}{{\em Phys.
  Rev. D} {\bfseries 108} no.~5, (2023) 055031},
  \href{http://arxiv.org/abs/2302.13697}{{\ttfamily arXiv:2302.13697
  [hep-ph]}}.

\bibitem{Eberhardt:2020dat}
O.~Eberhardt, A.~P. Mart\'\i{}nez, and A.~Pich, ``{Global fits in the Aligned
  Two-Higgs-Doublet model},''
  \href{http://dx.doi.org/10.1007/JHEP05(2021)005}{{\em JHEP} {\bfseries 05}
  (2021) 005}, \href{http://arxiv.org/abs/2012.09200}{{\ttfamily
  arXiv:2012.09200 [hep-ph]}}.

\bibitem{Karan:2023kyj}
A.~Karan, V.~Miralles, and A.~Pich, ``{Updated global fit of the aligned
  two-Higgs-doublet model with heavy scalars},''
  \href{http://dx.doi.org/10.1103/PhysRevD.109.035012}{{\em Phys. Rev. D}
  {\bfseries 109} no.~3, (2024) 035012},
  \href{http://arxiv.org/abs/2307.15419}{{\ttfamily arXiv:2307.15419
  [hep-ph]}}.

\bibitem{Karan:2023xze}
A.~Karan, V.~Miralles, and A.~Pich, ``{Aligned two Higgs doublet model and the
  global fits},'' \href{http://dx.doi.org/10.22323/1.449.0053}{{\em PoS}
  {\bfseries EPS-HEP2023} (2024) 053},
  \href{http://arxiv.org/abs/2312.00514}{{\ttfamily arXiv:2312.00514
  [hep-ph]}}.

\bibitem{Karan:2024kgr}
A.~Karan, A.~M. Coutinho, V.~Miralles, and A.~Pich, ``{Status of the Aligned
  two Higgs doublet model in the low mass region},''
  \href{http://dx.doi.org/10.22323/1.476.0075}{{\em PoS} {\bfseries ICHEP2024}
  (2025) 075}, \href{http://arxiv.org/abs/2409.14934}{{\ttfamily
  arXiv:2409.14934 [hep-ph]}}.

\bibitem{Coutinho:2024vzm}
A.~M. Coutinho, A.~Karan, V.~Miralles, and A.~Pich, ``{Bayesian analyses of the
  Aligned-Two-Higgs-Doublet Model with low-mass scalars},''
  \href{http://dx.doi.org/10.22323/1.478.0238}{{\em PoS} {\bfseries LHCP2024}
  (2025) 238}, \href{http://arxiv.org/abs/2410.22274}{{\ttfamily
  arXiv:2410.22274 [hep-ph]}}.

\bibitem{Banik:2024ugs}
S.~Banik, G.~Coloretti, A.~Crivellin, and H.~E. Haber, ``{Correlating
  A\textrightarrow{}\ensuremath{\gamma}\ensuremath{\gamma} with electric dipole
  moments in the two Higgs doublet model in light of the diphoton excesses at
  95~GeV and 152~GeV},''
  \href{http://dx.doi.org/10.1103/PhysRevD.111.075021}{{\em Phys. Rev. D}
  {\bfseries 111} no.~7, (2025) 075021},
  \href{http://arxiv.org/abs/2412.00523}{{\ttfamily arXiv:2412.00523
  [hep-ph]}}.

\bibitem{Coutinho:2024zyp}
A.~M. Coutinho, A.~Karan, V.~Miralles, and A.~Pich, ``{Light scalars within the
  $ \mathcal{CP} $-conserving Aligned-two-Higgs-doublet model},''
  \href{http://dx.doi.org/10.1007/JHEP02(2025)057}{{\em JHEP} {\bfseries 02}
  (2025) 057}, \href{http://arxiv.org/abs/2412.14906}{{\ttfamily
  arXiv:2412.14906 [hep-ph]}}.

\bibitem{Heo:2025vkz}
Y.~Heo, J.~S. Lee, and C.~B. Park, ``{Higgs boson precision analysis of
  two-Higgs-doublet models: Full LHC run 1 and run 2 data},''
  \href{http://dx.doi.org/10.1103/PhysRevD.111.075026}{{\em Phys. Rev. D}
  {\bfseries 111} no.~7, (2025) 075026},
  \href{http://arxiv.org/abs/2502.02992}{{\ttfamily arXiv:2502.02992
  [hep-ph]}}.

\bibitem{Eeg:2019eei}
J.~O. Eeg, ``{Electric dipole moment of the neutron in two-Higgs-doublet models
  with flavor changing couplings},''
  \href{http://dx.doi.org/10.1103/PhysRevD.102.095009}{{\em Phys. Rev. D}
  {\bfseries 102} no.~9, (2020) 095009},
  \href{http://arxiv.org/abs/1911.07291}{{\ttfamily arXiv:1911.07291
  [hep-ph]}}.

\bibitem{Altmannshofer:2025nsl}
W.~Altmannshofer, B.~Assi, J.~Brod, N.~Hamer, J.~Julio, P.~Uttayarat, and
  D.~Volkov, ``{Electron EDM and {\ensuremath{\Gamma}}({\ensuremath{\mu}}
  {\textrightarrow} e{\ensuremath{\gamma}}) in the 2HDM},''
  \href{http://dx.doi.org/10.1007/JHEP06(2025)156}{{\em JHEP} {\bfseries 06}
  (2025) 156}, \href{http://arxiv.org/abs/2410.17313}{{\ttfamily
  arXiv:2410.17313 [hep-ph]}}.

\bibitem{Huet:1994jb}
P.~Huet and E.~Sather, ``{Electroweak baryogenesis and standard model CP
  violation},'' \href{http://dx.doi.org/10.1103/PhysRevD.51.379}{{\em Phys.
  Rev. D} {\bfseries 51} (1995) 379--394},
  \href{http://arxiv.org/abs/hep-ph/9404302}{{\ttfamily arXiv:hep-ph/9404302}}.

\bibitem{deLima:2024lfc}
C.~H. de~Lima and H.~E. Logan, ``{Is the Real Two-Higgs-Doublet Model
  Consistent?},'' \href{http://dx.doi.org/10.1103/PhysRevLett.133.201801}{{\em
  Phys. Rev. Lett.} {\bfseries 133} no.~20, (2024) 201801},
  \href{http://arxiv.org/abs/2409.10603}{{\ttfamily arXiv:2409.10603
  [hep-ph]}}.

\bibitem{Fontes:2021znm}
D.~Fontes, M.~L\"oschner, J.~C. Rom\~ao, and J.~a.~P. Silva, ``{Leaks of CP
  violation in the real two-Higgs-doublet model},''
  \href{http://dx.doi.org/10.1140/epjc/s10052-021-09332-0}{{\em Eur. Phys. J.
  C} {\bfseries 81} no.~6, (2021) 541},
  \href{http://arxiv.org/abs/2103.05002}{{\ttfamily arXiv:2103.05002
  [hep-ph]}}.

\bibitem{deLima:2024hnk}
C.~H. de~Lima and H.~E. Logan, ``{Can CP be conserved in the two-Higgs-doublet
  model?},'' \href{http://dx.doi.org/10.1103/PhysRevD.110.095007}{{\em Phys.
  Rev. D} {\bfseries 110} no.~9, (2024) 095007},
  \href{http://arxiv.org/abs/2403.17052}{{\ttfamily arXiv:2403.17052
  [hep-ph]}}.

\bibitem{Dicus:1989va}
D.~A. Dicus, ``{Neutron Electric Dipole Moment From Charged Higgs Exchange},''
  \href{http://dx.doi.org/10.1103/PhysRevD.41.999}{{\em Phys. Rev. D}
  {\bfseries 41} (1990) 999}.

\bibitem{Egana-Ugrinovic:2015vgy}
D.~Egana-Ugrinovic and S.~Thomas, ``{Effective Theory of Higgs Sector Vacuum
  States},'' \href{http://arxiv.org/abs/1512.00144}{{\ttfamily arXiv:1512.00144
  [hep-ph]}}.

\bibitem{Crivellin:2016ihg}
A.~Crivellin, M.~Ghezzi, and M.~Procura, ``{Effective Field Theory with Two
  Higgs Doublets},'' \href{http://dx.doi.org/10.1007/JHEP09(2016)160}{{\em
  JHEP} {\bfseries 09} (2016) 160},
  \href{http://arxiv.org/abs/1608.00975}{{\ttfamily arXiv:1608.00975
  [hep-ph]}}.

\bibitem{Dawson:2022cmu}
S.~Dawson, D.~Fontes, S.~Homiller, and M.~Sullivan, ``{Role of dimension-eight
  operators in an EFT for the 2HDM},''
  \href{http://dx.doi.org/10.1103/PhysRevD.106.055012}{{\em Phys. Rev. D}
  {\bfseries 106} no.~5, (2022) 055012},
  \href{http://arxiv.org/abs/2205.01561}{{\ttfamily arXiv:2205.01561
  [hep-ph]}}.

\bibitem{Dawson:2023ebe}
S.~Dawson, D.~Fontes, C.~Quezada-Calonge, and J.~J. Sanz-Cillero, ``{Matching
  the 2HDM to the HEFT and the SMEFT: Decoupling and perturbativity},''
  \href{http://dx.doi.org/10.1103/PhysRevD.108.055034}{{\em Phys. Rev. D}
  {\bfseries 108} no.~5, (2023) 055034},
  \href{http://arxiv.org/abs/2305.07689}{{\ttfamily arXiv:2305.07689
  [hep-ph]}}.

\bibitem{DasBakshi:2024krs}
S.~Das~Bakshi, S.~Dawson, D.~Fontes, and S.~Homiller, ``{Relevance of one-loop
  SMEFT matching in the 2HDM},''
  \href{http://dx.doi.org/10.1103/PhysRevD.109.075022}{{\em Phys. Rev. D}
  {\bfseries 109} no.~7, (2024) 075022},
  \href{http://arxiv.org/abs/2401.12279}{{\ttfamily arXiv:2401.12279
  [hep-ph]}}.

\bibitem{Dermisek:2024ohe}
R.~Dermisek and K.~Hermanek, ``{Two-Higgs-doublet model effective field
  theory},'' \href{http://dx.doi.org/10.1103/PhysRevD.110.035026}{{\em Phys.
  Rev. D} {\bfseries 110} no.~3, (2024) 035026},
  \href{http://arxiv.org/abs/2405.20511}{{\ttfamily arXiv:2405.20511
  [hep-ph]}}.

\bibitem{Dermisek:2024btn}
R.~Dermisek and K.~Hermanek, ``{Feynman Rules in the Two-Higgs Doublet Model
  Effective Field Theory},'' \href{http://arxiv.org/abs/2411.07337}{{\ttfamily
  arXiv:2411.07337 [hep-ph]}}.

\bibitem{Davila:2023fkk}
J.~M. D\'avila, D.~Domenech, M.~J. Herrero, and R.~A. Morales, ``{Exploring
  correlations between HEFT Higgs couplings $\kappa _V$ and $\kappa _{2V}$ via
  HH production at $e^+e^-$ colliders},''
  \href{http://dx.doi.org/10.1140/epjc/s10052-024-12815-5}{{\em Eur. Phys. J.
  C} {\bfseries 84} no.~5, (2024) 503},
  \href{http://arxiv.org/abs/2312.03877}{{\ttfamily arXiv:2312.03877
  [hep-ph]}}.

\bibitem{Bowser-Chao:1997kjp}
D.~Bowser-Chao, D.~Chang, and W.-Y. Keung, ``{Electron electric dipole moment
  from CP violation in the charged Higgs sector},''
  \href{http://dx.doi.org/10.1103/PhysRevLett.79.1988}{{\em Phys. Rev. Lett.}
  {\bfseries 79} (1997) 1988--1991},
  \href{http://arxiv.org/abs/hep-ph/9703435}{{\ttfamily arXiv:hep-ph/9703435}}.

\bibitem{Jung:2013hka}
M.~Jung and A.~Pich, ``{Electric Dipole Moments in Two-Higgs-Doublet Models},''
  \href{http://dx.doi.org/10.1007/JHEP04(2014)076}{{\em JHEP} {\bfseries 04}
  (2014) 076}, \href{http://arxiv.org/abs/1308.6283}{{\ttfamily arXiv:1308.6283
  [hep-ph]}}.

\bibitem{EliasMiro:2020tdv}
J.~Elias~Mir\'o, J.~Ingoldby, and M.~Riembau, ``{EFT anomalous dimensions from
  the S-matrix},'' \href{http://dx.doi.org/10.1007/JHEP09(2020)163}{{\em JHEP}
  {\bfseries 09} (2020) 163}, \href{http://arxiv.org/abs/2005.06983}{{\ttfamily
  arXiv:2005.06983 [hep-ph]}}.

\bibitem{Egana-Ugrinovic:2018fpy}
D.~Egana-Ugrinovic and S.~Thomas, ``{Higgs Boson Contributions to the Electron
  Electric Dipole Moment},'' \href{http://arxiv.org/abs/1810.08631}{{\ttfamily
  arXiv:1810.08631 [hep-ph]}}.

\bibitem{Ellis:2008zy}
J.~R. Ellis, J.~S. Lee, and A.~Pilaftsis, ``{Electric Dipole Moments in the
  MSSM Reloaded},'' \href{http://dx.doi.org/10.1088/1126-6708/2008/10/049}{{\em
  JHEP} {\bfseries 10} (2008) 049},
  \href{http://arxiv.org/abs/0808.1819}{{\ttfamily arXiv:0808.1819 [hep-ph]}}.

\bibitem{Yamanaka:2017mef}
N.~Yamanaka, B.~K. Sahoo, N.~Yoshinaga, T.~Sato, K.~Asahi, and B.~P. Das,
  ``{Probing exotic phenomena at the interface of nuclear and particle physics
  with the electric dipole moments of diamagnetic atoms: A unique window to
  hadronic and semi-leptonic CP violation},''
  \href{http://dx.doi.org/10.1140/epja/i2017-12237-2}{{\em Eur. Phys. J. A}
  {\bfseries 53} no.~3, (2017) 54},
  \href{http://arxiv.org/abs/1703.01570}{{\ttfamily arXiv:1703.01570
  [hep-ph]}}.

\bibitem{Yanase:2018qqq}
K.~Yanase, N.~Yoshinaga, K.~Higashiyama, and N.~Yamanaka, ``{Electric dipole
  moment of $^{199}$Hg atom from $P$, $CP$-odd electron-nucleon interaction},''
  \href{http://dx.doi.org/10.1103/PhysRevD.99.075021}{{\em Phys. Rev. D}
  {\bfseries 99} no.~7, (2019) 075021},
  \href{http://arxiv.org/abs/1805.00419}{{\ttfamily arXiv:1805.00419
  [nucl-th]}}.

\bibitem{Ardu:2025rqy}
M.~Ardu and N.~Valori, ``{The equivalent Electric Dipole Moment in SMEFT},''
  \href{http://arxiv.org/abs/2503.21920}{{\ttfamily arXiv:2503.21920
  [hep-ph]}}.

\bibitem{Shifman:1978zn}
M.~A. Shifman, A.~I. Vainshtein, and V.~I. Zakharov, ``{Remarks on Higgs Boson
  Interactions with Nucleons},''
  \href{http://dx.doi.org/10.1016/0370-2693(78)90481-1}{{\em Phys. Lett. B}
  {\bfseries 78} (1978) 443--446}.

\bibitem{Donoghue:1990xh}
J.~F. Donoghue, J.~Gasser, and H.~Leutwyler, ``{The Decay of a Light Higgs
  Boson},'' \href{http://dx.doi.org/10.1016/0550-3213(90)90474-R}{{\em Nucl.
  Phys. B} {\bfseries 343} (1990) 341--368}.

\bibitem{Prades:1990vn}
J.~Prades and A.~Pich, ``{The Decay $\eta \to \pi^0 h^0$ in Two Higgs Doublet
  Models With a Light Scalar},''
  \href{http://dx.doi.org/10.1016/0370-2693(90)90174-5}{{\em Phys. Lett. B}
  {\bfseries 245} (1990) 117--121}.

\bibitem{Celis:2013xja}
A.~Celis, V.~Cirigliano, and E.~Passemar, ``{Lepton flavor violation in the
  Higgs sector and the role of hadronic $\tau$-lepton decays},''
  \href{http://dx.doi.org/10.1103/PhysRevD.89.013008}{{\em Phys. Rev. D}
  {\bfseries 89} (2014) 013008},
  \href{http://arxiv.org/abs/1309.3564}{{\ttfamily arXiv:1309.3564 [hep-ph]}}.

\bibitem{Cheung:2019bkw}
K.~Cheung, W.-Y. Keung, Y.-n. Mao, and C.~Zhang, ``{Constraining CP-violating
  electron-gluonic operators},''
  \href{http://dx.doi.org/10.1007/JHEP07(2019)074}{{\em JHEP} {\bfseries 07}
  (2019) 074}, \href{http://arxiv.org/abs/1904.10808}{{\ttfamily
  arXiv:1904.10808 [hep-ph]}}.

\bibitem{Bhattacharya:2015rsa}
T.~Bhattacharya, V.~Cirigliano, R.~Gupta, E.~Mereghetti, and B.~Yoon,
  ``{Dimension-5 CP-odd operators: QCD mixing and renormalization},''
  \href{http://dx.doi.org/10.1103/PhysRevD.92.114026}{{\em Phys. Rev. D}
  {\bfseries 92} no.~11, (2015) 114026},
  \href{http://arxiv.org/abs/1502.07325}{{\ttfamily arXiv:1502.07325
  [hep-ph]}}.

\bibitem{Bisal:2024nbb}
S.~Bisal, ``{Two-loop contributions to the anomalous chromomagnetic dipole
  moment of the top quark in two-Higgs-doublet models},''
  \href{http://dx.doi.org/10.1016/j.physletb.2024.138848}{{\em Phys. Lett. B}
  {\bfseries 855} (2024) 138848},
  \href{http://arxiv.org/abs/2404.14065}{{\ttfamily arXiv:2404.14065
  [hep-ph]}}.

\bibitem{Degenkolb:2024eve}
S.~Degenkolb, N.~Elmer, T.~Modak, M.~M\"uhlleitner, and T.~Plehn, ``{A Global
  View of the EDM Landscape},''
  \href{http://arxiv.org/abs/2403.02052}{{\ttfamily arXiv:2403.02052
  [hep-ph]}}.

\bibitem{Pospelov:1991zt}
M.~E. Pospelov and I.~B. Khriplovich, ``{Electric dipole moment of the W boson
  and the electron in the Kobayashi-Maskawa model},'' {\em Sov. J. Nucl. Phys.}
  {\bfseries 53} (1991) 638--640.

\bibitem{Yamaguchi:2020eub}
Y.~Yamaguchi and N.~Yamanaka, ``{Large long-distance contributions to the
  electric dipole moments of charged leptons in the standard model},''
  \href{http://dx.doi.org/10.1103/PhysRevLett.125.241802}{{\em Phys. Rev.
  Lett.} {\bfseries 125} (2020) 241802},
  \href{http://arxiv.org/abs/2003.08195}{{\ttfamily arXiv:2003.08195
  [hep-ph]}}.

\bibitem{Yamaguchi:2020dsy}
Y.~Yamaguchi and N.~Yamanaka, ``{Quark level and hadronic contributions to the
  electric dipole moment of charged leptons in the standard model},''
  \href{http://dx.doi.org/10.1103/PhysRevD.103.013001}{{\em Phys. Rev. D}
  {\bfseries 103} no.~1, (2021) 013001},
  \href{http://arxiv.org/abs/2006.00281}{{\ttfamily arXiv:2006.00281
  [hep-ph]}}.

\bibitem{Ng:1995cs}
D.~Ng and J.~N. Ng, ``{A Note on Majorana neutrinos, leptonic CKM and electron
  electric dipole moment},''
  \href{http://dx.doi.org/10.1142/S0217732396000254}{{\em Mod. Phys. Lett. A}
  {\bfseries 11} (1996) 211--216},
  \href{http://arxiv.org/abs/hep-ph/9510306}{{\ttfamily arXiv:hep-ph/9510306}}.

\bibitem{Archambault:2004td}
J.~P. Archambault, A.~Czarnecki, and M.~Pospelov, ``{Electric dipole moments of
  leptons in the presence of majorana neutrinos},''
  \href{http://dx.doi.org/10.1103/PhysRevD.70.073006}{{\em Phys. Rev. D}
  {\bfseries 70} (2004) 073006},
  \href{http://arxiv.org/abs/hep-ph/0406089}{{\ttfamily arXiv:hep-ph/0406089}}.

\bibitem{Roussy:2022cmp}
T.~S. Roussy {\em et~al.}, ``{An improved bound on the
  electron\textquoteright{}s electric dipole moment},''
  \href{http://dx.doi.org/10.1126/science.adg4084}{{\em Science} {\bfseries
  381} no.~6653, (2023) adg4084},
  \href{http://arxiv.org/abs/2212.11841}{{\ttfamily arXiv:2212.11841
  [physics.atom-ph]}}.

\bibitem{Pospelov:2013sca}
M.~Pospelov and A.~Ritz, ``{CKM benchmarks for electron electric dipole moment
  experiments},'' \href{http://dx.doi.org/10.1103/PhysRevD.89.056006}{{\em
  Phys. Rev. D} {\bfseries 89} no.~5, (2014) 056006},
  \href{http://arxiv.org/abs/1311.5537}{{\ttfamily arXiv:1311.5537 [hep-ph]}}.

\bibitem{Ema:2022yra}
Y.~Ema, T.~Gao, and M.~Pospelov, ``{Standard Model Prediction for Paramagnetic
  Electric Dipole Moments},''
  \href{http://dx.doi.org/10.1103/PhysRevLett.129.231801}{{\em Phys. Rev.
  Lett.} {\bfseries 129} no.~23, (2022) 231801},
  \href{http://arxiv.org/abs/2202.10524}{{\ttfamily arXiv:2202.10524
  [hep-ph]}}.

\bibitem{ACME:2018yjb}
{\bfseries ACME} Collaboration, V.~Andreev {\em et~al.}, ``{Improved limit on
  the electric dipole moment of the electron},''
  \href{http://dx.doi.org/10.1038/s41586-018-0599-8}{{\em Nature} {\bfseries
  562} no.~7727, (2018) 355--360}.

\bibitem{Abe:2013qla}
T.~Abe, J.~Hisano, T.~Kitahara, and K.~Tobioka, ``{Gauge invariant Barr-Zee
  type contributions to fermionic EDMs in the two-Higgs doublet models},''
  \href{http://dx.doi.org/10.1007/JHEP01(2014)106}{{\em JHEP} {\bfseries 01}
  (2014) 106}, \href{http://arxiv.org/abs/1311.4704}{{\ttfamily arXiv:1311.4704
  [hep-ph]}}. [Erratum: JHEP 04, 161 (2016)].

\bibitem{Papavassiliou:1989zd}
J.~Papavassiliou, ``{Gauge Invariant Proper Selfenergies and Vertices in Gauge
  Theories with Broken Symmetry},''
  \href{http://dx.doi.org/10.1103/PhysRevD.41.3179}{{\em Phys. Rev. D}
  {\bfseries 41} (1990) 3179}.

\bibitem{Degrassi:1992ue}
G.~Degrassi and A.~Sirlin, ``{Gauge invariant selfenergies and vertex parts of
  the Standard Model in the pinch technique framework},''
  \href{http://dx.doi.org/10.1103/PhysRevD.46.3104}{{\em Phys. Rev. D}
  {\bfseries 46} (1992) 3104--3116}.

\bibitem{Papavassiliou:1994pr}
J.~Papavassiliou, ``{Gauge independent transverse and longitudinal self
  energies and vertices via the pinch technique},''
  \href{http://dx.doi.org/10.1103/PhysRevD.50.5958}{{\em Phys. Rev. D}
  {\bfseries 50} (1994) 5958--5970},
  \href{http://arxiv.org/abs/hep-ph/9406258}{{\ttfamily arXiv:hep-ph/9406258}}.

\bibitem{Cornwall:2010upa}
J.~M. Cornwall, J.~Papavassiliou, and D.~Binosi,
  \href{http://dx.doi.org/10.1017/9781009402415}{{\em {The Pinch Technique and
  its Applications to Non-Abelian Gauge Theories}}}.
\newblock Cambridge University Press, 2011.

\bibitem{Weinberg:1996kr}
S.~Weinberg, {\em {The quantum theory of fields. Vol. 2: Modern applications}}.
\newblock Cambridge University Press, August, 2013.

\bibitem{ValeSilva:2022nzm}
L.~Vale~Silva, ``{Probing the Flavour of New Physics with Dipoles},''
  \href{http://dx.doi.org/10.3103/S0027134922021016}{{\em Moscow Univ. Phys.
  Bull.} {\bfseries 77} no.~2, (2022) 152--155}.

\bibitem{Jenkins:2013zja}
E.~E. Jenkins, A.~V. Manohar, and M.~Trott, ``{Renormalization Group Evolution
  of the Standard Model Dimension Six Operators I: Formalism and lambda
  Dependence},'' \href{http://dx.doi.org/10.1007/JHEP10(2013)087}{{\em JHEP}
  {\bfseries 10} (2013) 087}, \href{http://arxiv.org/abs/1308.2627}{{\ttfamily
  arXiv:1308.2627 [hep-ph]}}.

\bibitem{Bian:2014zka}
L.~Bian, T.~Liu, and J.~Shu, ``{Cancellations Between Two-Loop Contributions to
  the Electron Electric Dipole Moment with a CP-Violating Higgs Sector},''
  \href{http://dx.doi.org/10.1103/PhysRevLett.115.021801}{{\em Phys. Rev.
  Lett.} {\bfseries 115} (2015) 021801},
  \href{http://arxiv.org/abs/1411.6695}{{\ttfamily arXiv:1411.6695 [hep-ph]}}.

\bibitem{Fuyuto_2020}
K.~Fuyuto, W.-S. Hou, and E.~Senaha, ``Cancellation mechanism for the electron
  electric dipole moment connected with the baryon asymmetry of the universe,''
  \href{http://dx.doi.org/10.1103/physrevd.101.011901}{{\em Physical Review D}
  {\bfseries 101} no.~1, (Jan., 2020) }.
  \url{http://dx.doi.org/10.1103/PhysRevD.101.011901}.

\bibitem{Nowakowski:2004cv}
M.~Nowakowski, E.~A. Paschos, and J.~M. Rodriguez, ``{All electromagnetic
  form-factors},'' \href{http://dx.doi.org/10.1088/0143-0807/26/4/001}{{\em
  Eur. J. Phys.} {\bfseries 26} (2005) 545--560},
  \href{http://arxiv.org/abs/physics/0402058}{{\ttfamily
  arXiv:physics/0402058}}.

\bibitem{Kanemura:2024ezz}
S.~Kanemura and Y.~Mura, ``{Loop induced H$^{\pm}$W$^{\mp}$Z vertices in the
  general two Higgs doublet model with CP violation},''
  \href{http://dx.doi.org/10.1007/JHEP10(2024)041}{{\em JHEP} {\bfseries 10}
  (2024) 041}, \href{http://arxiv.org/abs/2408.06863}{{\ttfamily
  arXiv:2408.06863 [hep-ph]}}.

\end{thebibliography}\endgroup
